\documentclass[aps,prd,amsmath,floats,floatfix, twocolumn, superscriptaddress,nofootinbib,showpacs]{revtex4-1}
\usepackage[usenames, dvipsnames]{color}
\usepackage[colorlinks, pdfborder={0 0 0}, plainpages=false]{hyperref}
\usepackage{graphicx}
\usepackage{xspace}
\usepackage[usenames,dvipsnames]{color}
\usepackage{amssymb}
\usepackage[normalem]{ulem} 
\usepackage{letltxmacro}
\usepackage{amsfonts}
\usepackage{amsmath}
\usepackage{amssymb}
\usepackage{bm}
\usepackage{dcolumn}
\usepackage{epsfig}
\usepackage{graphicx}
\usepackage{graphics}
\usepackage[latin1]{inputenc}
\usepackage{latexsym}
\usepackage{rotating}
\usepackage{hyperref}
\usepackage[usenames]{color}
\usepackage{float}
\usepackage{xspace} 
\usepackage{mathrsfs}
\usepackage{subfigure}
\usepackage{multirow}

\def\prd{{\it Phys. Rev.} D~}

\def\apjl{{\it Astrophys. J.} Lett~}
\def\apj{{\it Astrophys. J.}}
\def\Msun{M_\odot}
\def\PRD{{\it Phys. Rev.} D~}

\def\mnras{{\it MNRAS~}} 

\def\aap{{\it A\&A~}}

\def\lesssim{\mathrel{\hbox{\rlap{\hbox{\lower4pt\hbox{$\sim$}}}\hbox{$<$}}}}
\def\gtrsim{\mathrel{\hbox{\rlap{\hbox{\lower4pt\hbox{$\sim$}}}\hbox{$>$}}}}
\def\alt{\mathrel{\hbox{\rlap{\hbox{\lower4pt\hbox{$\sim$}}}\hbox{$<$}}}}
\def\agt{\mathrel{\hbox{\rlap{\hbox{\lower4pt\hbox{$\sim$}}}\hbox{$>$}}}}

\def\gta{\ifmmode {\mathbin{\lower 3pt\hbox   
    {$\,\rlap{\raise 5pt\hbox{$\char'076$}}\mathchar"7218\,$}}}
    \else {${\mathbin{\lower 3pt\hbox
    {$\rlap{\raise 5pt\hbox{$\char'076$}}\mathchar"7218\,$}}}
    $}\fi}
\def\lta{\ifmmode {\,\mathbin{\lower 3pt\hbox   
    {$\,\rlap{\raise 5pt\hbox{$\char'074$}}\mathchar"7218\,$}}}
    \else {${\mathbin{\lower 3pt\hbox
    {$\rlap{\raise 5pt\hbox{$\char'074$}}\mathchar"7218\,$}}}
    $}\fi}

\newcommand{\beq}{\begin{equation}}
\newcommand{\eeq}{\end{equation}}
\newcommand{\bea}{\begin{eqnarray}}
\newcommand{\eea}{\end{eqnarray}}

\newcommand{\NCSA}{\affiliation{NCSA, University of Illinois at Urbana-Champaign, Urbana, Illinois 61801, USA}}
\newcommand{\ANCSA}{\affiliation{Department of Astronomy, University of Illinois at Urbana-Champaign, Urbana, Illinois 61801, USA}}
\newcommand{\PNCSA}{\affiliation{Department of Physics, University of Illinois at Urbana-Champaign, Urbana, Illinois 61801, USA}}
\newcommand{\CITA}{\affiliation{Canadian Institute for Theoretical Astrophysics, 60 St.~George Street, University of Toronto, Toronto, ON M5S 3H8, Canada}}
\newcommand{\SPIN}{\affiliation{Students Pushing Innovation (SPIN) intern at NCSA, University of Illinois at Urbana-Champaign, Urbana, Illinois 61801, USA}} 
\newcommand{\ECE}{\affiliation{Department of Electrical and Computer Engineering, University of Illinois at Urbana-Champaign, Urbana, Illinois 61801, USA}}
\newcommand{\AEI}{\affiliation{Max Planck Institute for Gravitational Physics (Albert Einstein Institute), Am M{\"u}hlenberg~1, 14476 Potsdam-Golm, Germany}}
\newcommand{\Caltech}{\affiliation{Theoretical Astrophysics 350-17, California Institute of Technology, Pasadena, CA 91125, USA}}
\newcommand{\Cornell}{\affiliation{Cornell Center for Astrophysics and Planetary Science, Cornell University, Ithaca, New York 14853, USA}}
 
\newcommand{\CIFAR}{\affiliation{Canadian Institute for Advanced Research,  180 Dundas St.~West, Toronto, ON M5G 1Z8, Canada}} 
 
\newcommand{\Princeton}{\affiliation{Department of Physics,  Princeton University, Jadwin Hall, Princeton, NJ 08544, USA}}
\newcommand{\JPL}{\affiliation{Jet Propulsion Laboratory, California Institute of Technology, 4800 Oak Grove Drive, Pasadena, CA 91109, USA}}

\begin{document}

\title{Complete waveform model for compact binaries on eccentric orbits}
\author{E. A. Huerta}
\email{elihu@illinois.edu}\NCSA
\author{Prayush Kumar}\CITA
\date{\today}
\author{Bhanu Agarwal}\NCSA\ECE\SPIN
\author{Daniel George}\NCSA\ANCSA
\author{Hsi-Yu Schive}\NCSA
\author{Harald P. Pfeiffer}\CITA\AEI\CIFAR
\author{Roland Haas}\NCSA
\author{Wei Ren}\NCSA\SPIN\PNCSA
\author{Tony Chu}\Princeton
\author{Michael Boyle}\Cornell
\author{Daniel A. Hemberger}\Caltech
\author{Lawrence E.~Kidder}\Cornell
\author{Mark A.~Scheel}\Caltech
\author{Bela~Szilagyi}\Caltech\JPL

\date{\today}

\begin{abstract}
We present a time domain waveform model that describes the inspiral, merger and ringdown of compact binary systems whose components are non-spinning, and which evolve on orbits with low to moderate eccentricity. The inspiral evolution is described using third order post-Newtonian equations both for the equations of motion of the binary, and its far-zone radiation field. This latter component also includes instantaneous, tails and tails-of-tails contributions, and a contribution due to non-linear memory.  This framework reduces to the post-Newtonian approximant \texttt{TaylorT4} at third post-Newtonian order in the zero eccentricity limit. To improve phase accuracy, we also incorporate higher-order post-Newtonian corrections for the energy flux of quasi-circular binaries and gravitational self-force corrections to the binding energy of compact binaries. This enhanced prescription for the inspiral evolution is combined with a fully analytical prescription for the merger-ringdown evolution constructed using a catalog of numerical relativity simulations. We show that this inspiral-merger-ringdown waveform model reproduces the effective-one-body model of Ref.~[Y. Pan \textit{et al.}, Phys. Rev. D 89, 061501 (2014)] for quasi-circular black hole binaries with mass-ratios between 1 to 15 in the zero eccentricity limit over a wide range of the parameter space under consideration. Using a set of eccentric numerical relativity simulations, not used during calibration, we show that our new eccentric model reproduces the true features of eccentric compact binary coalescence throughout merger. We use this model to show that the gravitational wave transients GW150914 and GW151226 can be effectively recovered with template banks of quasi-circular, spin-aligned waveforms if the eccentricity \(e_0\) of these systems when they enter the aLIGO band at a gravitational wave frequency of 14 Hz satisfies \(e_0^{\rm GW150914}\leq0.15\) and \(e_0^{\rm GW151226}\leq0.1\). We also find that varying the spin combinations of the quasi-circular, spin-aligned template waveforms does not improve the recovery of non-spinning, eccentric signals when \(e_0\geq0.1\). This suggests that these two signal manifolds are predominantly orthogonal.
\end{abstract}


\maketitle

\section{Introduction}
\label{intro}

The field of gravitational wave (GW) astronomy has been firmly inaugurated with the first direct detections of gravitational radiation from binary black hole (BBH) systems with the Advanced Laser Interferometer Gravitational-wave Observatory (aLIGO) detectors~\cite{DI:2016,secondBBH:2016,bbhswithligo:2016}. The growing sample of GW observations that is expected in aLIGO's next observing runs~\cite{D5:2016,bbhswithligo:2016} will enable an accurate census of the mass and angular momentum distribution of BHs and neutron stars (NSs), gaining insights into formation and evolution scenarios of compact object binaries, and the environments in which they reside~\cite{SathyaLRR:2009,bel:2010ApJ,Anto:2015arXiv,D9:2016,scenarioligo:2016LRR,Carl:2016arXiv,bel:2016Na,marchant:2016,deMink:2016MNRAS}. For instance, the detection of GWs from  eccentric compact binaries can provide important information of compact object populations in globular clusters and galactic nuclei~\cite{Anto:2015arXiv}. Any such analysis must start with the development of waveforms for eccentric compact binaries, which is the topic of this article.

GWs encode information about the properties of the astrophysical sources that generate them, and can be used to map the structure of spacetime in the vicinity of compact binary systems~\cite{ryan}. aLIGO is expected to detect a wide variety of GW sources, including: (i) compact binary systems that form in the galactic field and evolve through massive stellar evolution. These are expected to enter aLIGO's frequency band on nearly quasi-circular orbits because GWs are very effective at circularizing the orbits of compact binaries~\cite{Peters:1964,peters}; (ii) compact binaries formed in dense stellar environments, e.g., core-collapsed globular clusters and galactic nuclei. In these environments, compact systems can undergo a variety of N-body interactions that lead to the formation of compact binaries that retain eccentricity during their lifetime (see~\cite{Maccarone:2007,Strader:2012,cho:2013ApJ,Anto:2015arXiv,CR:2015PRL,Carl:2016arXiv} and references therein). 

The detection of stellar mass BHs in the galactic cluster M22~\cite{Strader:2012} led to the development of more accurate N-body algorithms to explore the formation and detectability of BBHs formed in globular clusters with aLIGO. These improved analyses indicate that about \(20\%\) of BBH mergers in globular clusters will have eccentricities \(e_0\gtrsim0.1\) when they first enter aLIGO band at 10Hz,  and that \(\sim 10\%\) may have eccentricities \(e\sim1\)~\cite{Anto:2015arXiv}. Furthermore, a fraction of galactic field binaries may retain significant eccentricity prior to the merger event~\cite{Samsing:2014}. BBHs formed in the vicinity of supermassive BHs may also merge with significant residual eccentricities~\cite{Van:2016}. Given the proven detecting capabilities of aLIGO, these results imply that we are now in a unique position to enhance the science reach of GW astronomy by targeting eccentric compact binary systems. The detection of these events requires the development of new waveform models and data analysis techniques because the imprint of eccentricity on GWs is multifold: it introduces modulations in the amplitude and frequency evolution of the waveforms, and it shortens their duration~\cite{pierro2001,Gopakumar:2002,Gopa:2004,Gopa:2004b,Gopakumar:2005b,GopakumarandK:2006,Blanchet:2006,Arun:2008,Arun:2009PRD,Brown:2010,Huerta:2013a,Huerta:2014}. GWs emitted by compact binaries that enter aLIGO band with moderate eccentricities, \(e_0\lesssim0.4\), can be modeled as continuous waves and searched for using matched-filtering algorithms. In contrast, systems that enter aLIGO band with  \(e_0\sim 1\) emit individual GW bursts at each periastron passage, most suitable searched by excess power algorithms utilizing time-frequency tiling~\cite{Tai:2014}.

In order to detect and characterize eccentric binary systems with aLIGO, we introduce an inspiral-merger-ringdown (IMR) waveform model that reproduces the dynamics of state-of-the-art non-spinning, quasi-circular waveform models~\cite{Tara:2014}. Using a set of non-spinning, eccentric numerical relativity (NR) simulations, we show that this new model can reproduce the dynamics of comparable mass-ratio, moderately eccentric binary systems throughout the merger. This model can be immediately used in the context of aLIGO to: (i) quantify the sensitivity of quasi-circular searches and burst searches to eccentric signals; (ii) study template bank construction for non-spinning, eccentric BBHs; (iii) estimate the eccentricity of detected BBH signals, under the assumption that the binary components are not spinning; (iv) explore the sensitivity of burst-like searches that have been tuned to detect highly eccentric systems (\(e_0\sim 1)\) to recover signals with moderate values of eccentricity~\cite{SKlimenko:2004CQGra,Klimenko:2004CQG,Sergey:2005K,Sergey:2008CQG,Sergey:2011PhRvD,Sergey:2016,Tiwari:2016}.

Previous work related to this particular subject includes the following: (i) frequency domain inspiral-only waveforms that include leading order post-Newtonian (PN~\footnote{When we state the accuracy of PN expansions below, a term of Nth PN order implies that the term of highest order in the weak-field expansion is proportional to \((v/c)^{2N}\), where \(v\) represents the orbital velocity~\cite{Blanchet:2006}.}) corrections in a post-circular or small eccentricity approximation~\cite{yunes-eccentric-2009,Mico:2015}; (ii) frequency and time domain waveforms that reduce to the PN-based approximants \texttt{TaylorF2}  and  \texttt{TaylorT4} at 2PN in the quasi-circular limit~\cite{Tanay:2016}; (iii) inspiral-only waveforms that include  2PN and 3PN corrections to the radiative and conservative pieces of the dynamics, respectively~\cite{Hinder:2010}; (iv) inspiral-only waveforms that include 3PN corrections to the radiative and conservative pieces of the dynamics~\cite{Arun:2009PRD,Mishra:2015,Moore:2016,lou:2016arXiv}; (v) inspiral-only frequency domain waveforms that reduce to the PN-based approximant \texttt{TaylorF2} 3.5PN at zero eccentricity, and to the post-circular approximation of Ref.~\cite{yunes-eccentric-2009} at small eccentricity~\cite{Huerta:2014}; (vi) hybrid waveforms that describe highly eccentric systems. These waveforms describe the inspiral evolution using geodesic equations of motion. The merger phase is modeled using a semi-analytical prescription that captures the features of NR simulations~\cite{East:2013}; (vii) self-force calculations for non-spinning BHs along eccentric orbits~\cite{Letiec:2014IJ,ackay:2015prd,hopper:2016PhRv,forseth:2016PhRvD,binida:2016PhRvD,akcavan:2016PhRvD,binidam:2016,Osburn:2016}; (viii) NR simulations that explore the dynamics of eccentric binary systems~\cite{ihh:2008PhRvD,Hinder:2010,east:2012a,east:2012,Gold:2012PG,Gold:2013,East:2015PRDa,East:2016PhRvD,Radice:2016MNRAS,2016arXiv161107531T,2016arXiv161103418L}.

Some of the aforementioned waveform models have been used in source detection~\cite{Huerta:2013a,Huerta:2014,MC:2014PhRvD,MC:2015PhRvD} and parameter estimation studies~\cite{Favata:2014,Sun:2015PRD} in the context of aLIGO. These studies have shown that detecting and characterizing eccentric binary systems will not be feasible using existing algorithms for quasi-circular binaries~\cite{Huerta:2013a,Favata:2014}. Furthermore, as discussed in~\cite{Huerta:2014}, to accurately model inspiral-dominated systems, i.e.,  binary systems with total mass \(M\lesssim 10\Msun\)~\cite{Prayush:2013a}, eccentric waveform models should reduce to high PN order approximants such as \texttt{TaylorT4} 3.5PN or \texttt{TaylorF2} 3.5PN~\cite{Huerta:2013a,Huerta:2014} in the zero eccentricity limit. On the other hand, for NSBH and BBH systems that require the inclusion of the merger and ringdown phase, eccentric waveforms models should reproduce the evolution rendered by IMR models such as~\cite{Pan:2013,khan:2016PhRvD,husacv:2016PhRvD} in the zero eccentricity limit.

In this paper we start addressing these important issues by developing an IMR waveform model valid for compact binaries with moderate eccentricities. The key features of our model are:

\begin{itemize}
\item It includes third order PN accurate expansions for eccentric orbits both for the equations of motion of the binary and its far-zone radiation field. The radiative evolution includes instantaneous, tails and tails-of-tails contributions, and a contribution due to non-linear memory. 
\item The accuracy of the inspiral evolution is improved by including 3.5PN corrections for quasi-circular orbits (at all powers of symmetric mass-ratio), improving on~\cite{Huerta:2014}.
\item To further improve phase accuracy especially for unequal mass systems, the 3PN accurate inspiral evolution for eccentric systems is corrected by including up to 6PN terms both for the energy flux of quasi-circular binaries and gravitational self-force corrections to the binding energy of compact binaries at first order in symmetric mass-ratio \(\eta\).
\item We combine the aforementioned enhanced inspiral evolution with a merger and ringdown treatment using the \emph{implicit rotating source} (IRS) formalism~\cite{Kelly:2011PRD}, fitted against NR simulations up to mass-ratio 10.
\end{itemize}

The eccentric model we develop in this article is the first model in the literature that combines all these features, and makes it a powerful tool to explore the detection of eccentric signals with aLIGO. To exhibit the reliability of our eccentric model, we show that it agrees well with the IMR effective-one-body model SEOBNRv2~\cite{Tara:2012,Tara:2014} in the non-spinning limit over a wide range of the BBH parameter space accessible to aLIGO. Furthermore, using non-spinning, eccentric NR simulations, we show that our model can reproduce the true accurate dynamics of moderately eccentric BBH mergers with mass-ratios \(q\in\{1,\,2\}\) throughout the merger. Having established the validity of our new eccentric model, we use it to shed light for the first time on the importance of including eccentricity in the detection of IMR systems, such as NSBH and BBH systems with asymmetric mass-ratios. We also show that our waveform model has a favorable computational cost, suitable for large scale data analysis studies. 

Throughout this article we use units \(G=c=1\). We denote the components masses by \(m_1\) and \(m_2\), where \(m_1\geq m_2\). Mass combinations used throughout the article include: total mass \(M=m_1+m_2\), reduced mass \(\mu = m_1\,m_2/M\),  mass-ratio \(q=m_1/m_2\), and symmetric mass-ratio \(\eta = \mu/M\). 

This paper is organized as follows: In Section~\ref{build} we describe the construction of our eccentric  waveform model. In Section~\ref{catch_me} we apply our eccentric waveform model to explore the detectability of eccentric compact binary systems with aLIGO. We summarize our results and discuss future directions of work in Section~\ref{disc}.


\section{Waveform model construction}
\label{build}

\subsection{Overview}
In this Section we introduce our new eccentric waveform model, named advanced \(x\)--model or `\(ax\)--model', since it extends the inspiral-only, low order PN eccentric \(x\)--model introduced in~\cite{Hinder:2010}. The construction of our model has several key ingredients that are described on an incremental basis. 

In the description below we refer to the conservative and radiative pieces of the dynamics. The conservative piece refers to the equations of motion of the binary that are derived from a PN Hamiltonian~\cite{Blanchet:2006}, whereas the radiative piece takes into account the energy and angular momentum that gravitational radiation carries away from the binary. 

\subsection{Eccentric orbit parametrization}

The model we introduce in this article aims to provide an improved description of the phase evolution of binaries moving on eccentric orbits. We do this by working in the adiabatic approximation. As extensively discussed in the literature, in this limit the radiation time scale would be much longer than the orbital time scale, and consequently we require an averaged description of the radiation reaction over an orbital period~\cite{Arun:2009PRD,ArunBlanchet:2008,Blanchet:2006}.

We parametrize the equations of motion in terms of the mean orbital frequency \(\omega\) through the gauge invariant quantity \(x=\left(M\omega\right)^{2/3}\), and the \textit{temporal} eccentricity \(e_t\equiv e\)~\cite{Hinder:2010}. Please note that in the context of eccentric binaries, \(\omega=\langle \dot{\phi}\rangle = K n\), where the average \(\langle \rangle\) is taken over an orbital period.  The mean motion \(n\) is related to the mean anomaly \(\ell\) through the relation \(M\dot{\ell} = Mn\) (see Eq.~\ref{conservative} below), \(\dot{\phi}\) is the instantaneous angular velocity, and the periastron precession \(K\) and relativistic precession \(k\) are related through  \(K=1+k\).  At 3PN order, the Keplerian parametrization of the orbit in terms of the magnitude of the relative separation vector \(r\), and the mean anomaly \(\ell\) is given by~\cite{Hinder:2010}:

\bea
\label{radial}
\frac{r}{M}&=&\frac{1-e\cos u}{x}+\sum^{i=3}_{i=1}r_{i\,\rm{PN}}x^{i-1}\,,\\
\label{ecc_anom}
\ell &=& u-e \sin u + \sum^{i=3}_{i=2}l_{i\,\rm{PN}}x^{i}\,.
\eea

\noindent The orbital evolution has two components. The conservative piece is derived from a PN Hamiltonian including corrections at 3PN order and has the form:

\bea
\label{conservative}
M \dot{\phi} &=& \dot{\phi}_{0\,\rm{PN}} x^{3/2} + \dot{\phi}_{1\,\rm{PN}} x^{5/2} + \dot{\phi}_{2\,\rm{PN}} x^{7/2}  \nonumber\\&+& \dot{\phi}_{3\,\rm{PN}} x^{9/2} + {\cal{O}}(x^{11/2}),\\
M \dot{\ell} &=& Mn = x^{3/2} +  n_{1\,\rm{PN}} x^{5/2} + n_{2\,\rm{PN}} x^{7/2} \nonumber\\&+& n_{3\,\rm{PN}} x^{9/2}+{\cal{O}}(x^{11/2}) \,,
\eea

\noindent where \(\phi\) represents the relative orbital phase. The PN coefficients \((r_{i\,\rm{PN}},\, l_{i\,\rm{PN}})\), \((\dot{\phi}_{i\,\rm{PN}},\,n_{i\,\rm{PN}} )\) are given in~\cite{Hinder:2010}. The radiative part of the orbital evolution takes into account the energy and angular momentum that gravitational radiation carries away from coalescing compact binaries. This effect implies that the gauge-invariant expansion parameter \(x\) and the eccentricity \(e\) are no longer conserved, but evolve as follows:

\bea
\label{radiative}
M\dot{x} &=&  \dot{x}_{0\,\rm{PN}} x^5 + \dot{x}_{1\,\rm{PN}} x^6 + \dot{x}_{2\,\rm{PN}} x^7 \nonumber\\&+& \dot{x}_{3\,\rm{PN}} x^8 + \dot{x}_{\rm HT} \,,\\
M\dot{e} &=&  \dot{e}_{0\,\rm{PN}} x^4 + \dot{e}_{1\,\rm{PN}} x^5 + \dot{e}_{2\,\rm{PN}} x^6 \nonumber\\&+& \dot{e}_{3\,\rm{PN}} x^7 + \dot{e}_{\rm HT}  \,.
\eea

\noindent \textit{In the above expressions we have derived 3PN corrections for \(\dot{x}\), and have also derived hereditary terms (HT) \(\dot{x}_{\rm HT}\). These new calculations are presented in Appendix~\ref{ap1}.} Hereditary terms are non-linear contributions that depend on the dynamics of the system in its entire past, and comprise tails, tails-of-tails and tail square terms for the energy and angular flux, but also a 2.5PN memory contribution for the angular momentum flux. These terms include fractional powers in \(x\) --- see Equations~\eqref{her} and~\eqref{ecc_ev_ht}. We provide a detailed discussion of the importance of including hereditary contributions in Appendix~\ref{importance_hereditary}.  

Regarding the time evolution of the eccentricity \(e\), we use 3PN calculations and the corresponding hereditary contributions derived in Ref.~\cite{Arun:2009PRD}. In constructing this model, we have ensured that the choice of coordinates is consistent throughout, i.e., we are using modified harmonic coordinates. We construct the PN waveform strain as follows:

\bea
\label{strain_ins}
h^{\rm inspiral}(t) = h^{\rm inspiral}_{+}(t) -i  h^{\rm inspiral}_{\times}(t)\,,
\eea
\noindent with the plus and cross polarizations given by~\cite{Hinder:2010}:
\bea
\label{h_plus}
h_{+} &=&-\frac{M\eta}{R}\Bigg\{\left(\cos^2\iota+1\right)\Bigg[\left(-\dot{r}^2+r^2\dot{\phi}^2+\frac{M}{r}\right) \cos2\Phi \nonumber\\&+& 2r\dot{r}\dot{\phi}\sin2\Phi\Bigg] + \left(-\dot{r}^2-r^2\dot{\phi}^2+\frac{M}{r}\right)\sin^2\iota\Bigg\}\,,\\
\label{h_cross}
h_{\times} &=&-\frac{2M\eta}{R}\cos\iota\Bigg\{\left(-\dot{r}^2+r^2\dot{\phi}^2+\frac{M}{r}\right)\sin2\Phi \nonumber\\&-& 2r\dot{r}\dot{\phi} \cos2\Phi\Bigg\}\,,
\eea

\noindent  where \(\Phi=\phi-\chi\), and \((\chi,\,\iota)\) represent the polar angles of the observer, and \(R\) the distance to the binary.

\subsection{Eccentricity decay}
\label{ho_ec}

In this Section we explore the importance of including 3PN accurate eccentricity corrections to the binary evolution. To do so we consider a population of BBH systems with component masses  \(m_{1,\,2}\in[5\Msun,\,50\Msun]\), and with an orbital eccentricity \(e_0\) when they enter aLIGO frequency band at a GW frequency \(f_{\rm GW} = 15{\rm Hz}\). 

\begin{figure*}[htp!]
\centerline{
\includegraphics[height=0.4\textwidth,  clip]{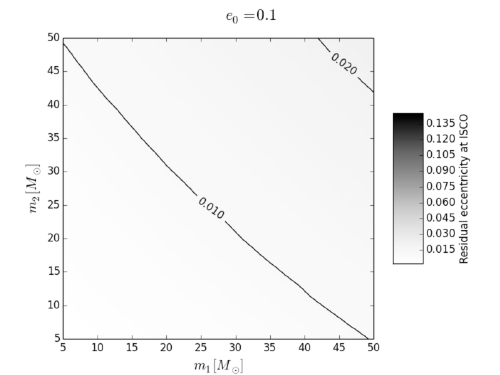}
\includegraphics[height=0.4\textwidth,  clip]{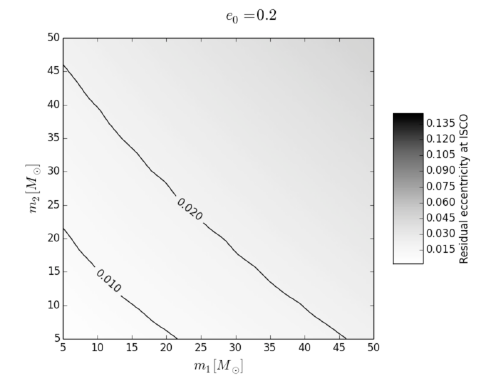}
}
\centerline{
\includegraphics[height=0.4\textwidth,  clip]{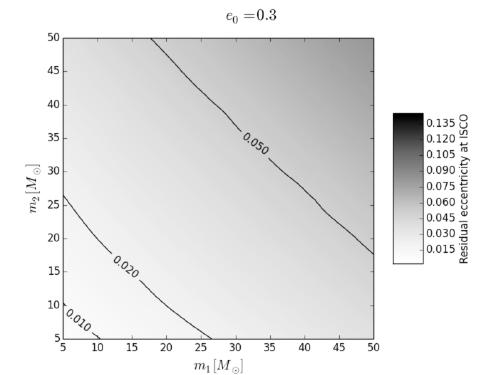}
\includegraphics[height=0.4\textwidth,  clip]{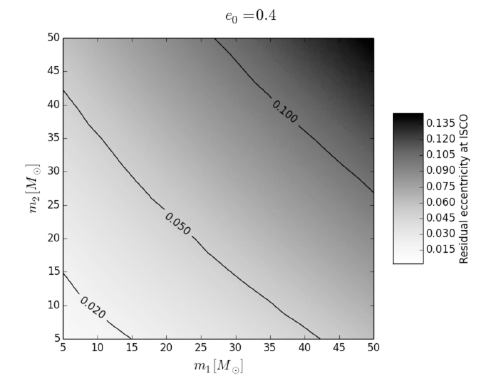}
}
\caption {The panels show the eccentricity at ISCO of black hole binaries prior to the merger event. We assume that the black hole population on each panel has an initial eccentricity \(e_0\) at a gravitational wave frequency \(f_{\rm GW}= 15{\rm Hz}\). These results have been obtained using PN equations of motion that include conservative and radiative corrections up to 3PN order.}
\label{mer_ecc_15hz_3pn}
\end{figure*}

Figure~\ref{mer_ecc_15hz_3pn} presents the residual eccentricity at the last stable circular orbit (ISCO) for the aforementioned BBH population using equations of motion that include conservative and radiative corrections up to 3PN order (cf. Eq.~\eqref{radiative}) at the ISCO frequency given by~\cite{yunes-eccentric-2009}

\beq
\label{final_gwf}
f_{\rm ISCO}= \frac{1}{\pi M}\left(\frac{1+e}{6+2e}\right)^{3/2}\,.
\eeq

\noindent Figure~\ref{mer_ecc_15hz_3pn} includes contour lines of residual eccentricity at ISCO, namely: \(e_{\rm ISCO} =\{0.01,\, 0.02,\,0.05, \,0.1\}\). A key assumption in the construction of our eccentric model is that moderately eccentric binaries attain circularization prior to the merger event. In practice, we consider compact binary systems whose residual eccentricity at ISCO satisfies \(e_{\rm ISCO}\lesssim0.05\). Figure~\ref{mer_ecc_15hz_3pn} indicates that this assumption covers a wide range of the parameter space for moderately eccentric systems. We note that \(e_0=0.4\) at \(f_{\rm GW} =15{\rm Hz}\) is already a very high value for astrophysically motivated systems. 

Figure~\ref{mer_ecc_15hz_3pn} also indicates that the largest value of residual eccentricity in all cases corresponds to the most massive BBH systems under consideration, which merge at lower frequencies and have less time to circularize under GW emission. On the other hand, BBH systems with less massive components merge at higher frequencies, and therefore undergo further circularization within the aLIGO frequency band. 

\begin{figure*}[htp!]
\centerline{
\includegraphics[height=0.4\textwidth,  clip]{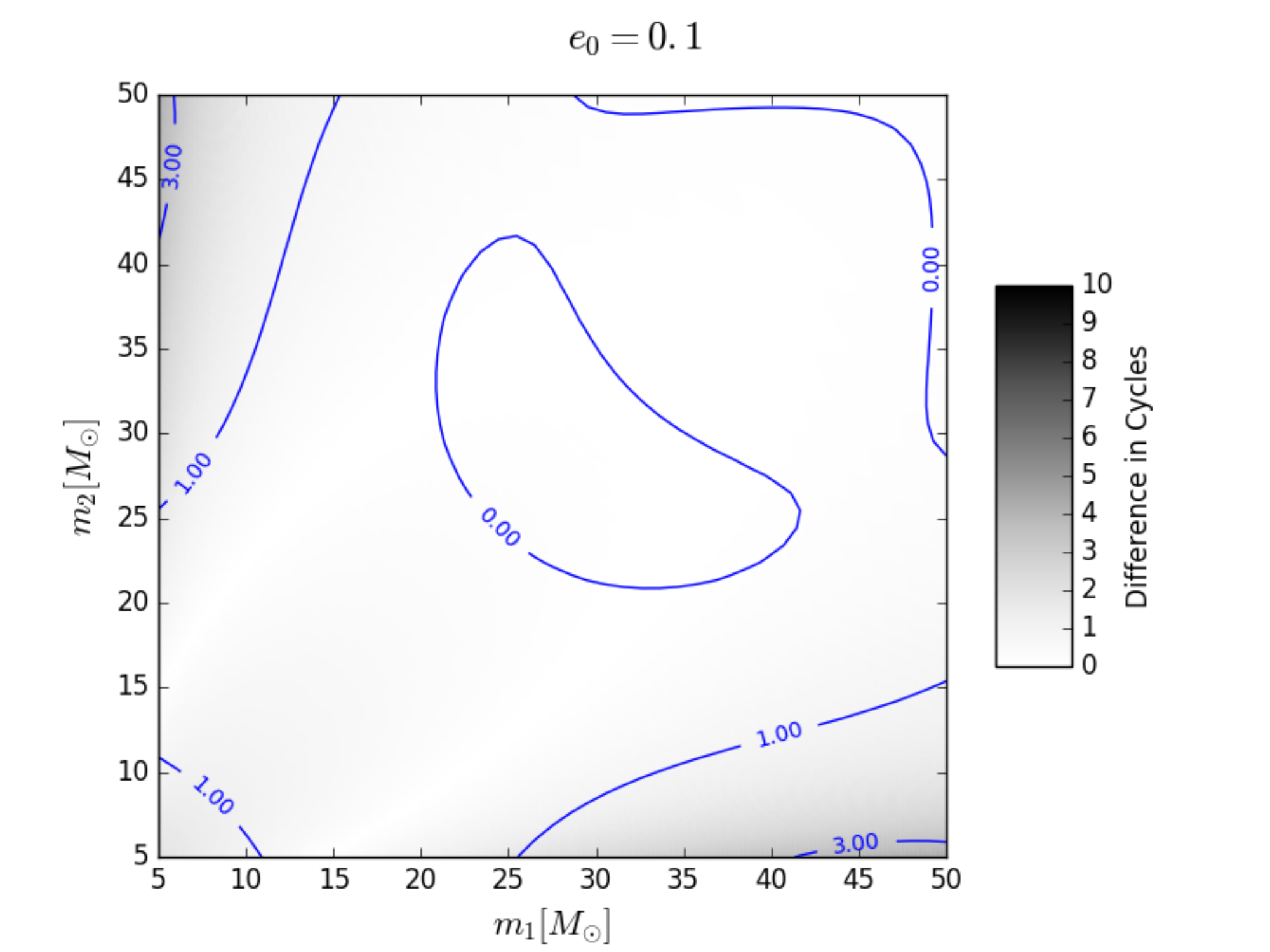}
\includegraphics[height=0.4\textwidth,  clip]{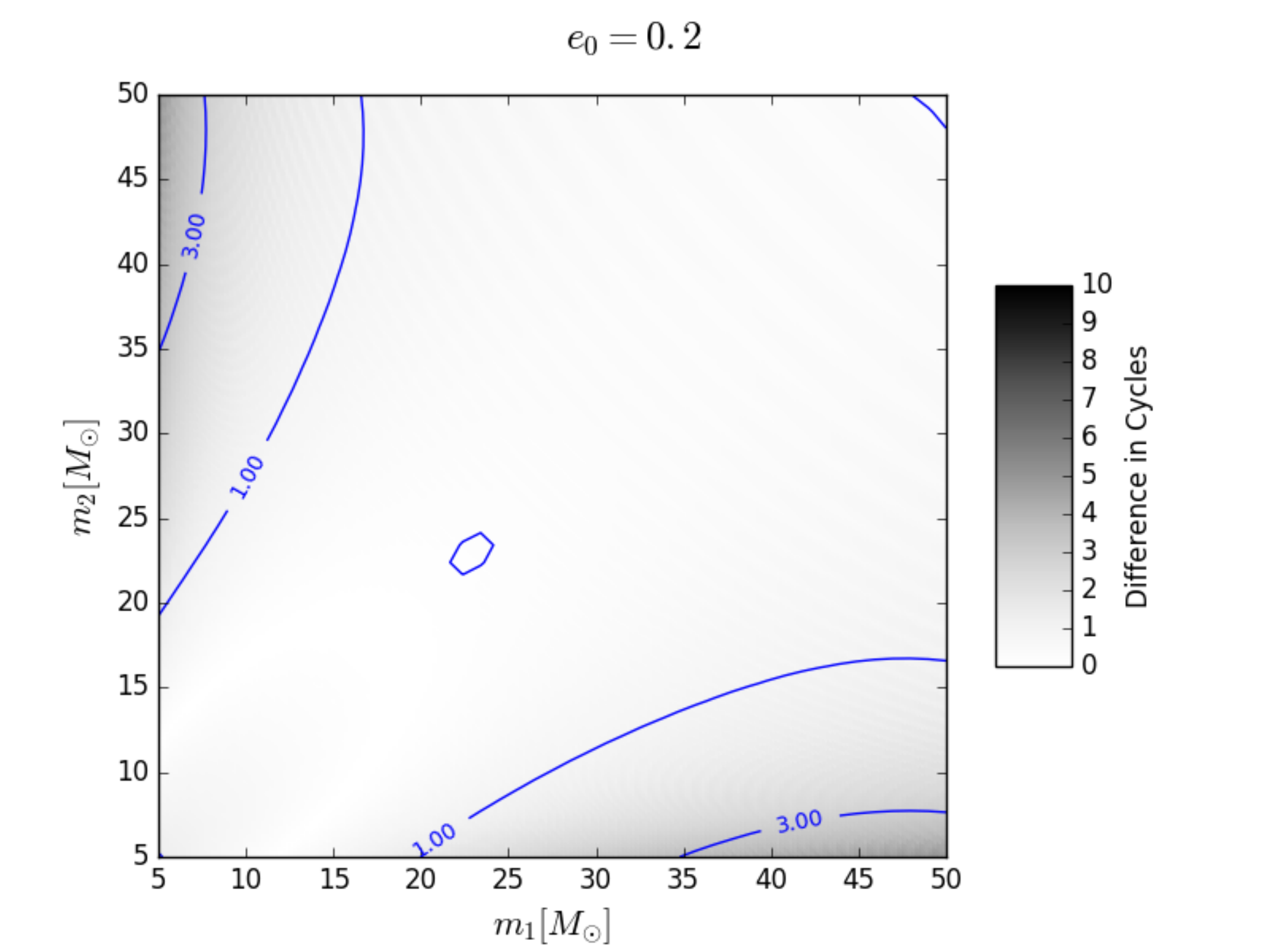}
}
\centerline{
\includegraphics[height=0.4\textwidth,  clip]{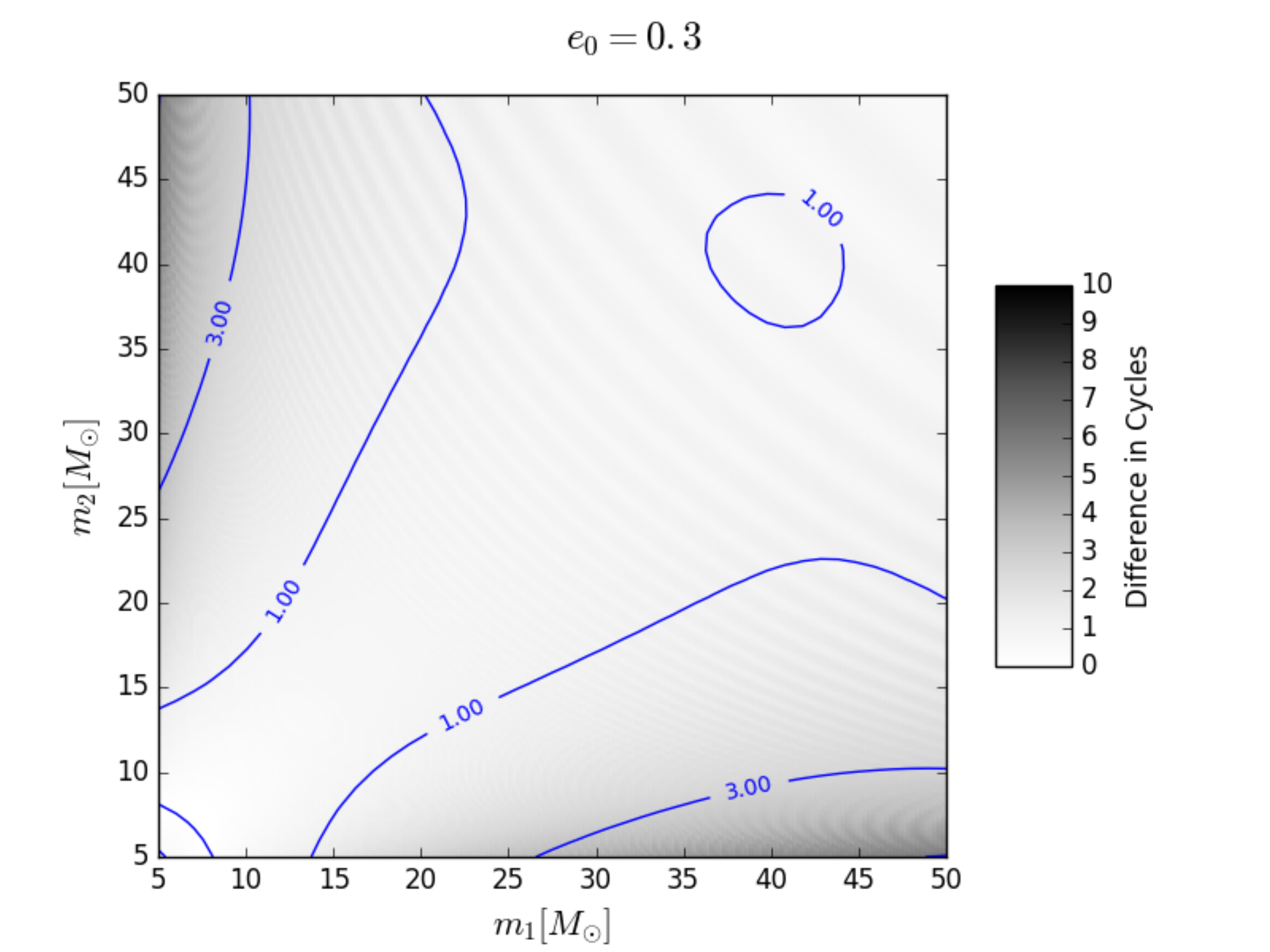}
\includegraphics[height=0.4\textwidth,  clip]{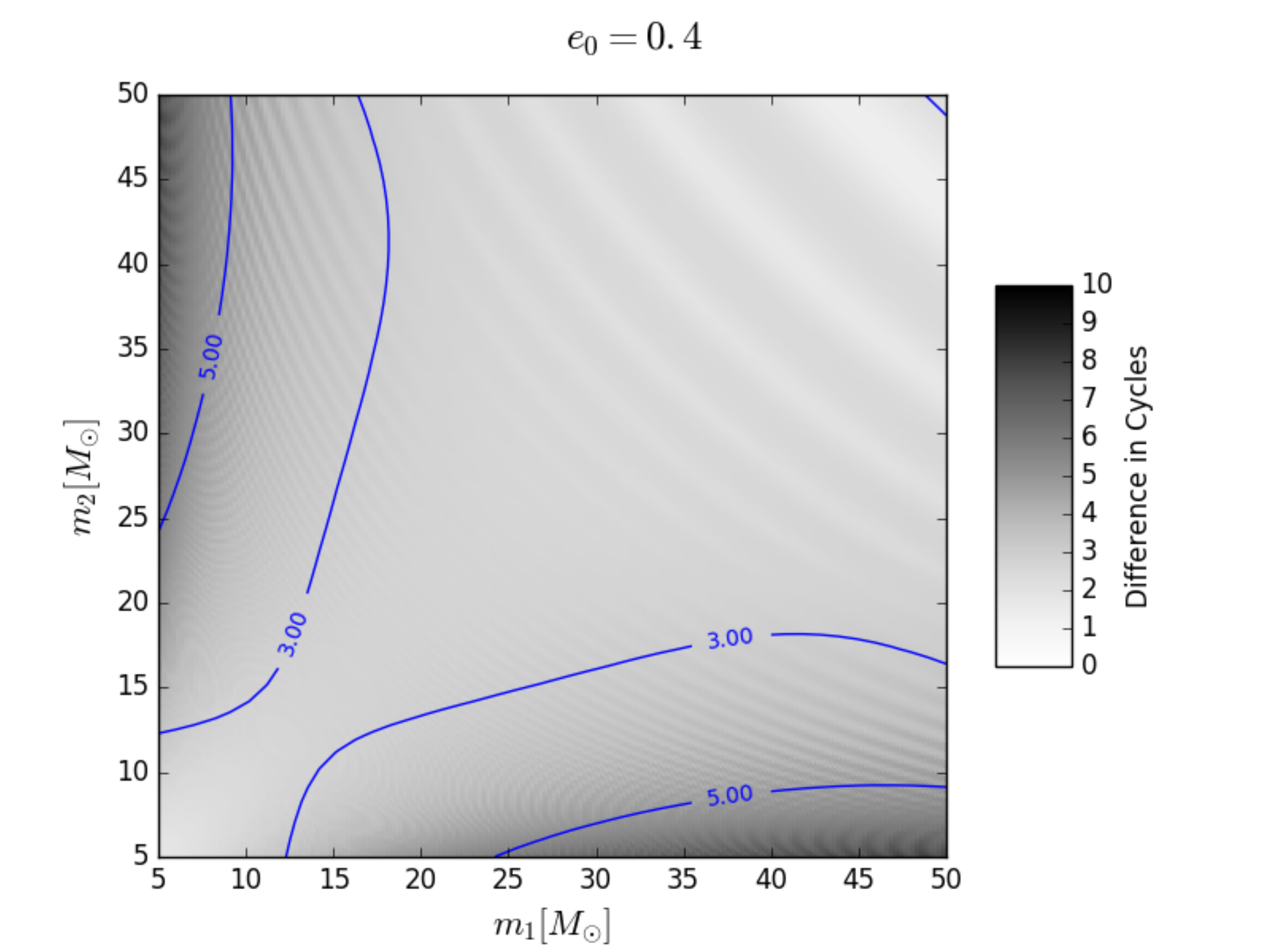}
}
\caption {The panels show the difference in the number of GW cycles when we use a waveform model that includes conservative corrections up to 3PN order and radiative corrections up to 2PN or 3PN order.  The binary black hole population on each panel has an initial eccentricity \(e_0\) at \(f_{\rm GW}= 15{\rm Hz}\). Note that the discrepancy between the two approximate models becomes very noticeable when \(e_0\gtrsim0.3\) for systems with asymmetric mass-ratios.}
\label{cycles_15hz}
\end{figure*}

In the previous study~\cite{Huerta:2014} we emphasized the importance of developing waveform models that encode higher-order PN corrections. We showed that waveform templates that include only 2PN corrections for the radiative piece of the dynamics will significantly reduce the ability to observe eccentric compact binaries. To further explore the effect of including higher-order PN corrections, Figure~\ref{cycles_15hz} presents the difference in the number of GW cycles \({\cal{N}}\) when we use a waveform model that includes conservative corrections up to 3PN order and radiative corrections up to 2PN or 3PN order. \({\cal{N}}\) is defined as 

\beq
\label{cycles}
N= \frac{1}{\pi}\bigg[\langle\phi\rangle\left(f_{\rm ISCO}\right) - \langle\phi\rangle\left(f_{\rm min}\right) \bigg]\,,
\eeq 

\noindent and \(f_{\rm min}=15{\rm Hz}\). The color bar in Figure~\ref{cycles_15hz} describes \(\Delta {\cal{N}} =  |{\cal{N}} (3{\rm PN})-  {\cal{N}} (2{\rm PN})|\). These results demonstrate that waveform templates using only 2PN radiative corrections will significantly deviate from waveform models that include all known eccentric corrections up to 3PN order when \(e_0\gtrsim0.2\), particularly for asymmetric mass-ratio systems.

In summary, the results of this Section indicate that an astrophysically motivated population of moderately eccentric compact binaries will circularize prior to the merger event. For these systems, it is physically motivated to add a non-eccentric merger waveform to the inspiral evolution. Finally, we have discussed the importance of including all known eccentric PN calculations to provide the most accurate description of these systems. 

\subsection{Improved non-eccentric terms}
\label{ax_zero_e_limit}

The inspiral evolution of the \(ax\)--model reduces to the PN-based approximant \texttt{TaylorT4} 3PN in the quasi-circular limit. To explicitly show this feature, we simplify the equations we derived in Appendix~\ref{ap1} in the \(e\rightarrow 0\) limit. Please note that to obtain the following results it is necessary to include the hereditary corrections presented in equation~\eqref{radiative}, since these cancel out gauge-dependent quantities that are present in the instantaneous part of the fluxes. To be precise, this cancellation takes place because we include the tails-of-tails contributions in the fluxes --- see Appendix~\ref{importance_hereditary}. After including these non-linear contributions, equation~\eqref{radiative} takes the form: 

\begin{eqnarray}
\label{dxdtT4}
M\frac{\mathrm{d}x}{\mathrm{d}t}\bigg|^{\rm 3PN}_{e\rightarrow 0} &=& \frac{64}{5} \eta\,x^5\, \Bigg\{
1 +
\left( -\frac{743}{336} - \frac{11}{4} \eta \right) x
+ 4 \pi x^{3/2}
\nonumber\\&+& \left( \frac{34\,103}{18\,144} +\frac{13\,661}{2016} \eta + \frac{59}{18} \eta^2  \right) x^2
\nonumber \\
&+& \left( -\frac{4159 \pi}{672} -\frac{189 \pi}{8} \eta  \right) x^{5/2}
\nonumber \\
&+& \Bigg[ \frac{16\,447\,322\,263}{139\,708\,800} - \frac{1712 \gamma}{105} + \frac{16 \pi^2}{3} \nonumber\\&-& \frac{856}{105} \log (16 x) 
+ \left(-\frac{56\,198\,689}{217\,728} + \frac{451 \pi ^2}{48} \right) \eta \nonumber\\&+& \frac{541}{896} \eta^2 - \frac{5605}{2592} \eta^3  \Bigg] x^3 
\Bigg\}\,,
\end{eqnarray}

\noindent where \(\gamma\) is Euler's constant. Furthermore, the equations of the time evolution of the eccentricity \(e\), relative orbital phase \(\phi\), and  the mean anomaly \(\ell\) reduce to~\cite{Arun:2009PRD,Hinder:2010}:
 
\bea
\label{mphidot}
M\frac{\mathrm{d}\phi}{\mathrm{d}t}\bigg|_{e\rightarrow 0} &=& x^{3/2}\,,\\
\label{edot}
M\frac{\mathrm{d}e}{\mathrm{d}t}\bigg|_{e\rightarrow 0} &=& 0\,,\\
\label{ldot}
M\frac{\mathrm{d}\ell}{\mathrm{d}t}\bigg|_{e\rightarrow 0} &=& x^{3/2}\Bigg\{ 1+3x + \left(7\eta-\frac{9}{2}\right)x^2 \nonumber\\&+&\left(-\frac{27}{2}+ \left(\frac{481}{4}-\frac{123}{32}\pi^2\right)\eta-7\eta^2\right)x^3\Bigg\}\,.\nonumber\\
\eea

\noindent Note that Equation~\eqref{ldot} describes the periastron advance in the \(e\rightarrow 0\) limit. In order to further increase the reliability of our waveform model for inspiral dominated systems, we include 3.5PN corrections to the radiative equations of motion in the quasi-circular limit:

\bea
\label{def_35pn}
M\frac{\mathrm{d}x}{\mathrm{d}t}\bigg|^{\rm 3.5PN}_{e\rightarrow 0} &=&  M\frac{\mathrm{d}x}{\mathrm{d}t}\bigg|^{\rm 3PN}_{e\rightarrow 0}+ \frac{64\pi}{5} \eta\,x^5\bigg[-\frac{4415}{4032}\nonumber\\&+&\frac{358675}{6048}\eta +\frac{91945}{1512}\eta^2\bigg]x^{7/2}\,,
\end{eqnarray}

\noindent where the first term on the right hand side of Eq.~\eqref{def_35pn} is given  by Eq.~\eqref{dxdtT4}. Several studies argue that 3.5PN corrections are not sufficient for many applications, such as parameter estimation~\cite{pnbuo,boyle,Prayush:2013a}. Therefore, to improve phase accuracy for asymmetric mass-ratio systems, in this article we use the energy flux, \(\dot{E}^{\rm 6PN}\left(x,\,\eta\right)\), derived in Ref.~\cite{Fujita:2012} up to 6PN order and amend it by including all known finite mass-ratio corrections for the energy flux of quasi-circular compact binaries. We then combine this prescription for the energy flux with the 6PN expression for the binding energy \(E(x,\,\eta)^{\rm 6PN}\) of compact systems derived in~\cite{barus,BiniDa:2014PRD}\footnote{The results presented in this Section were computed using the expressions for the binding energy of Refs.~\cite{barus,BiniDa:2014PRD}. We found that the results from both prescriptions rendered very similar results. In the rest of this paper we quote results obtained using \(E(x,\,\eta)^{\rm 6PN}\) from Ref.~\cite{barus}.}, i.e.,:

\bea
\label{pre_hyb}
M\frac{\mathrm{d}x}{\mathrm{d}t}\bigg|^{\rm 6PN}_{e\rightarrow 0} &=& M\frac{\mathrm{d}x}{\mathrm{d}t}\bigg|^{\rm 3.5PN}_{e\rightarrow 0} +M \dot{E}^{\rm 6PN}\frac{dx}{\mathrm{d}E(x,\,\eta)^{\rm 6PN}}\,,\\
\label{dxdtT4_byho}
M\frac{\mathrm{d}x}{\mathrm{d}t}\bigg|^{\rm 6PN}_{e\rightarrow 0} &=& M\frac{\mathrm{d}x}{\mathrm{d}t}\bigg|^{\rm 3.5PN}_{e\rightarrow 0} + \frac{64\,\eta\, x^5}{5}\Bigg[a_4 x^4 + a_{9/2} x^{9/2} \nonumber\\&+& a_5 x^5 + a_{11/2} x^{11/2} + a_6 x^6 \Bigg]\,,
\eea

\noindent and the coefficients \(a_4,\,a_{9/2},\, a_5,\, a_{11/2},\, a_6\) are presented in Appendix~\ref{ho_pn_corrections}. We have found that a model that combines Eq.~\eqref{dxdtT4_byho} with the merger-ringdown model presented in the following Section agrees well with SEOBNRv2 up to mass-ratios \(q=15\). We present a quantitative discussion of this result in Section~\ref{perform}. Regarding the accuracy of this hybrid scheme to describe the dynamics of eccentric binary systems throughout merger, in Section~\ref{perform_eccentric} we validate our model against a set of NR simulations that describe moderately eccentric BBHs with mass-ratios \(q\in\{1,\,2\}\).

\subsection{Merger and ringdown evolution}
\label{liv}

We now turn our attention to the late time dynamical evolution. To construct the merger phase of our \(ax\)-model, we assume that the system circularizes prior to the merger event, i.e., the eccentricity at ISCO \(e_{\rm ISCO}\lesssim 0.05\). Under this assumption, we complement the inspiral evolution of the \(ax\)--model with a non-eccentric merger waveform. This stand alone merger waveform is constructed by calibrating the IRS model introduced by Kelly \textit{et al}~\cite{Kelly:2011PRD} with a catalog of NR simulations~\cite{Chu:2015} obtained with the Spectral Einstein Code~\cite{AAA}. These simulations describe non-spinning, quasi-circular compact binary systems with mass-ratios between \(q=\)2.5 and \(q=\)10~\cite{Mroue:2013,Chu:2015}. To ensure that our merger waveform reproduces the expected behavior of extreme-mass ratio binaries, we also utilize an SEOBNRv2 waveform with mass-ratio \(q=1000\), since the SEOBNRv2 model is tuned to black hole perturbation theory calculations.

\begin{figure*}[htp!]
\centerline{
\includegraphics[height=0.35\textwidth,  clip]{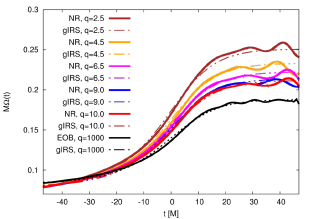}
\includegraphics[height=0.35\textwidth,  clip]{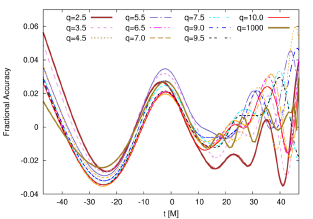}
}
\caption{Left panel: time evolution of the orbital frequency evolution, \(M\Omega(t)\), of NR simulations and an SEOBNRv2 waveform (\(q=1000\)) compared with our gIRS model, Eq.~\eqref{ome}. The right panel shows that the analytical expressions given by Eqs.~\eqref{for_b}--\eqref{for_kappa} can accurately reproduce the orbital frequency evolution of NR waveforms at mass-ratios that were \textit{not} used for their calibration.}
\label{irs_generic}
\end{figure*}

\noindent The IRS model encapsulates the evolution of the orbital frequency evolution, \(\omega(t)\), and the waveform amplitude, \(A(t)\), using the prescription~\cite{Eche:1989,Baker:2008,Kelly:2011PRD,East:2013}:

\bea
\label{ome}
\omega(t)&=&\omega_{\rm QNM}\left(1-\hat{f}\right)\,,\\
\label{omqnm}
\omega_{\rm QNM}&=& 1 - 0.63\left(1 - \hat{s}_{\rm fin}\right)^{0.3}\,,\\
\label{amp}
A(t)&=&\frac{A_0}{\omega(t)}\left[\frac{\big|\dot{\hat{f}}\big|}{1+\alpha\left(\hat{f}^2-\hat{f}^4\right)}\right]\,,
\eea

\noindent where \(\hat{s}_{\rm fin}\) is the spin of the BH remnant. Furthermore, \(\hat{f}\) and \(\dot{\hat{f}}\) are given by~\cite{East:2013}:

\bea
\label{fhat}
\hat{f} &=& \frac{c}{2}\left(1+\frac{1}{\kappa}\right)^{1+\kappa}\left[1-\left(1+\frac{1}{\kappa}e^{-2t/b}\right)^{-\kappa}\right]\,, \\
\label{fdot}
\dot{\hat{f}}&=&\frac{\mathrm{d} \hat{f}}{\mathrm{d}t}\,.
\eea

\noindent \textit{Using the NR catalog previously described, we have derived the following analytical fit for \(\hat{s}_{\rm fin}\):}

\beq
\label{fin_s}
\hat{s}_{\rm fin} = 2 \sqrt{3}\, \eta - \frac{390}{79} \eta^2 + \frac{2379}{287} \eta^3 - \frac{4621}{276} \eta^4\,.
\eeq

\noindent This prescription for \(\hat{s}_{\rm fin}\) reproduces NR results with an accuracy better than \(0.02\%\). Using this prescription, Eq.~\eqref{omqnm} is fully determined. The other free parameters in Eq.~\eqref{ome} are \(b, \,c\) and \( \kappa\). Please note that in previous studies with the IRS model, these parameters have been determined for a few mass-ratio values using a catalog of NR simulations~\cite{East:2013}. In this article, we develop a merger waveform that is reliable for systems with mass-ratios up to \(q=10\), and which also reproduces the correct behavior of extreme-mass ratio binaries. \textit{The novelty of our approach is that we now provide the free parameters \(b, \,c\) and \( \kappa\) as smooth functions of the symmetric mass-ratio \(\eta\)}. To do this, we have used five NR simulations with mass-ratios \(q= \{2.5,\, 4.5,\, 6.5,\, 9,\,10\}\) and an SEOBNRv2 waveform of mass-ratio \(q=1000\). Using these waveforms as input data, we have constructed the following functions:

\begin{figure*}[htp!]
\centerline{
\includegraphics[height=0.35\textwidth,  clip]{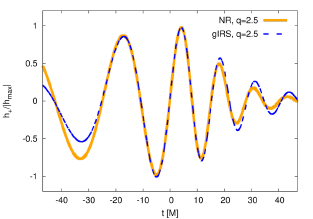}
\includegraphics[height=0.35\textwidth,  clip]{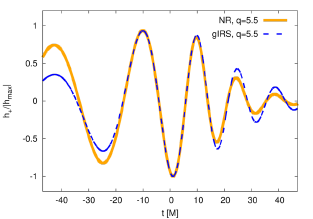}
}
\centerline{
\includegraphics[height=0.35\textwidth,  clip]{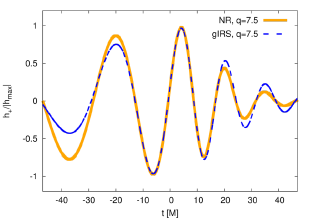}
\includegraphics[height=0.35\textwidth,  clip]{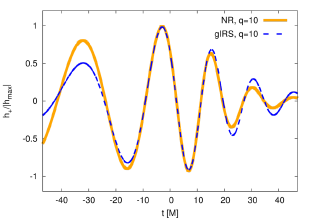}
}
\caption{The panels present the merger model introduced in the main text.  We present a direct comparison with waveforms used to calibrate the model, i.e., those with mass-ratio \(q=\{\)2.5, 10\(\}\), and two additional cases to exhibit the performance of this approach. For reference, \(h_{+}= \Re{[h_{\rm merger}]}\), where \(h_{\rm merger}\) is given by Eq.~\eqref{m_model} in the main text.}
\label{mer_fig}
\end{figure*}

\bea
\label{for_b}
b\left(\eta\right)&=&\frac{16014}{979} - \frac{29132}{1343}\eta^2\,,\\
\label{for_c}
c\left(\eta\right)&=&\frac{ 206}{903} + \frac{180}{1141}\sqrt{\eta} + \frac{424}{1205}\frac{\eta^2}{\log\left(\eta\right)}\,,\\
\label{for_kappa}
\kappa\left(\eta\right)&=&\frac{713}{1056}-\frac{23}{193}\eta\,.
\eea

\noindent Note that these expressions are well behaved throughout the whole range of the symmetric mass-ratio \(\eta\). Turning to the amplitude \(A(t)\), see Eq.~\eqref{amp}, we need to fix the extra parameter \(\alpha\). Following the approach outlined above we have found the parametrization:

\bea
\label{alpha}
\alpha\left(\eta\right)&=&\frac{1}{Q^2\left(\hat{s}_{\rm fin}\right)}\left(\frac{16313}{562} + \frac{21345}{124}\eta\right)\,, \\
\label{q_eche}
Q\left(\hat{s}_{\rm fin}\right) & = & \frac{2}{\left(1-\hat{s}_{\rm fin}\right)^{0.45}}\,.
\eea

\noindent \textit{Note that Eq.~\eqref{alpha} is an extension to all symmetric mass-ratio \(\eta\) values of the fit quoted in~\cite{East:2013}}. The fit for the quality factor  \(Q\left(\hat{s}_{\rm fin}\right)\) in Eq.~\eqref{q_eche} was proposed in Ref.~\cite{Eche:1989}. Having determined the analytical expressions for the free parameters \(b(\eta),\, c(\eta)\), \(\kappa(\eta)\) and \(\alpha(\eta)\), we are equipped to provide a robust description of the merger phase for compact binaries with non-spinning components and mass-ratios \(q \leq 10\). \textit{Since this framework enables us to describe in a unified framework the merger of non-spinning compact binaries over a wide range of mass-ratios, we label this formalism `generic IRS' (gIRS) model.} In the left panel of Figure~\ref{irs_generic} we show the suite of numerical simulations used to obtain Eqs.~\eqref{for_b}--\eqref{for_kappa}. The right panel of Figure~\ref{irs_generic} shows that this simple prescription accurately reproduces the evolution of NR waveforms that were \textit{not} used in the calibration of the free parameters \(b, \,c\) and \( \kappa\), i.e., numerical simulations for compact binary systems with \(q=\{3.5,\, 5.5,\, 7.0,\, 7.5,\, 9.5\}\). 

Finally, we obtain the merger waveform by

\bea
\label{m_model}
h^{\rm merger}(t) &=&h^{\rm merger}_{+} - i h^{\rm merger}_{\times}= A(t)\,e^{-i\Phi_{\rm gIRS}(t)}\,,\\
\label{int_phase}
\Phi_{\rm gIRS}(t) &=& \int_{t_0}^{t}\omega(t)\mathrm{d}t\,,
\eea

\noindent where \(\omega(t)\), \(A(t)\) are given by Eq.~\eqref{ome},~\eqref{amp}, respectively, and \(t_0\) is a fiducial value within the range of applicability of the gIRS model. Figure~\ref{mer_fig} shows the regime of applicability of the gIRS model for a variety of compact binary systems, including waveforms that were used for its calibration, and NR simulations that we only use to test the reliability of this scheme. 

To combine the inspiral model from Equation~\eqref{strain_ins} with the gIRS model given by Equation~\eqref{m_model}, we proceed as follows:

\begin{itemize}
\item For the inspiral evolution, \(h^{\rm inspiral}(t)\), we define \(t=0\) at 15Hz.
\item In the merger waveform, we introduce the free parameters \(\Delta t\) and \(\Phi_0\) in Equations~\eqref{m_model} and~\eqref{int_phase}, i.e., \(t\rightarrow t+\Delta t\) and \(\Phi_{\rm gIRS}\rightarrow \Phi_{\rm gIRS} + \Phi_{0}\).
\item To compute \(\Delta t\) and \(\Phi_{0}\), we construct a polynomial using the last three data samples of the inspiral waveform prior to the merger attachment, and require that at the attachment time \(t^*\):
\begin{itemize}
\renewcommand\labelitemii{\textasteriskcentered}
\item the inspiral and merger waveform are continous:  \(h^{\rm inspiral}(t^*) = h^{\rm merger}(t^*)\),
\item the inspiral and merger waveform are differentiable: \(\dot{h}^{\rm inspiral}(t^*) = \dot{h}^{\rm merger}(t^*)\).
\end{itemize}
\item To find the optimal value of attachment, \(t^*_{\rm opt}\), we do the following:
\begin{itemize}
\item For a given \((m_1,\, m_2)\) system we consider a frequency window that includes the quasi-circular ISCO: \(r_{\rm window}=[5M,\, 8M]\). We then sample this window using 200 points, and compute the overlap between our IMR \(ax\) model and its SEOBNRv2 counterpart for each point. We repeat this procedure for the \(m_{1,\,2}\in[5\Msun,\,50\Msun]\) space with a grid that samples the total mass in steps of \(\Delta M=1\) and the mass-ratio in steps of \(\Delta q =0.25\).   
\item Gathering the above information, we construct a map \((M, q)\) that provides the transition point  \(t^*\) that maximizes the overlap between a given \(ax\)--waveform in the zero eccentricity limit and its SEOBNRv2 counterpart. We label this optimized attachment point as \(t^*_{\rm opt}\). 
\end{itemize}
\end{itemize}

The aforementioned attachment procedure covered the window \(r_{\rm window}=[5M,\, 8M]\) because, according to~\cite{Boyle:2007PhRvD}, the quasi-circular 3.5PN calculations can reproduce equal-mass NR simulations with excellent accuracy in the GW frequency range between \(M\Omega \in[0.035,\, 0.15]\). Since our enhanced inspiral evolution includes quasi-circular corrections up to 6PN order, we decided to explore a wide region of parameter space that goes slightly beyond the quasi-circular ISCO \(r_{\rm ISCO}=6M\). We have found, however, that the optimal transition point occurs before the quasi-circular ISCO in all cases.

\begin{figure}[htp!]
\centerline{
\includegraphics[height=0.35\textwidth,  clip]{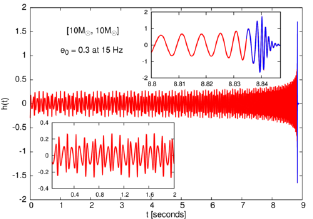}
}
\centerline{
\includegraphics[height=0.35\textwidth,  clip]{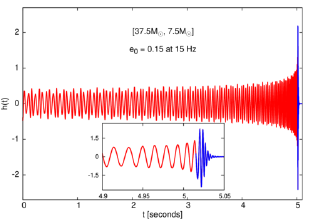}
}
\caption{The panels present the inspiral (red) and merger-ringdown (blue) evolution of two binary black hole systems (see Eq.~\eqref{full_waveform}). The top panel presents a BBH system with component masses \((10\Msun,\, 10 \Msun)\) with an initial orbital eccentricity \(e_0=0.3\) at a GW frequency \(f_{\rm GW}=15{\rm Hz}\). This panel has two insets that show the imprint of eccentricity at low frequencies, and the late-time evolution of this system when the eccentricity has been radiated away. Bottom: BBH system with mass-ratio \(q=5\), total mass \(M=45\Msun\) and \(e_0=0.15\) at \(f_{\rm GW}=15{\rm Hz}\). }
\label{imr_sample}
\end{figure}

An additional comment is in order regarding the validity of this approach in the case of eccentric binaries. In order to ensure that the aforementioned algorithm works for moderately eccentric systems, we have implemented a condition in our waveform code that only attaches a quasi-circular merger waveform to the eccentric inspiral evolution if and only if the residual eccentricity at the attachment point satisfies \(e_{\rm transition}\leq0.05\)  --- see Figure~\ref{mer_ecc_15hz_3pn}. The fact that the optimal attachment point \(t^*_{\rm opt}\) is robust, i.e, we can choose another transition point in the vicinity of \(t^*_{\rm opt}\) that provides a high overlap between IMR \(ax\) and SEOBNRv2 waveforms, implies that this algorithm will remain reliable for systems that meet the condition  \(e_{\rm transition}\leq0.05\) prior to the merger event. In Figure~\ref{imr_sample} we provide two sample waveforms that satisfy this condition. We note that prior to merger the binary systems have circularized, and therefore the attachment procedure that we describe above still applies. In Section~\ref{perform_eccentric} we directly compare our IMR \(ax\) model against eccentric NR simulations and show that this approach performs well.

Under the above considerations the full IMR waveform is written as follows:

\beq
\label{full_waveform}
h(t)=\begin{cases}
h^{\rm inspiral}(t)\quad t\leq t^*_{\rm opt}\,,\\
h^{\rm merger}(t+\Delta t, \Phi_{\rm gIRS} + \Phi_{0})\quad t\geq t^*_{\rm opt}\,.
\end{cases}
\eeq

\noindent Figure~\ref{imr_sample} shows two sample waveforms. The top panel shows a BBH system with component masses \((10\Msun,\, 10 \Msun)\). The total mass of this system is such that it merges at a time when the system has undergone circularization due to GW emission. The bottom left inset in this panel shows the signatures of eccentricity at low frequency, whereas the top right inset shows that the system has undergone circularization prior to merger. The bottom panel shows a BBH system with component masses \((37.5\Msun,\, 7.5 \Msun)\). Since this system is heavier than the previous one, it merges at lower frequencies but still circularizes before merger.

\subsection{Computational Cost}
\label{performance}

Another important aspect of the \(ax\)--model is its computational efficiency. We have benchmarked the performance of the code introduced in this article using the Campus Cluster of the University of Illinois at Urbana-Champaign (CCUIUC). The specifications of the processors used to carry out this work are: Intel(R) Xeon(R) CPUs E5-2660 at 2.20 GHz. 

\begin{figure*}[htp!]
\centerline{
\includegraphics[height=0.4\textwidth,  clip]{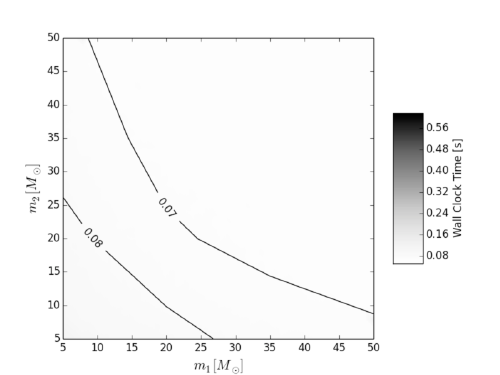}
\includegraphics[height=0.4\textwidth,  clip]{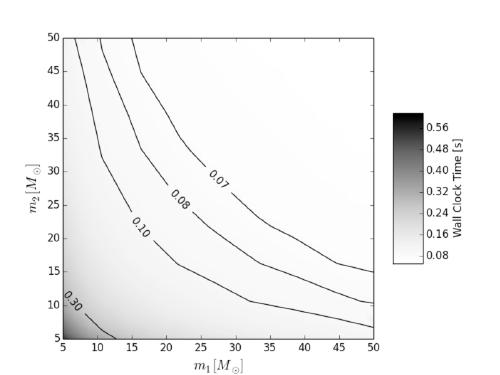}
}
\caption{The panels present the time our code takes to generate a waveform, averaged over 15 iterations, for a given set of parameters. We assume that the binary systems have an initial eccentricity \(e_0\) at a gravitational wave frequency of 15Hz. Left panel: \(e_0=0\). The contour lines indicate how fast we can generate IMR waveforms in different regions of the BBH parameter space under consideration. Right panel: as left panel but for systems with \(e_0=0.4\).}
\label{bench_results_15hz}
\end{figure*}

In order to take into account the fluctuation in performance of compute nodes at the CCUIUC, we compute a waveform for a given set of parameters fifteen times, and quote the average time in Figure~\ref{bench_results_15hz}. Assuming an initial frequency of 15Hz, Figure~\ref{bench_results_15hz} indicates that the time taken by our code to generate a waveform for binaries with total mass \(M=10\Msun\) and \(e_0=0.4\) is about 0.5 seconds, and ten times faster for quasi-circular systems. Reducing the starting evolution frequency to 10Hz increases the computational cost by about a factor of two.

\noindent These results indicate that the \(ax\)--waveform model is fast enough to perform large scale parameter estimation studies over the BBH parameter space that can be detected with aLIGO. We are currently implementing this code in the LIGO Algorithms Library~\cite{LAL}. 

\subsection{Data analysis toolkit}
\label{dat}
In order to quantify the performance of the \(ax\)--model in the zero eccentricity limit, we introduce basic GW data analysis tools. Given two signals \(h\) and \(s\), the noise-weighted inner product is defined as

\beq
\left( h | s\right) = 2 \int^{f_{\rm max}}_{f_{\rm min}} \frac{\tilde{h}^{*}(f)\tilde{s}(f) + \tilde{h}(f)\tilde{s}^{*}(f) }{S_n(f)}\mathrm{d}f\,,
\label{inn_pro}
\eeq

\noindent where \(S_n(f)\) represents the power spectral density (PSD) of the detector noise, and \(\tilde{h}(f)\) is the Fourier transform of \(h(t)\). We take the lower limit of the integral to be \(f_{\rm min}=15\,{\rm Hz}\), and \(f_{\rm max}=4096\,{\rm Hz}\). We generate the waveforms using a sample rate of \(8192\,{\rm Hz}\). The matched-filter signal-to-noise ratio (SNR) is given by

\beq
\label{snr}
\rho = \frac{\left(s | h\right)}{\sqrt{\left(h | h\right)}}\,.
\eeq

\noindent Using Eq.~\eqref{inn_pro} we construct the normalized waveform:

\beq
\hat{h}=h\,\left(h | h\right)^{-1/2}\,,
\label{nor_ov}
\eeq

\noindent and the normalized overlap

\beq
\label{over}
{\cal{O}}  (h,\,s)= \underset{ t_c\, \phi_c}{\mathrm{max}}\left(\hat{h}|\hat{s}_{ t_c,\,  \phi_c}\right)\,,
\eeq

\noindent where \(\hat{s}_{ t_c,\,  \phi_c}\)  indicates that the normalized waveform \(\hat{s}\)  has been time- and phase-shifted. The Fitting Factor \(({\cal{FF}})\) is defined as the maximum value of maximized normalized overlaps between a GW signal \(h^e\) and all members \(h_b^{T}\) of a bank of template waveforms~\cite{FittingFactorApostolatos}

\beq
\label{ff}
{\cal{FF}}=  \underset{ b\in{\rm bank}}{\mathrm{max}}{\cal{O}} \left(h^e,\, h_b^{T}\right)\,.
\eeq

\noindent The observed SNR \(\rho'\) is related to the optimal SNR \(\rho\) and the fitting factor through the relation:

\beq
\label{opt_vs_obs}
\rho' = {\cal{FF}}\rho\,.
\eeq

\noindent The waveforms detected by the aLIGO detectors are a combination of the two independent GW polarizations \(h_{+}(t)\) and \(h_{\times}(t)\) through the relation~\cite{SathyaLRR:2009}:

\bea
\label{strain}
H(t)&=& F_{+}\left(\theta,\, \varphi,\, \psi\right) h_{+}(t)\nonumber\\&+&F_{\times}\left(\theta,\, \varphi,\, \psi\right) h_{\times}(t)\,,\\
\label{f+}
F_{+}\left(\theta,\, \varphi,\, \psi\right)  &=& -\frac{1}{2}\left(1+\cos^2\theta\right)\cos2\varphi\cos2\psi\nonumber\\&-&\cos\theta\sin2\varphi\sin2\psi\,,\\
\label{fx}
F_{\times}\left(\theta,\, \varphi,\, \psi\right)&=& \frac{1}{2}\left(1+\cos^2\theta\right)\cos2\varphi\sin2\psi\nonumber\\&-&\cos\theta\cos2\varphi\cos2\psi\,,
\eea

\noindent where \((\theta, \, \varphi)\) represent the Euler angles of the detector, and \(\psi\) is the Euler angle of the polarization plane. 

\subsection{Behavior in the zero eccentricity limit}
\label{perform}

\begin{figure}[htp!]
\centerline{
\includegraphics[height=0.35\textwidth,  clip]{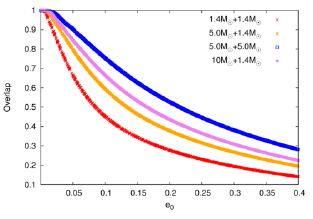}
}
\caption {Overlap between the \(ax\)--model and the approximant \texttt{TaylorT4} including 3.5PN corrections. We have used the Zero Detuned High Power sensitivity configuration for Advanced LIGO and a lower frequency cut--off of 15Hz.}
\label{overlap_with_t4}
\end{figure}

In order to exhibit that the \(ax\)--model renders the expected evolution for inspiral dominated systems in the quasi-circular limit, in Figure~\ref{overlap_with_t4} we present the results of overlap calculations between the \(ax\)--model and \texttt{TaylorT4} at 3.5PN order. Please note that we have used Eq.~\eqref{def_35pn} for this study. Comparisons with \texttt{TaylorT4} at 2PN, 2.5PN and 3PN render a similar behavior, and have the correct asymptotic behavior in the zero eccentricity limit. In these calculations we assume that the binaries are optimally oriented, i.e., \(F_{+}=1,\, F_{\times} =0\). We use the Zero Detuned High Power sensitivity configuration for aLIGO~\cite{ZDHP:2010} and a low frequency cut--off of 15Hz. 

\begin{figure*}[htp!]
\centerline{
\includegraphics[height=0.4\textwidth,  clip]{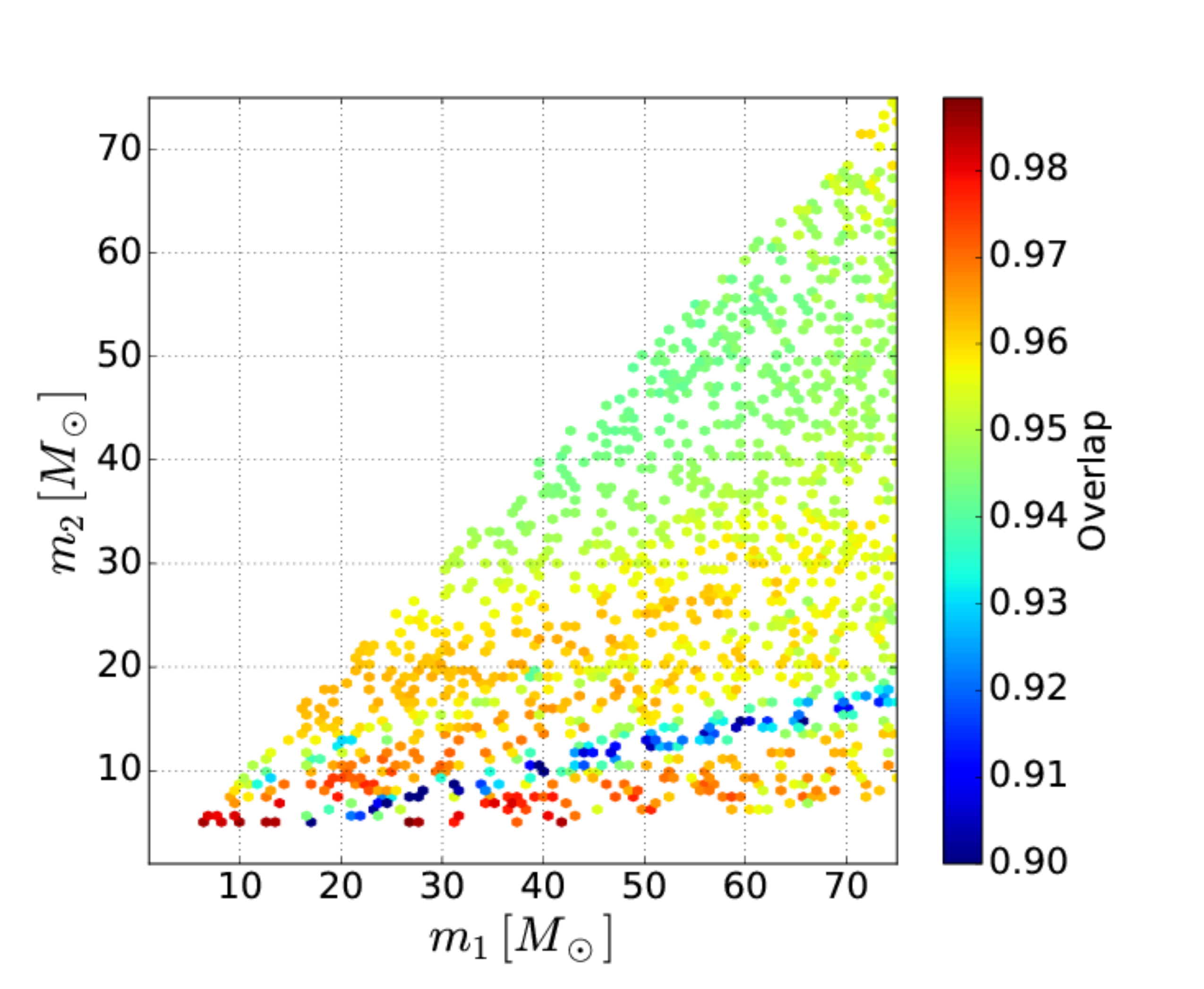}
\includegraphics[height=0.4\textwidth,  clip]{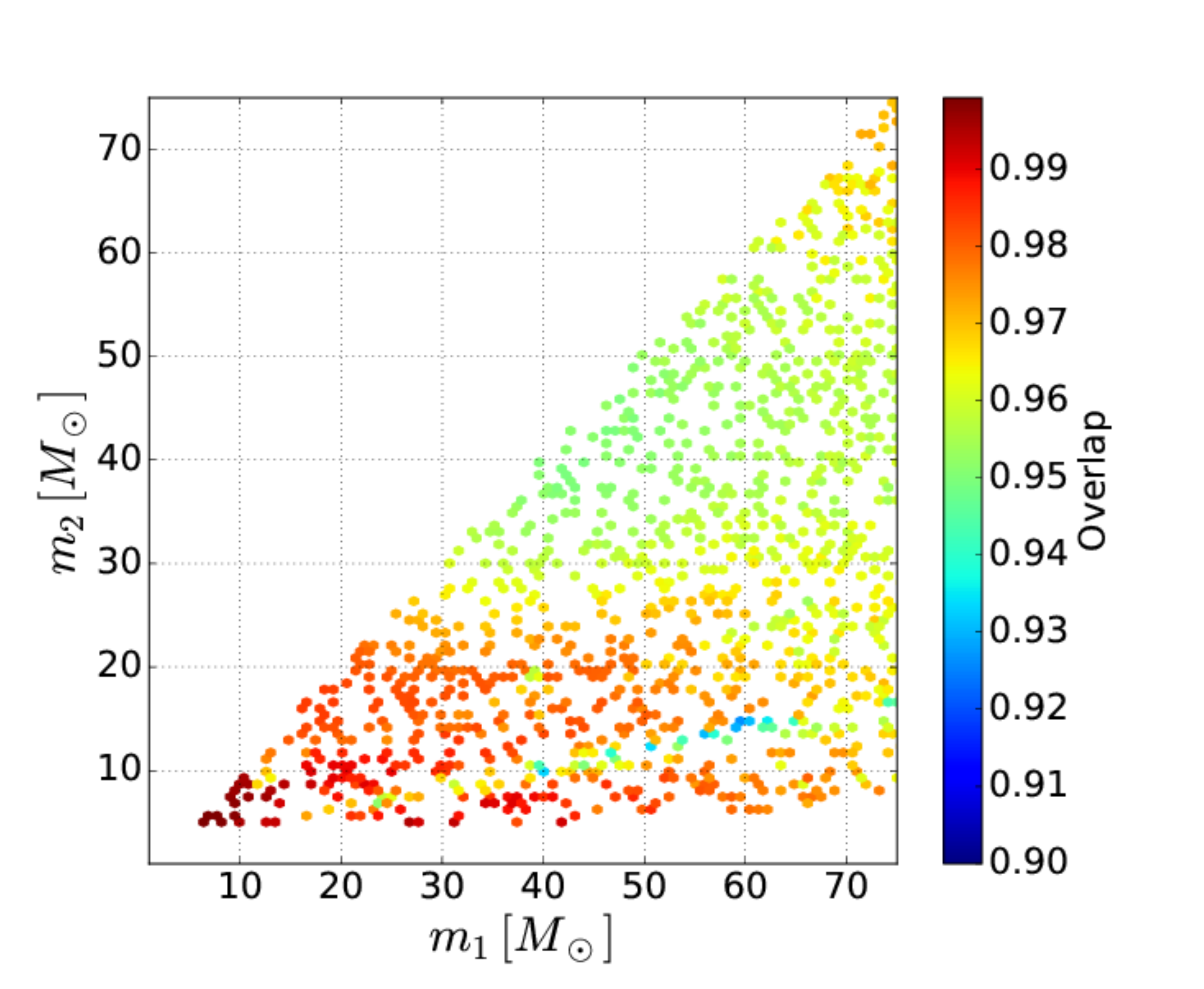}
}
\caption{Overlap between the \(ax\)--model in the zero eccentricity limit and the SEOBNRv2 model. Left panel: the overlap is computed from an initial gravitational wave frequency of 15Hz using the Zero Detuned High Power sensitivity configuration for aLIGO. Right panel: the overlap is computed from an initial gravitational wave frequency of 25 Hz using the mid aLIGO sensitivity curve described in Ref.~\cite{scenarioligo:2016LRR}. We note that in both cases a large portion of the parameter space under consideration is accurately reproduced by the \(ax\)--model with  \({\cal{O}}\gtrsim 0.95\). }
\label{nice_fig}
\end{figure*}

\noindent We have also explored the performance of the \(ax\)--model in the quasi-circular limit using non-spinning IMR SEOBNRv2 waveforms. For this study we have used the improved inspiral evolution of the \(ax\) model given by Eq.~\eqref{dxdtT4_byho}. We consider a BBH population with components masses  \(m_{1,\,2}\in[5\Msun,\,75\Msun]\), i.e., mass-ratios up to \(q=15\). In Figure~\ref{nice_fig} we present the overlap between the IMR \(ax\)--model with \(e=0\) and SEOBNRv2 for two different scenarios of aLIGO sensitivity~\cite{scenarioligo:2016LRR}. The left panel corresponds to the Zero Detuned High Power sensitivity configuration for aLIGO, using a lower frequency cut-off of 15Hz. The right panel represents the `mid aLIGO' sensitivity configuration, which serves as proxy for the upcoming observing runs O2/O3~\cite{scenarioligo:2016LRR}, using a low frequency cut--off of 25Hz. 

In the left panel of Figure~\ref{nice_fig} we find overlaps \({\cal{O}}\gtrsim 0.95\), indicating that the quasi-circular limit of the IMR \(ax\)--model can reproduce the dynamical evolution predicted by the SEOBNRv2 model over a wide region of the BBH parameter space.  We should take these results, even if they are positive, with a grain of salt since neither of these models have been calibrated with NR simulations that represent systems with \(q>\)10.  The right panel of Figure~\ref{nice_fig} indicates that agreement between the IMR \(ax\)--model in the zero eccentricity limit and SEOBNRv2 is better when we consider the `mid aLIGO' sensitivity configuration, which is expected given its narrower sensitive frequency band.

\noindent Figure~\ref{nice_fig} is the first comparison of an IMR eccentric waveform with a state-of-the-art quasi-circular IMR waveform model such as SEOBNRv2. The panels in Figure~\ref{nice_fig} indicate that our IMR \(ax\) model can reproduce non-spinning SEOBNRv2's dynamics with an average overlap \({\cal{O}}_{\rm average} \sim 0.95\) and that some regions of parameter space have \({\cal{O}}_{\rm max} \sim 0.99\). \textit{This is the first IMR eccentric model in the literature that has this level of agreement with SEOBNRv2 for BBH systems with mass-ratios \(1\leq q \leq 15\).} We notice, however, that the model has anomalously low overlaps, \({\cal{O}}_{\rm min}\sim 0.9\), for binaries in a narrow band of mass-ratios centered at \(q\sim4\). We can understand this undesirable feature by taking a closer look at the construction of our waveform model. In Figure~\ref{imr_sample} we see that we combine a PN-based eccentric inspiral model with a merger waveform very late in the inspiral evolution. This late-time attachment, however, does not work uniformly well in the binary parameter space, and introduces anomalous features in the model for \(q \sim 4\). While the gIRS model provides a good description of the merger dynamics in the vicinity of the light-ring, its accuracy deteriorates rapidly several cycles before merger, cf. Figure~\ref{mer_fig}. In different words, we are pushing the PN equations of motion to the limit of their applicability to ensure we get the best possible overlap with SEOBNRv2. We comment in Section~\ref{disc} on possible improvements to the gIRS model.

Figure~\ref{nice_fig} demonstrates the importance of the amended inspiral dynamics in Eq.~\eqref{dxdtT4_byho}. Without those amendments, typical overlap values between the \(ax\) model and SEOBNRv2 are \({\cal{O}}\lesssim 0.5\) for systems with mass-ratios \(q\gtrsim4\). Thus, the corrections that have to be implemented to ensure that the minimum overlap between the IMR \(ax\)--model and SEOBNRv2 satisfies \({\cal{O}}\gtrsim 0.99\) over the whole BBH space are within reach with additional work that we describe in Section~\ref{disc}. Furthermore, as we show in Section~\ref{perform_eccentric}, this approach can reproduce the dynamics of comparable mass-ratio, moderately eccentric NR simulations.

At present, the \(ax\)--model presented in this article can be used to: (i) explore how well eccentricity can be measured in parameter estimation studies. We can do this by injecting \(ax\) signals in real data and do a parameter estimation analysis with \(ax\) templates; (ii) study the bias incurred in parameter estimation studies caused by the intrinsic inaccuracies of the \(ax\)--model. We can do this by injecting NR waveforms and doing a parameter estimation study with \(ax\) templates; (iii)  furthermore, we can study how well eccentric BBH signals can be recovered with non-eccentric waveform templates by using eccentric \(ax\)--waveforms as injections to be recovered with a template bank consisting of non-eccentric waveform templates.  This will be the topic of Section~\ref{catch_me}.

\subsection{Comparison to eccentric numerical relativity simulations}
\label{perform_eccentric}

In this Section we directly compare IMR \(ax\) waveforms with a set of eccentric NR simulations that we have generated with the \texttt{Einstein Toolkit}~\cite{ETL:2012CQGra,etweb,2004CQGra..21..743T,2004CQGra..21.1465S}. To translate the NR relativity orbital eccentricity parameter into the PN version that is used in our IMR \(ax\) waveform, we use the fitting procedure described in Section II of~\cite{Hinder:2010}, but now using higher-order eccentric and quasi-circular PN corrections. 

\begin{figure}[htp]
\centerline{
\includegraphics[height=0.35\textwidth,  clip]{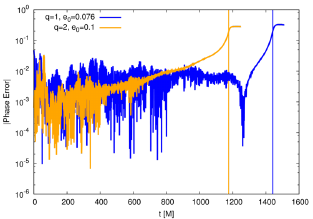}
}
\caption{Using the \texttt{Richardson Extrapolation} we provide a phase error estimate for each of our highest resolution eccentric NR simulations. The vertical lines indicate the merger time of each of the BBH systems under consideration.}
\label{convergence}
\end{figure}

The two simulations we use to assess the accuracy of our IMR \(ax\) model correspond to BBH systems with the following properties: (i) equal mass compact binary system with initial orbital eccentricity \(e_0=0.076\) and mean anomaly \(\ell_0=3.09\) at \(x_0=0.074\); (ii) compact binary system with mass-ratio \(q=2\), eccentricity \(e_0=0.1\) and mean anomaly \(\ell_0=3.11\) at \(x_0=0.076\). For each of these simulations we run three different resolutions. The convergence order of the numerical scheme used by the \texttt{Einstein Toolkit} for vacuum BBH simulations is 8. We have found that our simulations have convergence orders consistent with this value, namely: 8 and 9 for the \(q=\{1,\,2\}\) BBH simulations, respectively. We expect that the slight deviation from the nominal convergence order of 8 for the scheme in the \(q=2\) BBH simulation is due to either still unresolved effects near the punctures, and interpolation artifacts in the mesh refinement and curvilinear grid boundaries. The observed convergence order becomes less well defined near merger when phase errors accumulate rapidly --- see Figure~\ref{convergence}, where we use the \texttt{Richardson Extrapolation} to provide an estimate of the phase error of the highest resolution run of each set of our NR simulations. In the analysis below, we use the highest resolution run of each mass-ratio case.

\begin{figure*}[htp!]
\centerline{
\includegraphics[height=0.35\textwidth,  clip]{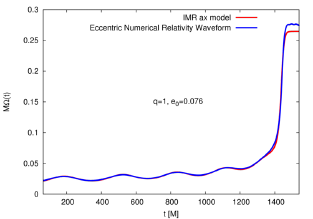}
\includegraphics[height=0.35\textwidth,  clip]{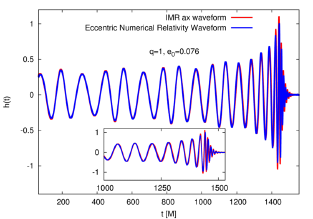}
}
\caption{For an equal mass BBH system with initial eccentricity \(e_0=0.076\) and mean anomaly \(\ell_0=3.09\) at a gauge-invariant frequency value \(x_0=0.074\), we present a direct comparison of the dynamics predicted by our IMR \(ax\) model and an eccentric NR simulation. Left panel: our IMR \(ax\) predicts with very good accuracy the orbital frequency evolution throughout late inspiral, merger and ringdown. Right panel: our IMR \(ax\) model can accurately reproduce the true NR features of the amplitude and phase evolution of an equal mass, eccentric BBH merger.}
\label{here_it_is}
\end{figure*}

\noindent In the left panel of Figure~\ref{here_it_is} we show the time evolution of the orbital frequency for an equal mass BBH system that has an initial orbital eccentricity \(e_0=0.076\) and mean anomaly \(\ell_0=3.09\) at \(x_0=0.074\). We notice that our IMR \(ax\) model reproduces the orbital evolution throughout the entire evolution of the eccentric NR simulation. The final orbital frequency asymptotes to the values \(M\Omega^{\rm ringdown}_{\rm NR}=0.275\), whereas \(M\Omega^{\rm ringdown}_{ax}=0.265\), i.e., our model has a \(\sim4\%\) discrepancy from the eccentric NR value. The right panel of Figure~\ref{here_it_is} shows a direct comparison between the corresponding IMR \(ax\) waveform and its NR counterpart. This comparison exhibits two important features: our IMR \(ax\) model reproduces with excellent accuracy the amplitude modulations of eccentric mergers, and the waveform remains in phase throughout the length of the eccentric NR evolution. These results indicate that the strategy we have followed to compute higher-order eccentric PN corrections for the instantaneous and hereditary terms is the right approach to reproduce the true evolution of eccentric compact binary coalescence. 

\begin{figure*}[htp!]
\centerline{
\includegraphics[height=0.35\textwidth,  clip]{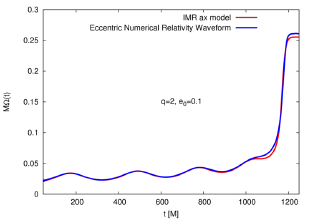}
\includegraphics[height=0.35\textwidth,  clip]{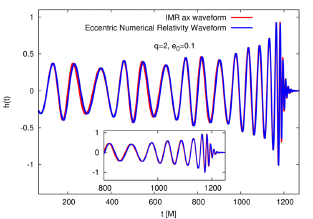}
}
\caption{As Figure~\ref{here_it_is}, but now for a BBH system with mass-ratio \(q=2\), initial eccentricity \(e_0=0.1\), mean anomaly \(\ell_0=3.11\) and gauge-invariant frequency parameter \(x_0=0.076\).}
\label{here_it_is_two}
\end{figure*}

\noindent In Figure~\ref{here_it_is_two} we perform a similar exercise for a BBH merger with mass-ratio \(q=2\), eccentricity \(e_0=0.1\) and mean anomaly \(\ell_0=3.11\) at \(x_0=0.076\). We notice that the ringdown frequency of our IMR \(ax\) model and the eccentric NR counterpart differ by \(\sim 3\%\). These results further confirm that our IMR \(ax\) model renders a good description of eccentric compact binary coalescence for compact mass-ratio systems throughout the merger. 

It is worth mentioning that the discrepancy on the predicted values for the ringdown frequency between our IMR \(ax\) model and our eccentric NR simulations can be accounted for by the numerical error of our numerical simulations. Future work should include a larger set of eccentric NR simulations for calibration and validation of new eccentric waveform models.

\section{Detectability of eccentric unequal mass binaries}
\label{catch_me}

A previous study~\cite{Huerta:2013a} of the importance of eccentricity to model and detect BNSs that have moderate values of residual eccentricity used the \(x\)--model of Ref.~\cite{Hinder:2010}. Now that we have developed the IMR \(ax\)--model, we are equipped to extend that analysis to systems that have asymmetric mass-ratios including the inspiral, merger and ringdown phases. \textit{To the best of our knowledge, this is the first analysis of this nature in the literature.} 

\begin{figure*}[htp!]
\centerline{
\includegraphics[height=0.4\textwidth,  clip]{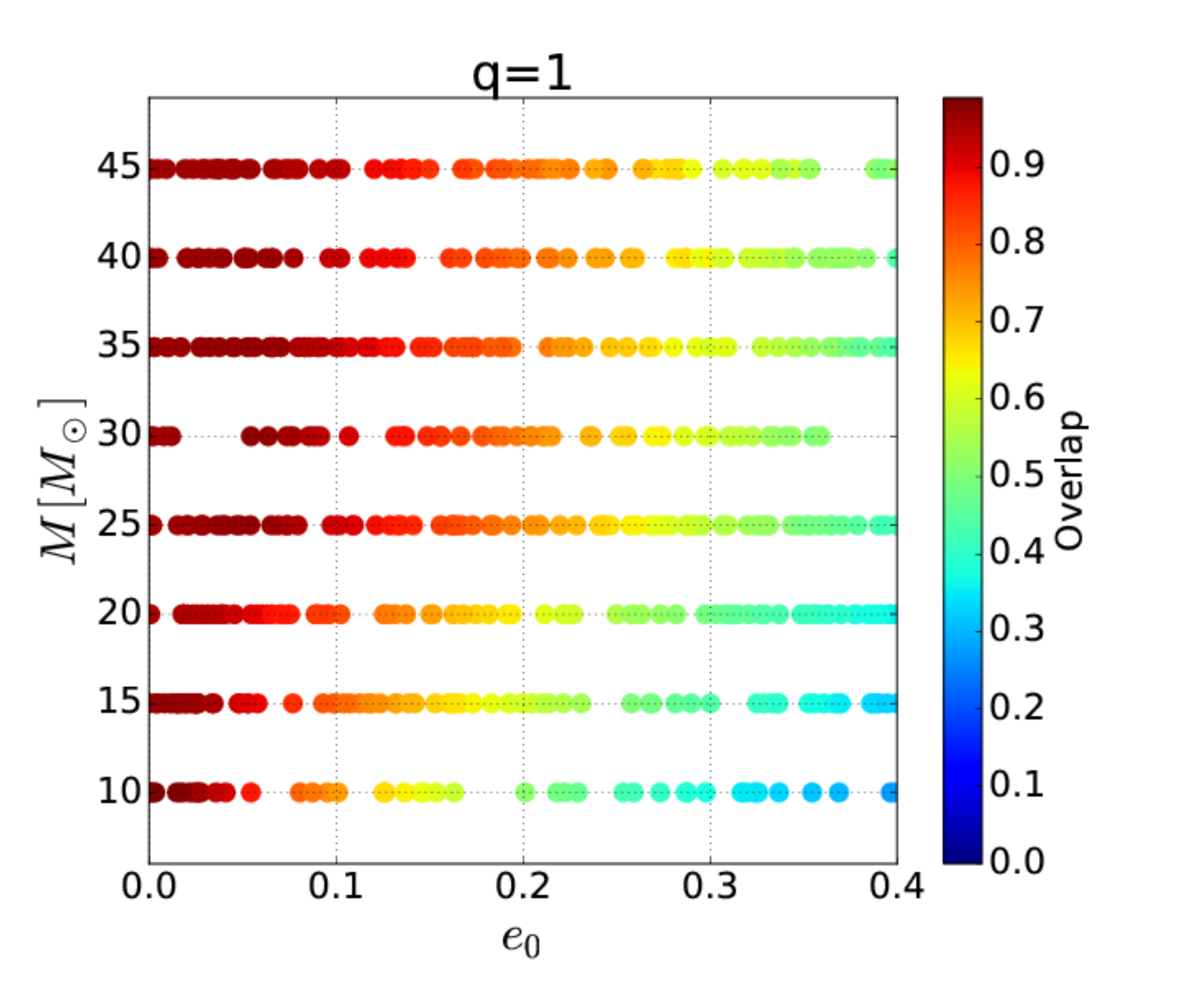}
\includegraphics[height=0.4\textwidth,  clip]{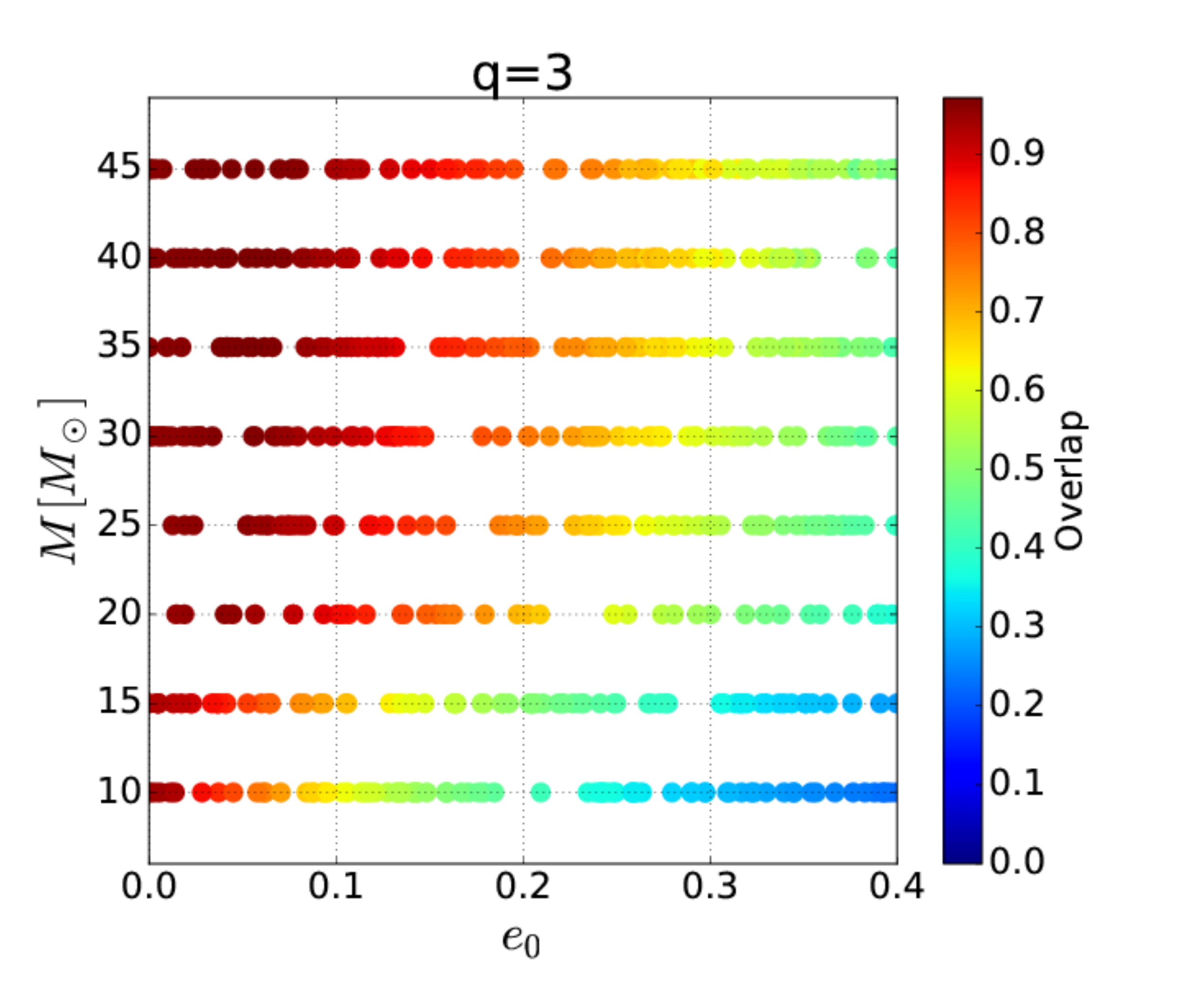}
}
\centerline{
\includegraphics[height=0.4\textwidth,  clip]{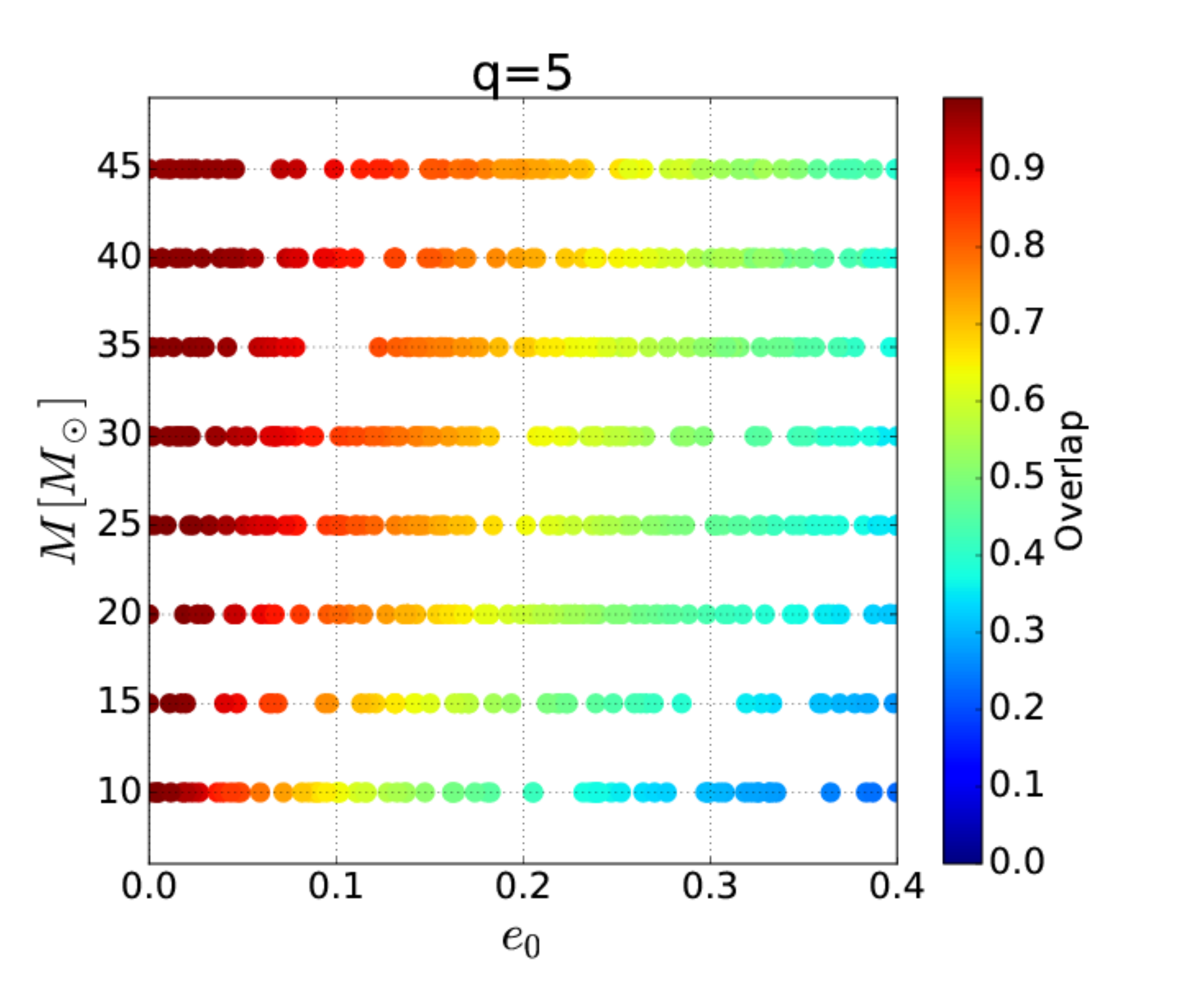}
\includegraphics[height=0.4\textwidth,  clip]{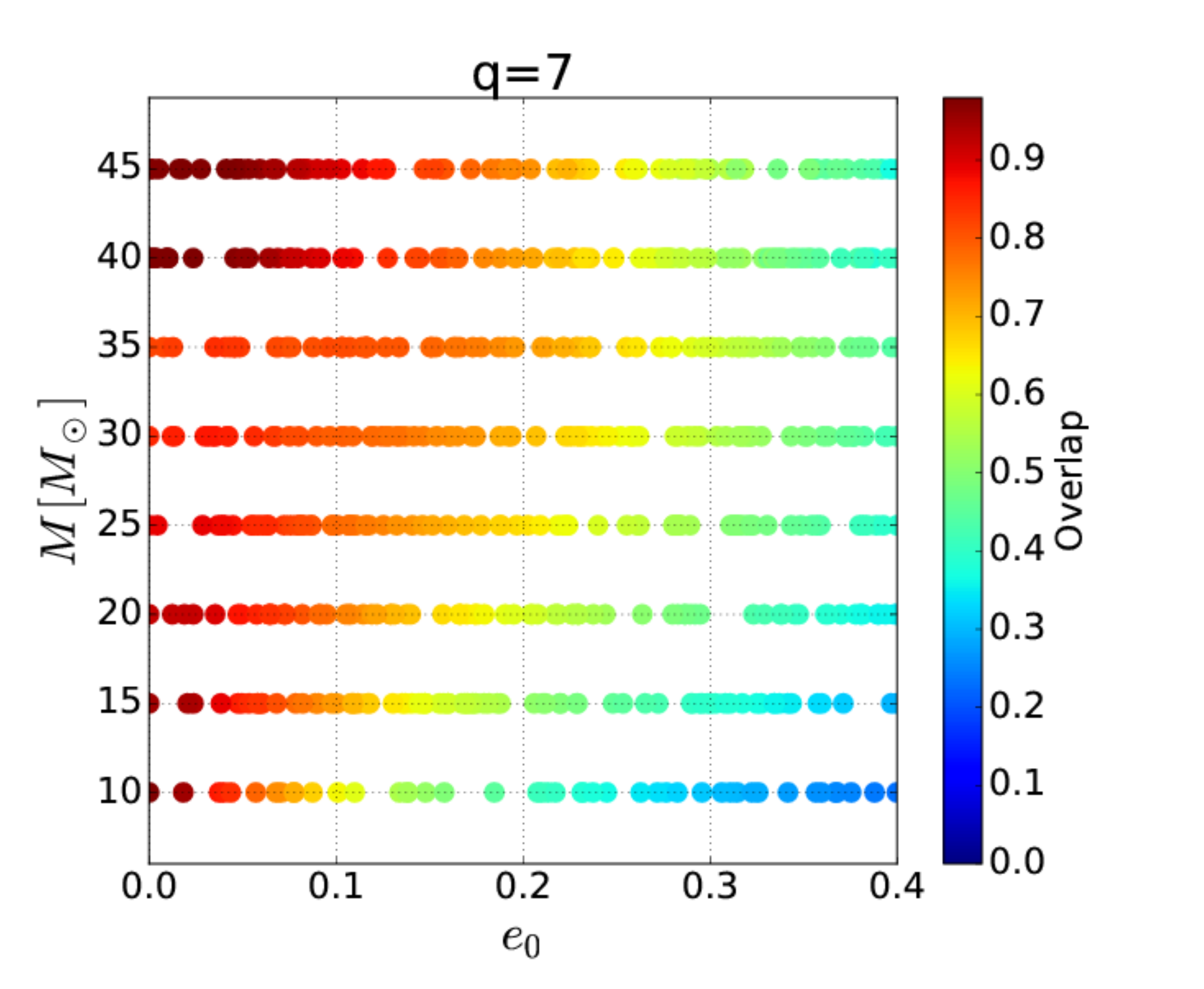}
}
\caption{The panels present overlap calculations between IMR \(ax\) and SEOBNRv2 waveforms. The IMR \(ax\) waveforms are generated for binaries that enter aLIGO band with eccentricities \(e_0\in[0,\,0.4]\) at \(f_{\rm GW}= 14{\rm Hz}\). The overlaps are computed from an initial gravitational wave frequency \(f_{\rm min}= 15{\rm Hz}\) (see Eqs.~\eqref{inn_pro}) using the Zero Detuned High Power sensitivity configuration for aLIGO. }
\label{nice_fig_two}
\end{figure*}

The first part of this analysis is related to quantifying the effect of eccentricity in the dynamical evolution of stellar mass BBH and NSBH systems. We carry out this study by directly comparing SEOBNRv2 waveforms against IMR \(ax\) waveforms using astrophysically motivated values of eccentricity, i.e., \(e_0\in[0,\,0.4]\), where \(e_0\) is defined at \(f_{\rm GW}= 14{\rm Hz}\). The results of this study are presented in Figure~\ref{nice_fig_two} for compact binaries with mass-ratios \(q=\{1,\,3,\,5,\,7\}\). We restrict this study to systems with total mass \(M \leq 45M_\odot\), since such binaries will effectively circularize by the time they reach their ISCO, i.e., \(e_{\rm ISCO}\leq 0.05\) --- see Figure~\ref{mer_ecc_15hz_3pn}. These results were obtained using \(f_{\rm min}= 15{\rm Hz}\) (see Eq.~\eqref{inn_pro}), and the Zero Detuned High Power PSD of aLIGO.

\noindent The results presented in Figure~\ref{nice_fig_two} indicate that low mass binaries with very asymmetric mass-ratios are the systems that differ the most from their quasi-circular counterparts. For instance, the overlaps between SEOBNRv2 and IMR \(ax\)--waveforms for a \((5\Msun,\, 5\Msun)\) BBH and a \((8.75\Msun,\, 1.25\Msun)\) NSBH binary that enter the aLIGO band with \(e_0=0.1\) at \(f_{\rm GW}=14{\rm Hz}\) are: \({\cal{O}}\sim 0.75\) and  \({\cal{O}}\sim 0.6\), respectively. This significant drop in overlap is caused by several factors: (i) eccentricity corrections have a cumulative effect in the orbital phase of waveform signals. Therefore, the orbital phase of long lived eccentric signals will significantly deviate from their quasi-circular counterparts. In a population of binaries with total mass \(M\), those with the most asymmetric mass-ratios have the longest lifespan. Therefore, we expect that the most significant drop in overlap between eccentric and quasi-circular systems should correspond to NSBHs and BBHs with asymmetric mass-ratios, as shown in Figure~\ref{nice_fig_two}; (ii) eccentricity reduces the lifespan of waveform signals. Signals with \(e_0=0.4\) are a factor \(\sim 2\) shorter than their quasi-circular counterparts. Therefore, it is no surprise that the overlap between these type of signals and SEOBNRv2 is \({\cal{O}}\sim0.2\). Putting (i) and (ii) together, we can understand that this effect is exacerbated for low mass, asymmetric mass-ratio systems. On the other hand, more massive systems spend less time in the aLIGO band, preventing eccentricity corrections to accumulate. As a result, the overlap between quasi-circular templates and eccentric binaries with  \(M\sim45\Msun\) and \(e_0\leq0.1\)  is  \({\cal{O}}\geq 0.9\). 

Having developed a basic understanding on the effect of eccentricity in terms of the total mass and mass-ratio of compact binaries, we now turn our attention to the detectability of eccentric signals using template banks of quasi-circular waveforms. We can quantify the recovery of non-spinning, eccentric signals using two types of template banks of quasi-circular waveforms: (i) SEOBNRv2 template banks allow us to do recovery with non-spinning and aligned-spin templates. Therefore, we can test whether aligned-spin templates do capture the effect of eccentricity.  This is important, because GW searches with aLIGO utilize aligned-spin templates, so this is a relevant question when assessing aLIGO's sensitivity to eccentric systems. Unfortunately, this yields ambiguities at small \(e_0\), because small \(e_0\) \(ax\)--injections do not perfectly agree with SEOBNRv2. This information is conveyed in Figure~\ref{nice_fig_two}: overlaps with \(e_0\lesssim0.05\) are comparable to their quasi-circular counterparts, i.e., it is not possible to make clear cut statements about the effect of eccentricity for systems with \(e_0\lesssim0.05\). Rather, these overlap calculations provide information about the accuracy of the \(ax\)--waveforms in the zero eccentricity limit. On the other hand, overlaps for systems with \(e_0\gtrsim0.1\) significantly  drop from the quasi-circular case, which indicates that for \(e_0\gtrsim0.1\) we are probing predominantly the effect of eccentricity. As discussed above, these boundaries depend on the total mass and mass-ratio of the systems, with high masses being less sensitive to eccentricity. (ii)  Conversely, with \(ax\)--template banks we can make rigorous statements about recovery efficiency of small eccentricity injections with \textit{non-spinning} templates, but cannot make statements about recovery with aligned-spin templates. For the present study, we choose the first approach and consider two scenarios: (\texttt{a}) we set the spin of the binary components to zero and construct a template bank that describes binaries with non-spinning components on quasi-circular orbits, (\texttt{b}) we construct a template bank that describes binaries on quasi-circular orbits whose components have spin in the \(z\) direction only. To quantify the effectualness with which these template banks recover eccentric signals, we computed  \({\cal{FF}}{\rm 's}\) from an initial \(f_{\rm min}=15{\rm Hz}\) using the Zero Detuned High Power sensitivity configuration for aLIGO. The simulated eccentric signals enter the aLIGO band with initial eccentricity \(e_0\) at \(f_{\rm GW}=14{\rm Hz}\).

\begin{figure}[htp!]
\centerline{
\includegraphics[height=0.39\textwidth,  clip]{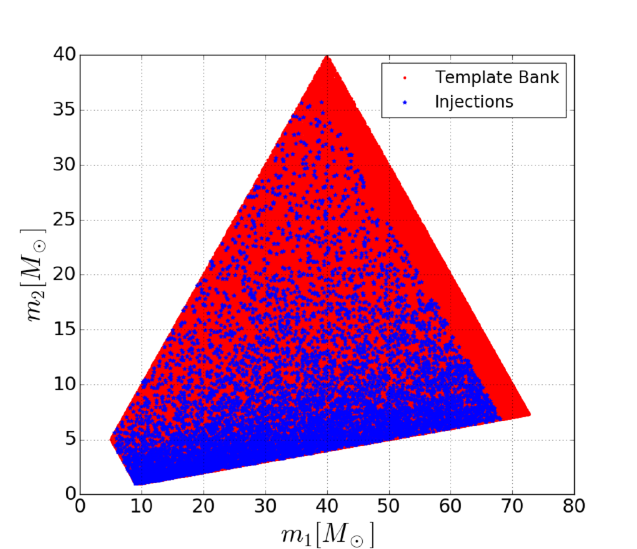}
}
\caption{The panel shows the coverage of the mass parameter space \((m_1,\,m_2)\) using \(10^6\) quasi-circular template waveforms. We also show the mass distribution of the \(8\times10^3\) simulated eccentric signals or `injections'.}
\label{bank_and_injection}
\end{figure}

In order to ensure that the template bank discreteness does not affect the recovery of simulated eccentric signals, we constructed template banks with \(5\times10^5\), \(10^6\) and \(1.5\times10^6\) waveforms, and tested the convergence of the \({\cal{FF}}{\rm 's}\) presented below. We found that the bank constructed with non-spinning waveforms is a proper subset of the spin-aligned bank when we densely sample the parameter space using 1M waveforms. We compared the \({\cal{FF}}{\rm 's}\) obtained using the spin-aligned bank with \(10^6\) and \(1.5\times10^6\) waveforms and confirmed that the \({\cal{FF}}{\rm 's}\) were exactly the same. In different words, this consistency check indicates that the results we present below represent the true maximum \({\cal{FF}}{\rm 's}\), which surpass the effect of template bank discreteness. In Figure~\ref{bank_and_injection}, we show the coverage of the mass parameter space \((m_1,\,m_2)\) used for the construction of these template banks, and the \(8\times10^3\) eccentric simulated signals or `injections'.

\begin{figure*}[htp!]
\centerline{
\includegraphics[height=0.33\textwidth,  clip]{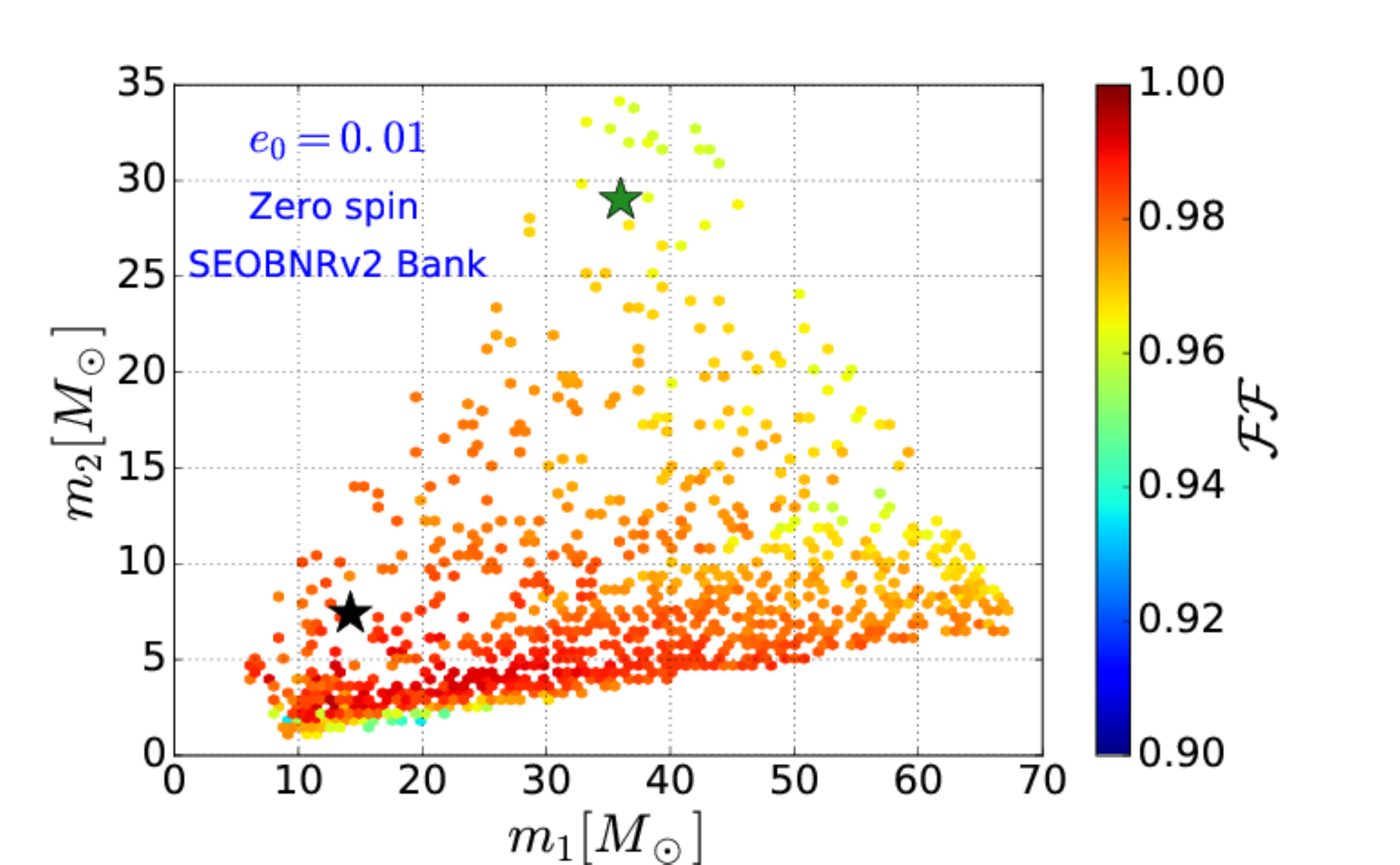}
\includegraphics[height=0.33\textwidth,  clip]{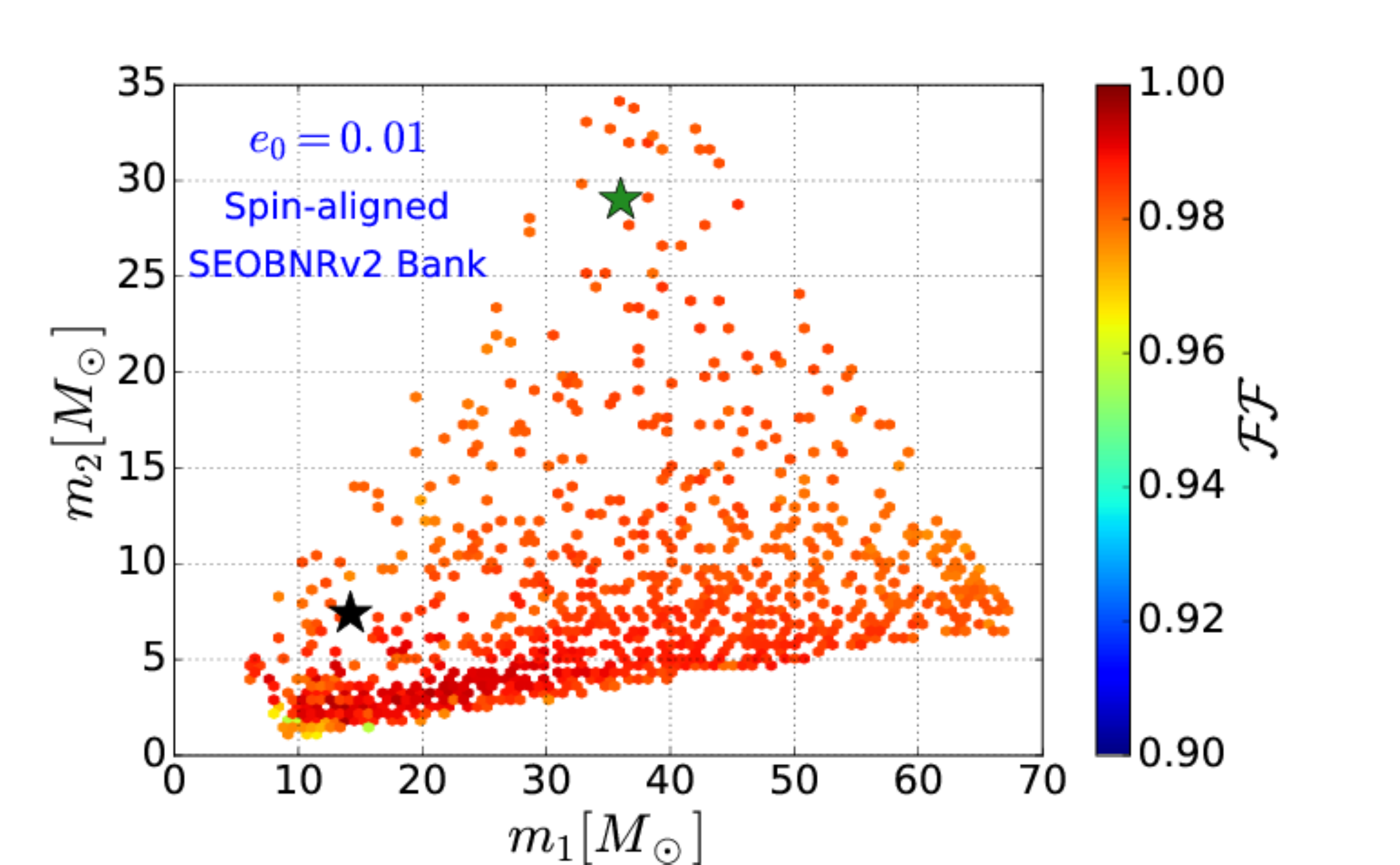}
}
\centerline{
\includegraphics[height=0.33\textwidth,  clip]{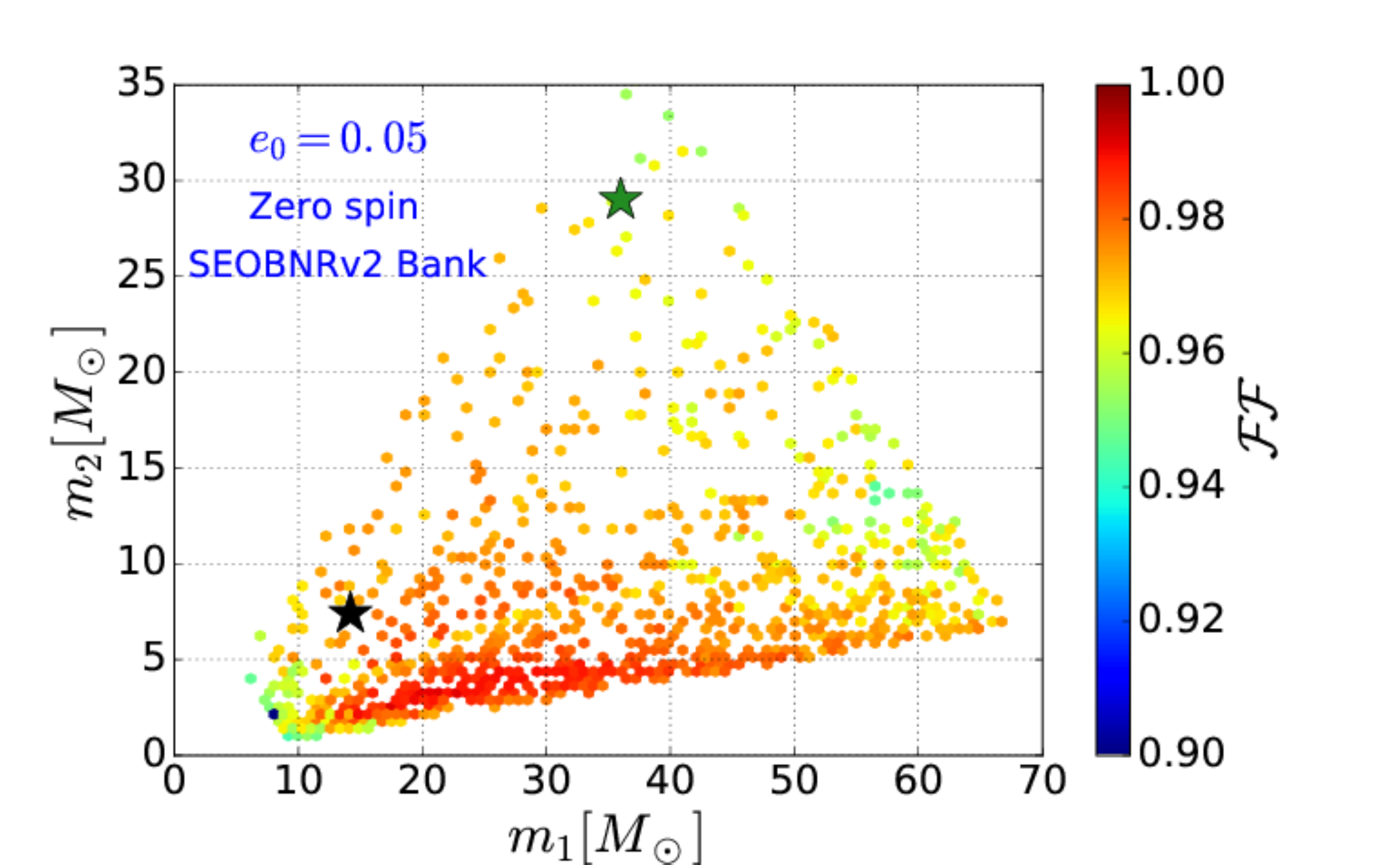}
\includegraphics[height=0.33\textwidth,  clip]{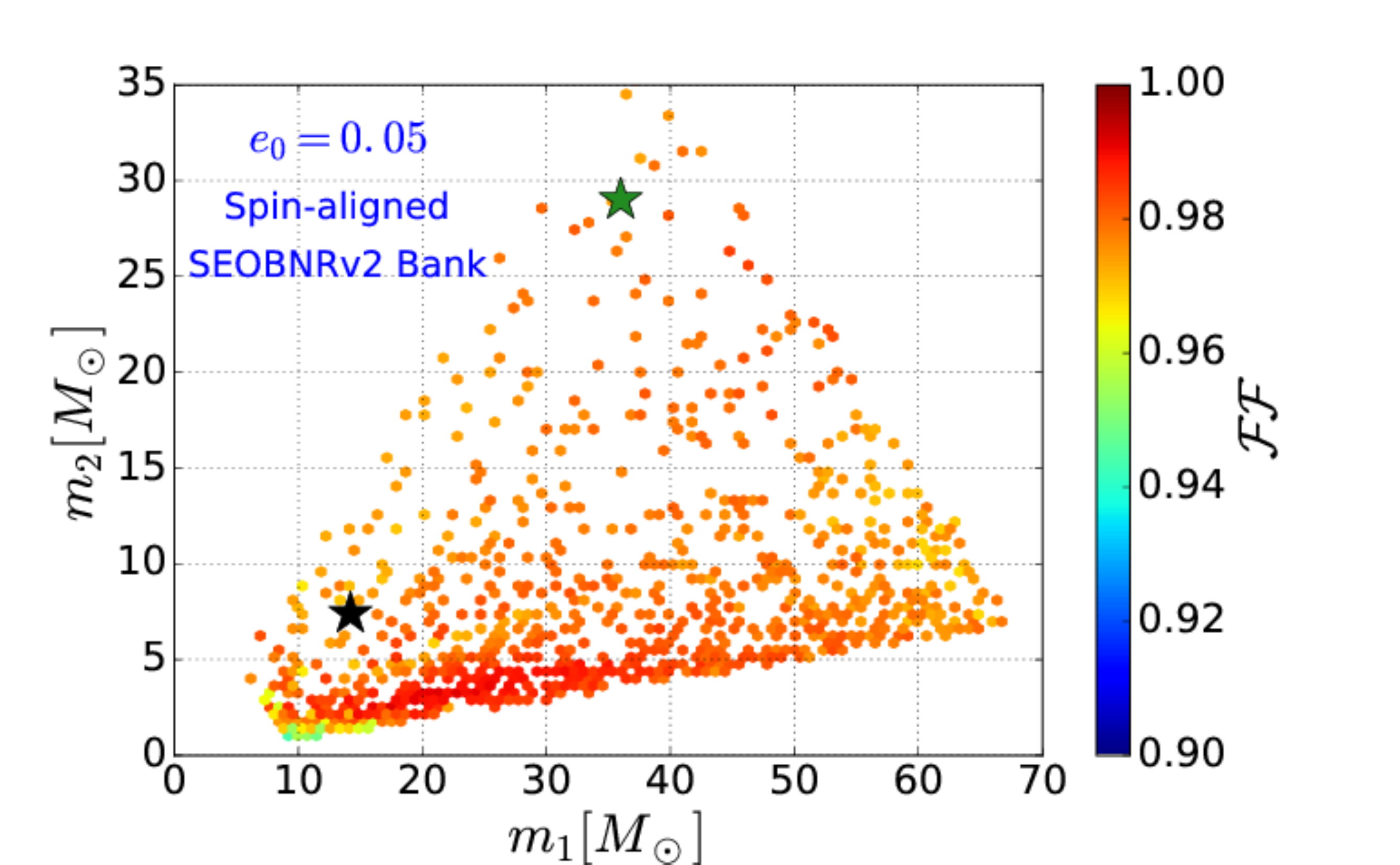}
}
\caption{Left panels: effectualness of a bank of quasi-circular, non-spinning SEOBNRv2 templates to recover a population of eccentric, non-spinning signals. Right panels: recovery of non-spinning eccentric injections with an \emph{aligned-spin} template bank of SEOBNRv2 waveforms. Each panel indicates the eccentricity \(e_0\) at which these systems enter aLIGO band at \(f_{\rm GW}=14{\rm Hz}\). The Fitting Factors \(({\cal{FF}})\) are computed using \(f_{\rm min}= 15{\rm Hz}\) (see Eqs.~\eqref{inn_pro} and~\eqref{ff}), and the Zero Detuned High Power sensitivity configuration for aLIGO. The green and black stars represent the GW transients GW150914 and GW151226, respectively.}
\label{nice_fig_v1}
\end{figure*}

\noindent In Figure~\ref{nice_fig_v1} we investigate recovery of \(ax\)--waveforms with \(e_0\leq0.05\) using non-spinning and spin-aligned SEOBNRv2 banks. Figure~\ref{nice_fig_v1} indicates that template bank maximization does increase the overlap results presented in Figure~\ref{nice_fig_two}. Furthermore, template banks of spin-aligned SEOBNRv2 waveforms recover non-spinning, \emph{mildly} eccentric \(ax\)--waveforms with higher \({\cal{FF}}{\rm 's}\) than their quasi-circular, non-spinning counterparts. This is because the additional degrees of freedom of spin-aligned waveforms can be optimally combined to reproduce the dynamical evolution of non-spinning, weakly eccentric \(ax\)--waveforms. The panels in this Figure include a black and a green star, which represent the GW transients detected by aLIGO: GW150914 with \(M^{\star} = 67\Msun\), and GW151226 with \(M^{{\color{ForestGreen}{\star}}}=22\Msun\), respectively. Our results show that a template bank of spin-aligned SEOBNRv2 waveforms can recover GW150914 with \({\cal{FF}}\geq0.98\) and GW151226 with \({\cal{FF}}\geq0.97\) if \(e_0\leq0.05\). As discussed before, we should take these results with a grain of salt because in this low eccentricity regime \({\cal{FF}}{\rm's}\) may be dominated by the modeling errors of \(ax\)--waveforms in the zero eccentricity limit. 

\begin{figure*}[htp!]
\centerline{
\includegraphics[height=0.33\textwidth,  clip]{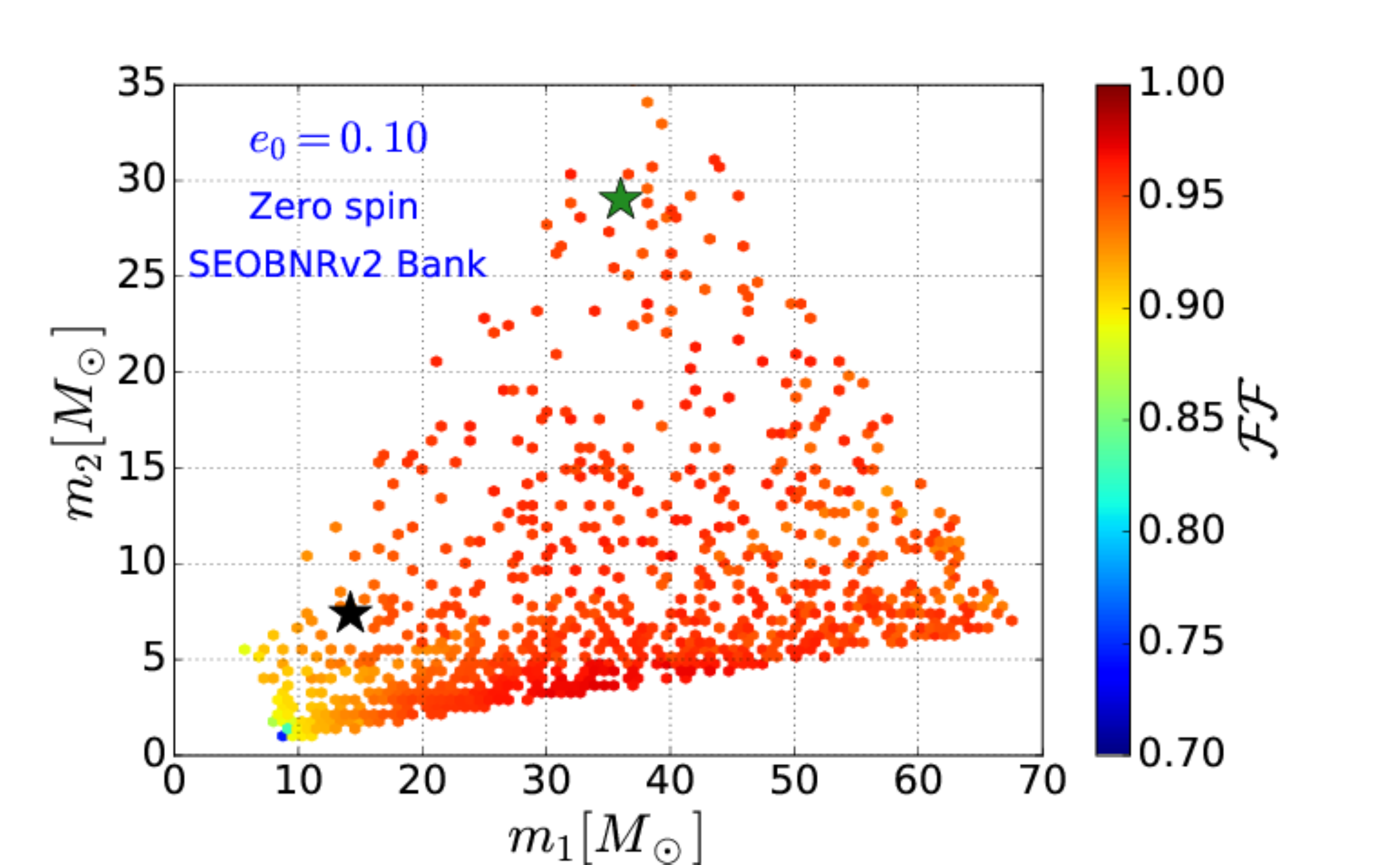}
\includegraphics[height=0.33\textwidth,  clip]{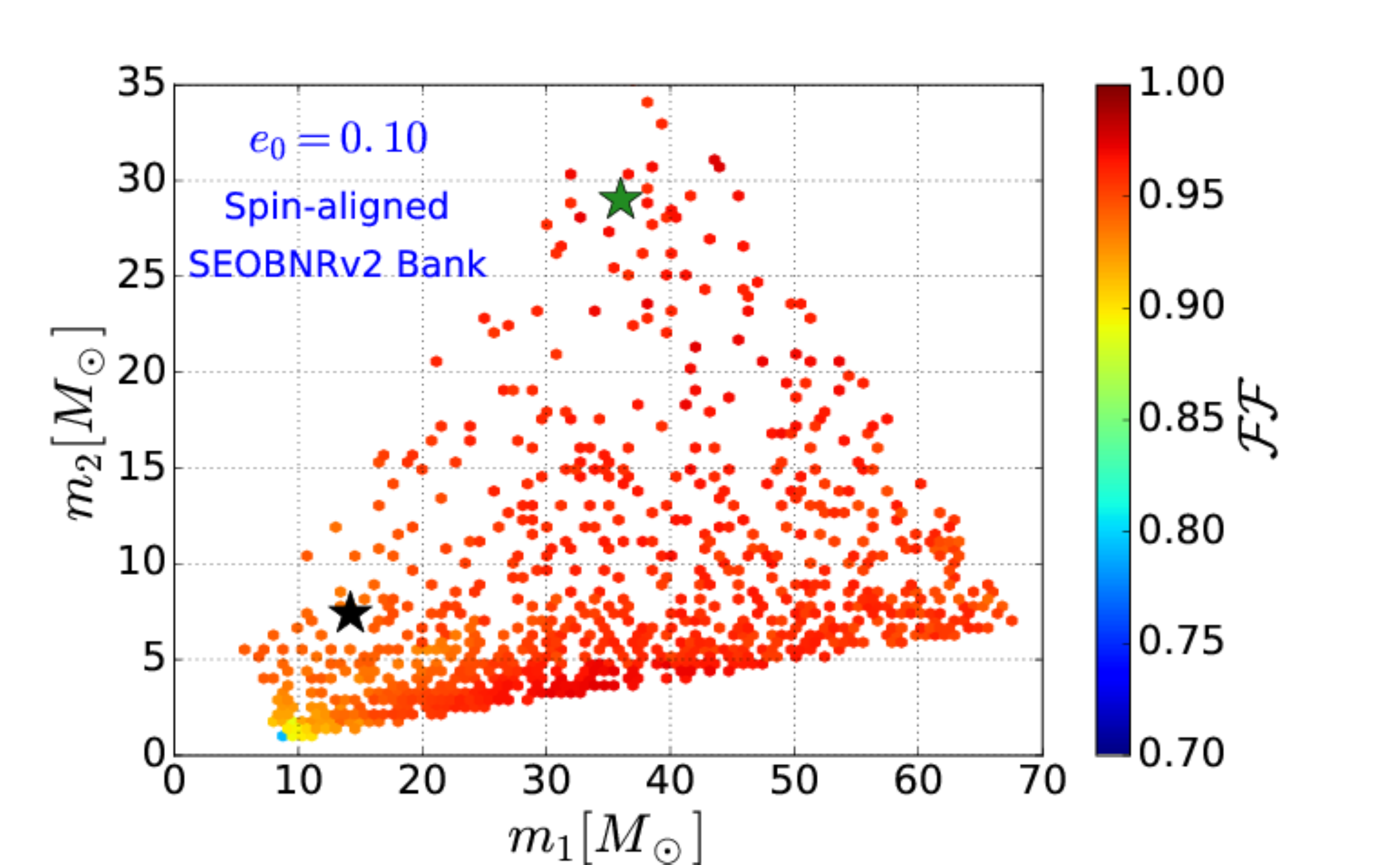}
}
\centerline{
\includegraphics[height=0.33\textwidth,  clip]{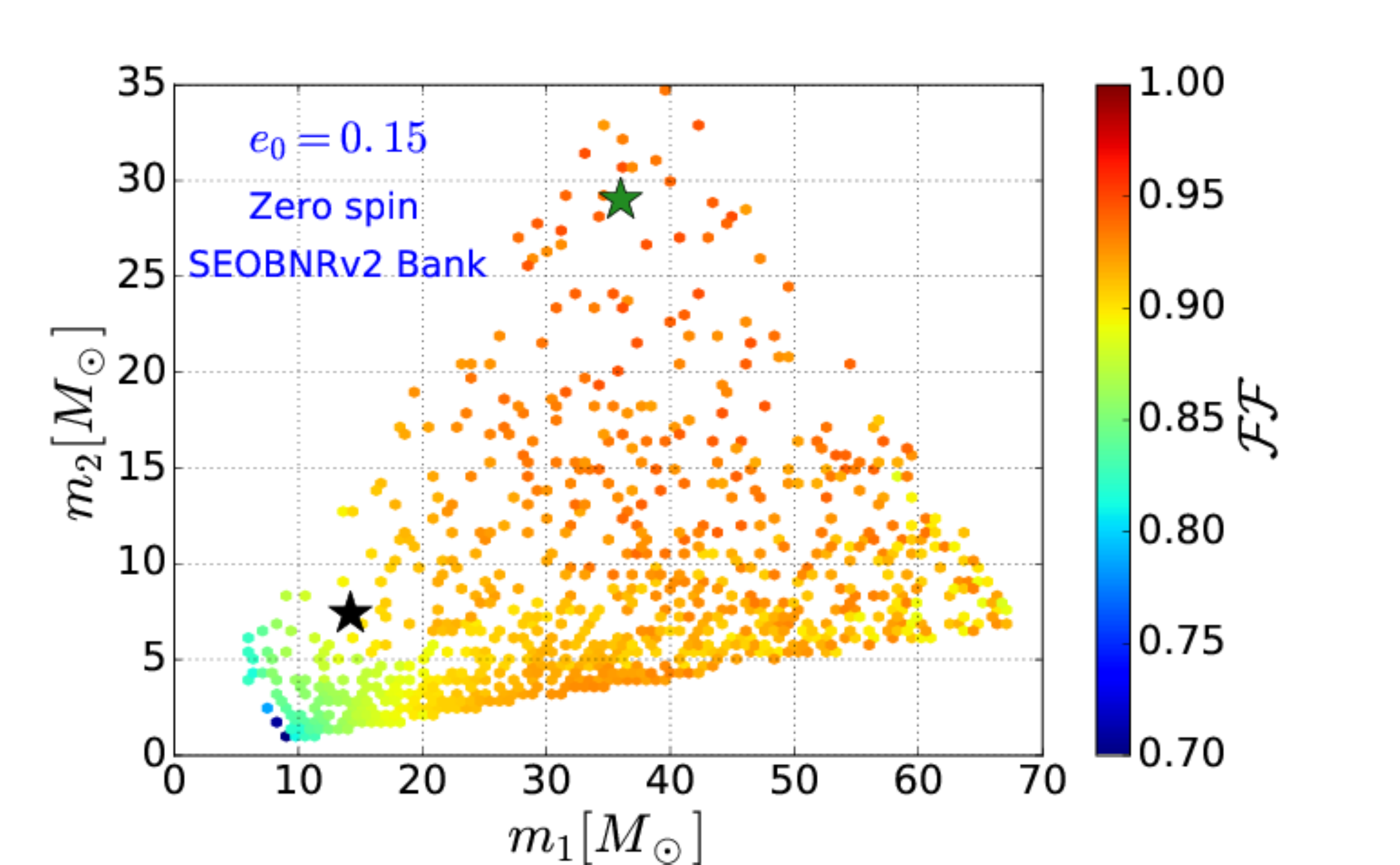}
\includegraphics[height=0.33\textwidth,  clip]{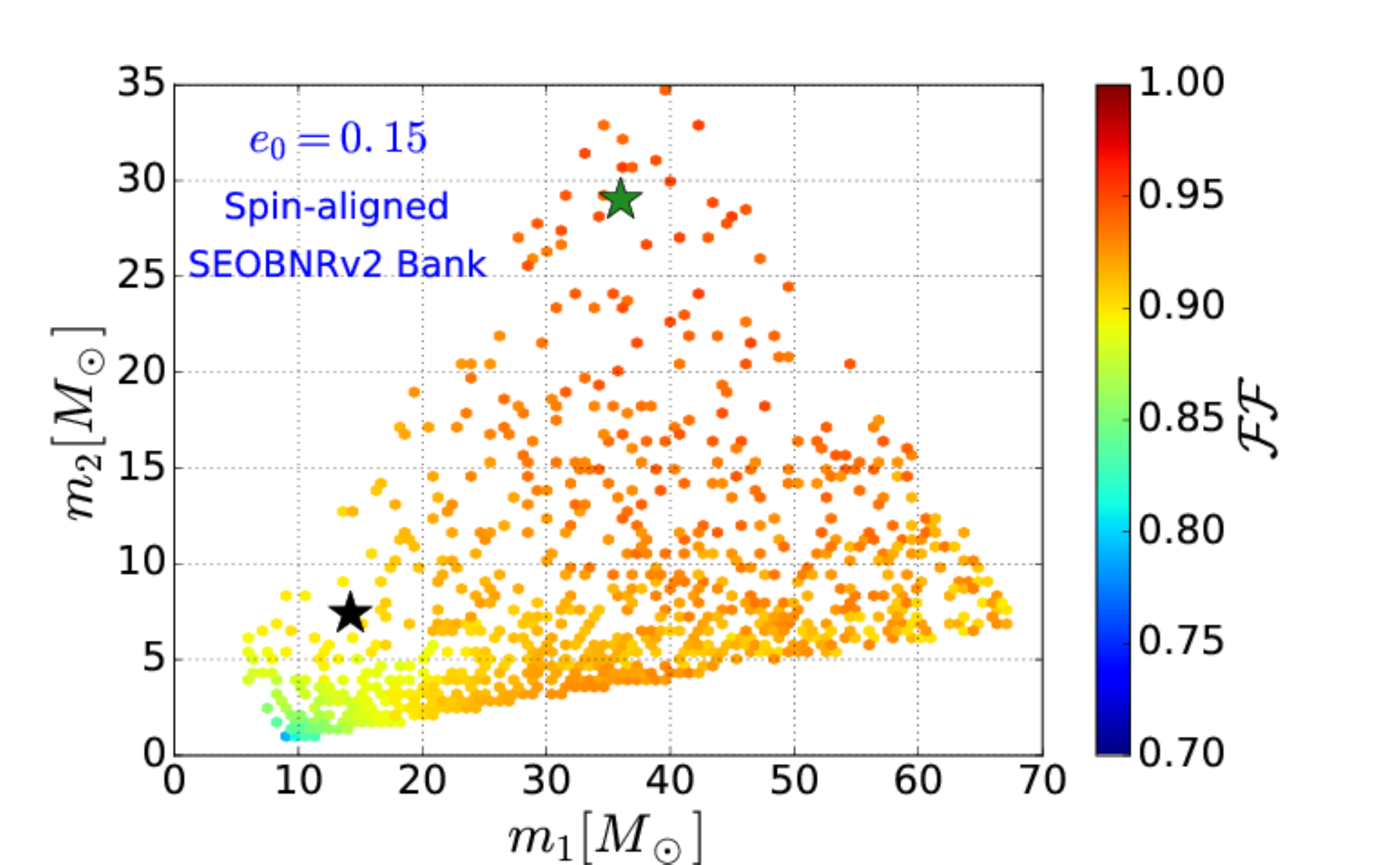}
}
\caption{As Figure~\ref{nice_fig_v1} but now for compact binary populations with \(e_0=\{0.1,\,0.15\}\). Note that the color bar has been adjusted to the range \([0.7,\,1]\) to exhibit additional structure for low \({\cal{FF}}\) values.}
\label{nice_fig_v2}
\end{figure*}

Let us now consider astrophysically realistic eccentricities, \(e_0=0.1\) and \(e_0=0.15\).  At these eccentricities, \(ax\) vs  SEOBNRv2 overlaps have already significantly deteriorated relative to the \(e_0=0\) comparison (cf.~Figure~\ref{nice_fig_two}), so we expect that we are really probing the effect of eccentricity in our comparisons. We notice that template bank maximization, given by the \({\cal{FF}}\) results in Figure~\ref{nice_fig_v2}, does not significantly improve the overlap calculations presented in Figure~\ref{nice_fig_two}. This suggests that the manifold generated by the eccentric signals is orthogonal to the usual quasi-circular manifold. Furthermore, recovery with the spin-aligned SEOBNRv2 template bank does not render significantly better results than its non-spinning counterpart. This implies that the spin-aligned degrees of freedom of the template bank are orthogonal to the eccentric degree of freedom of the injection manifold. Regarding the recovery of GW150914 and GW151226, we notice that these transients can be recovered with spin-aligned SEOBNRv2 templates with \({\cal{FF}}\geq 0.95\) if \(e_0\leq0.15\) and  \({\cal{FF}}\geq 0.94\) if \(e_0\leq0.1\), respectively. Furthermore, the impact of total mass and mass-ratio in the recovery of eccentric signals is significant in this regime. For \(e_0=0.1\), an equal mass \(10\Msun\) BBH and a \(10\Msun\) NSBH with \(q=8\) are recovered with \({\cal{FF}}=0.90\) and \({\cal{FF}}=0.86\), respectively. These results indicate that BBH and NSBH systems with astrophysically motivated values of eccentricity (\(e_0\sim0.1\)) will not be recovered with matched-filtering algorithms based on quasi-circular waveforms~\cite{Anto:2015arXiv}. In general, we find that systems with \(e_0\geq0.15\) are poorly recovered, \({\cal{FF}}\leq0.93\).

\begin{figure*}[htp!]
\centerline{
\includegraphics[height=0.33\textwidth,  clip]{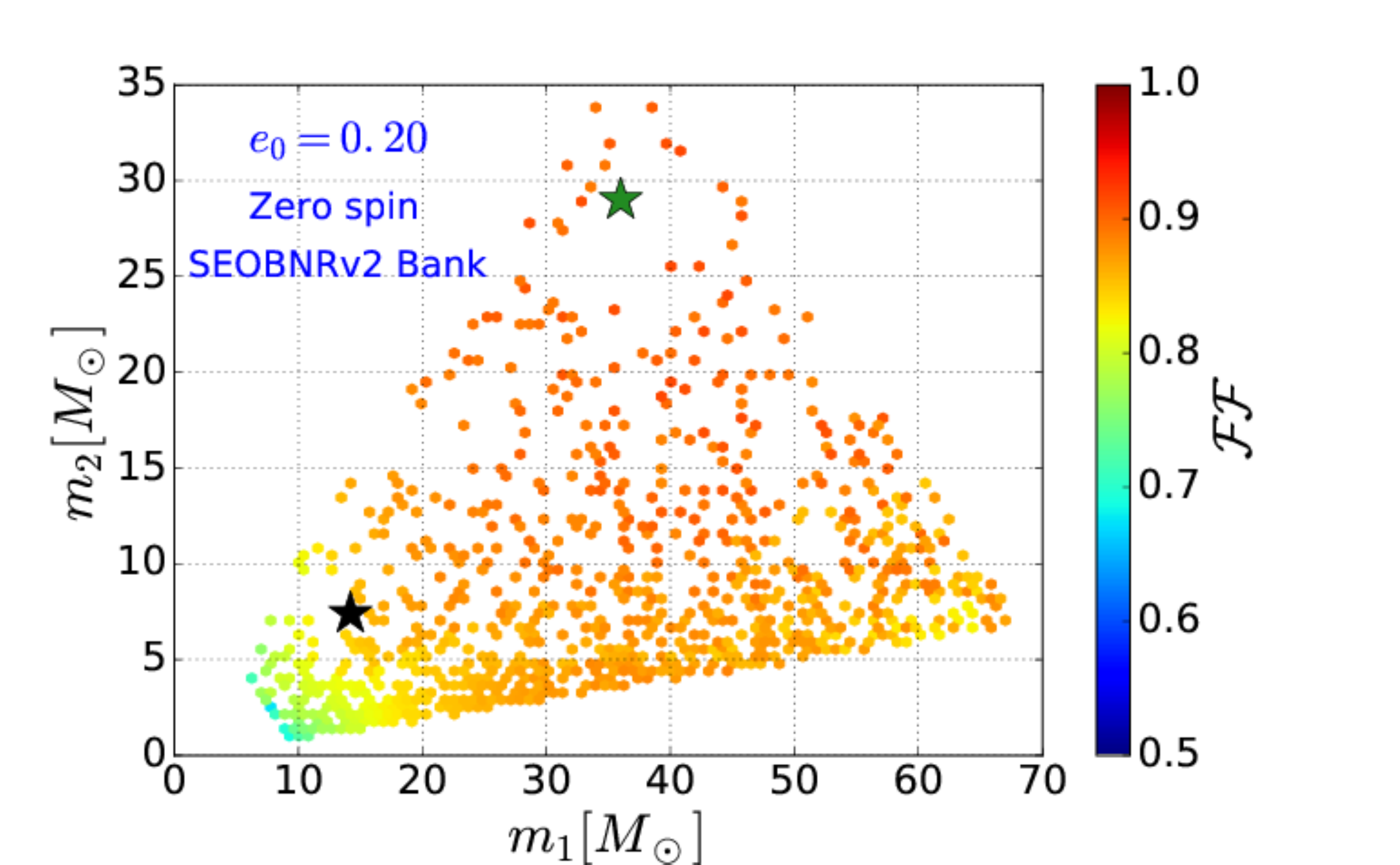}
\includegraphics[height=0.33\textwidth,  clip]{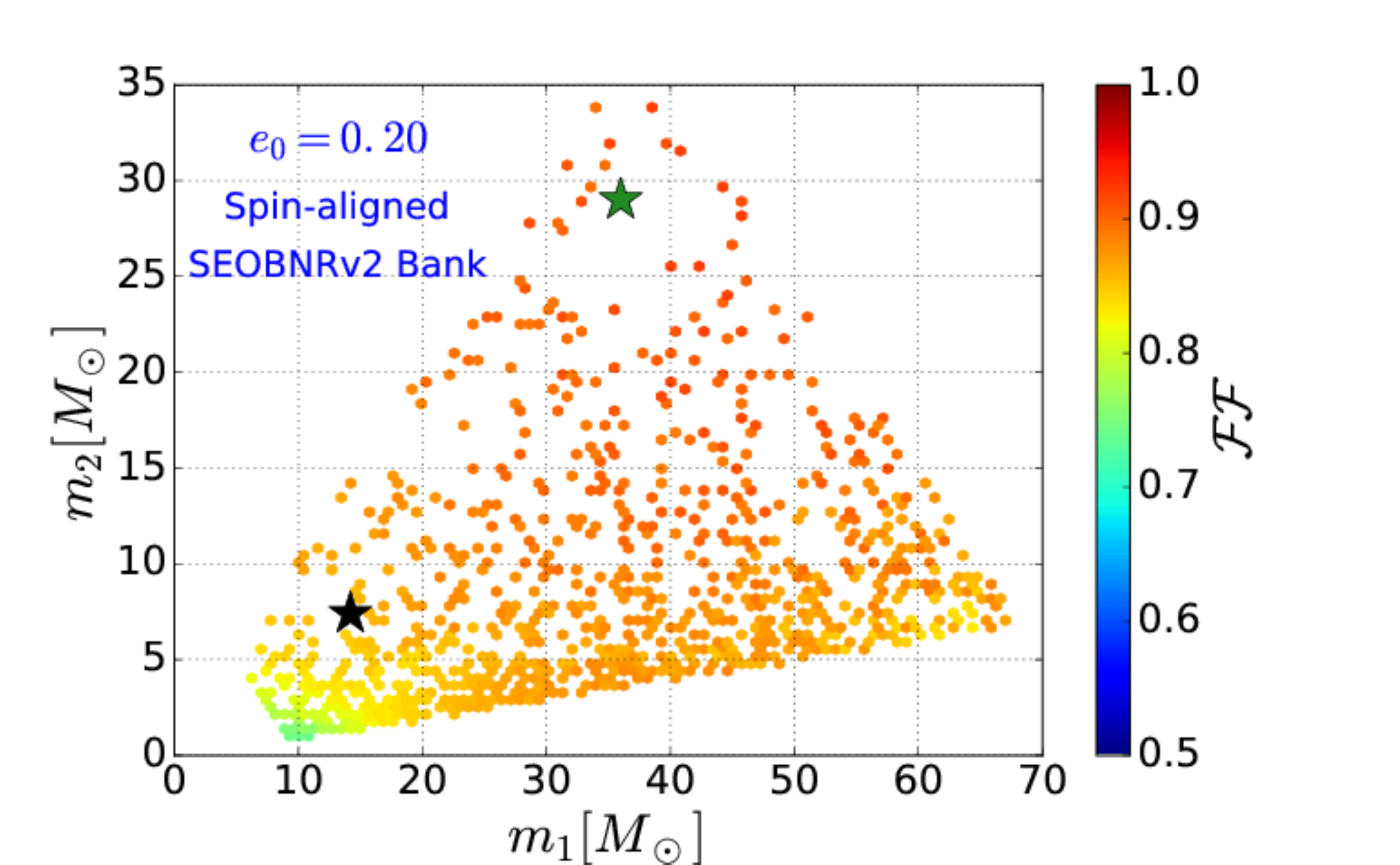}
}
\centerline{
\includegraphics[height=0.33\textwidth,  clip]{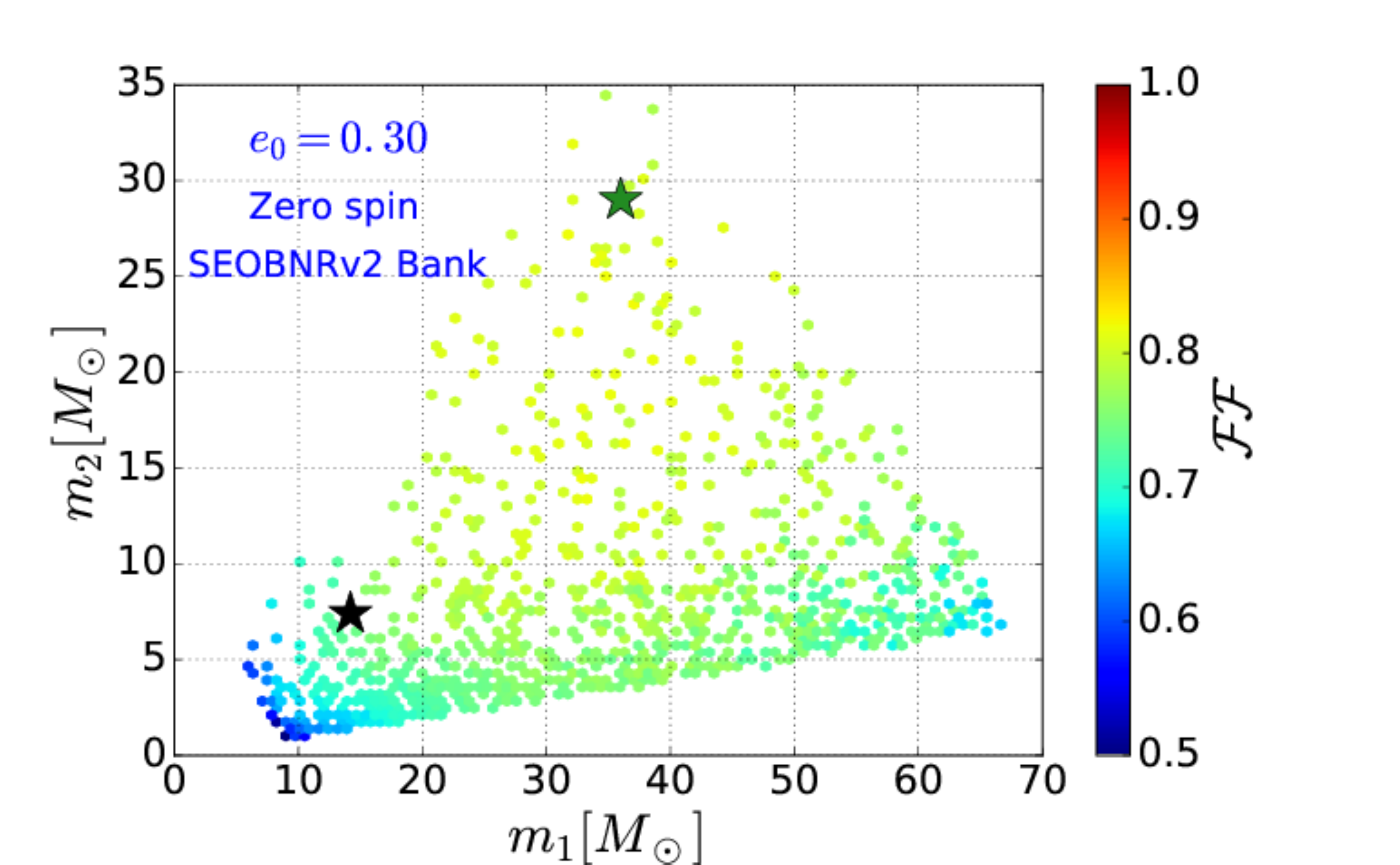}
\includegraphics[height=0.33\textwidth,  clip]{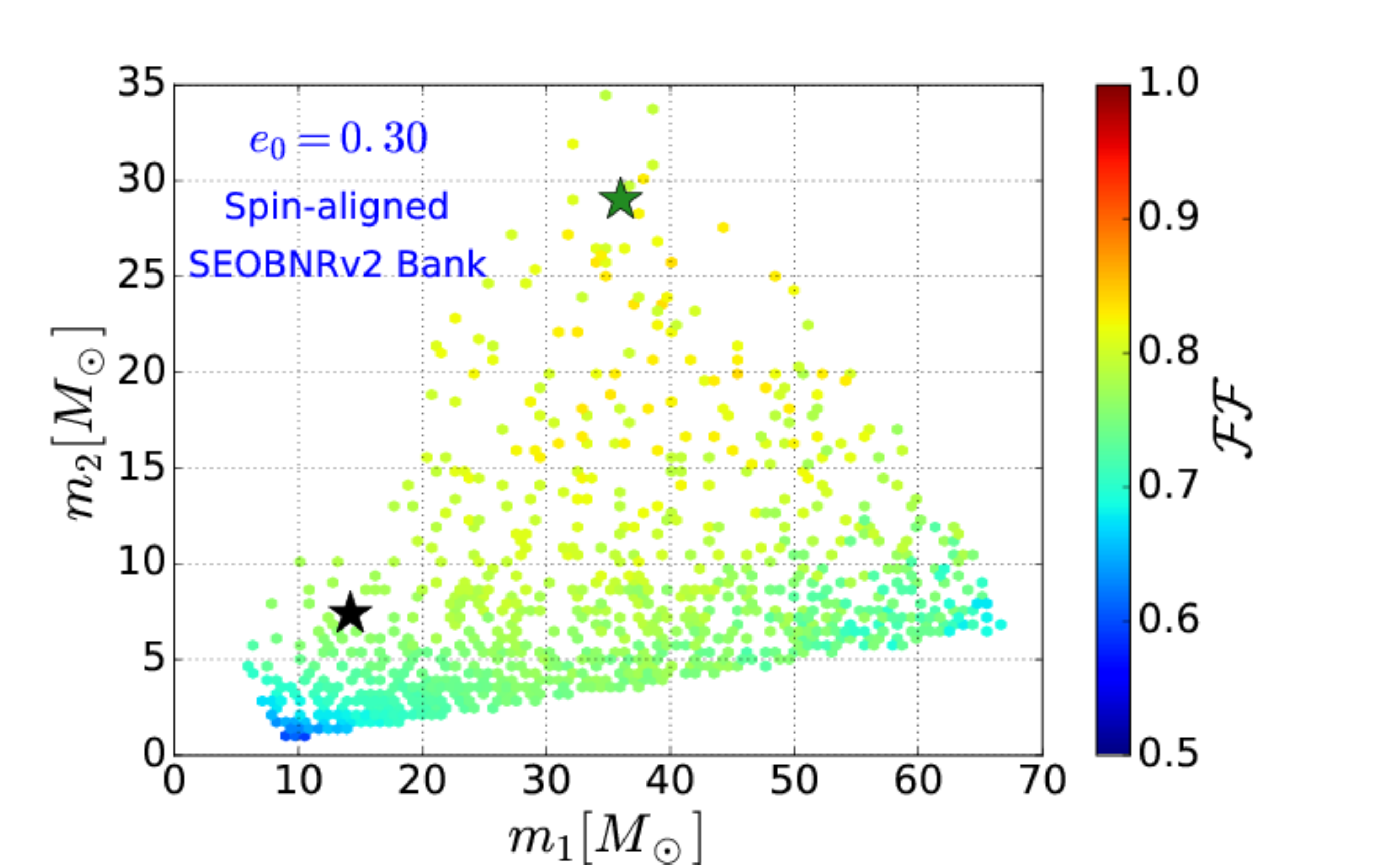}
}
\caption{As Figure~\ref{nice_fig_v1} but now for compact binary populations with \(e_0=\{0.2,\,0.3\}\). Note that the color bar range is \([0.5,\,1]\).}
\label{nice_fig_v3}
\end{figure*}

For completeness, let us finally investigate large eccentricities, \(e_0=0.2\) and \(e_0=0.3\), which --- according to present astrophyical understanding --- are hard to achieve~\cite{Anto:2015arXiv,Carl:2016arXiv,CR:2015PRL}.  Nevertheless, it is important to know how sensitive aLIGO is to such eccentric binaries, to independently verify astrophysical theory. As in the case of astrophysically motivated values of eccentricity (cf.~Figure~\ref{nice_fig_v2}), Figure~\ref{nice_fig_v3} indicates that the eccentric signal manifold is orthogonal to the non-spinning and spin-aligned template bank manifolds. Furthermore, since the recovery with both types of SEOBNRv2 banks is similar, we infer that the eccentricity degree of freedom of the signal manifold cannot be captured with the additional degrees of freedom of the spin-aligned SEOBNRv2 bank. These results also indicate that it will be unfeasible for quasi-circular searches to capture GW signals with eccentricities \(e_0\geq 0.2\). For \(e_0=0.2\), an equal mass \(10\Msun\) BBH and a \(10\Msun\) NSBH with \(q=8\) have \({\cal{FF}}=0.81\) and \({\cal{FF}}=0.73\), respectively. Recovery deteriorates very significantly for \(e_0\geq0.3\) --- most eccentric signals are recovered with \({\cal{FF}}\leq0.8\), and typical NSBH systems have \({\cal{FF}}\leq0.6\).

\begin{figure*}[htp!]
\centerline{
\includegraphics[height=0.33\textwidth,  clip]{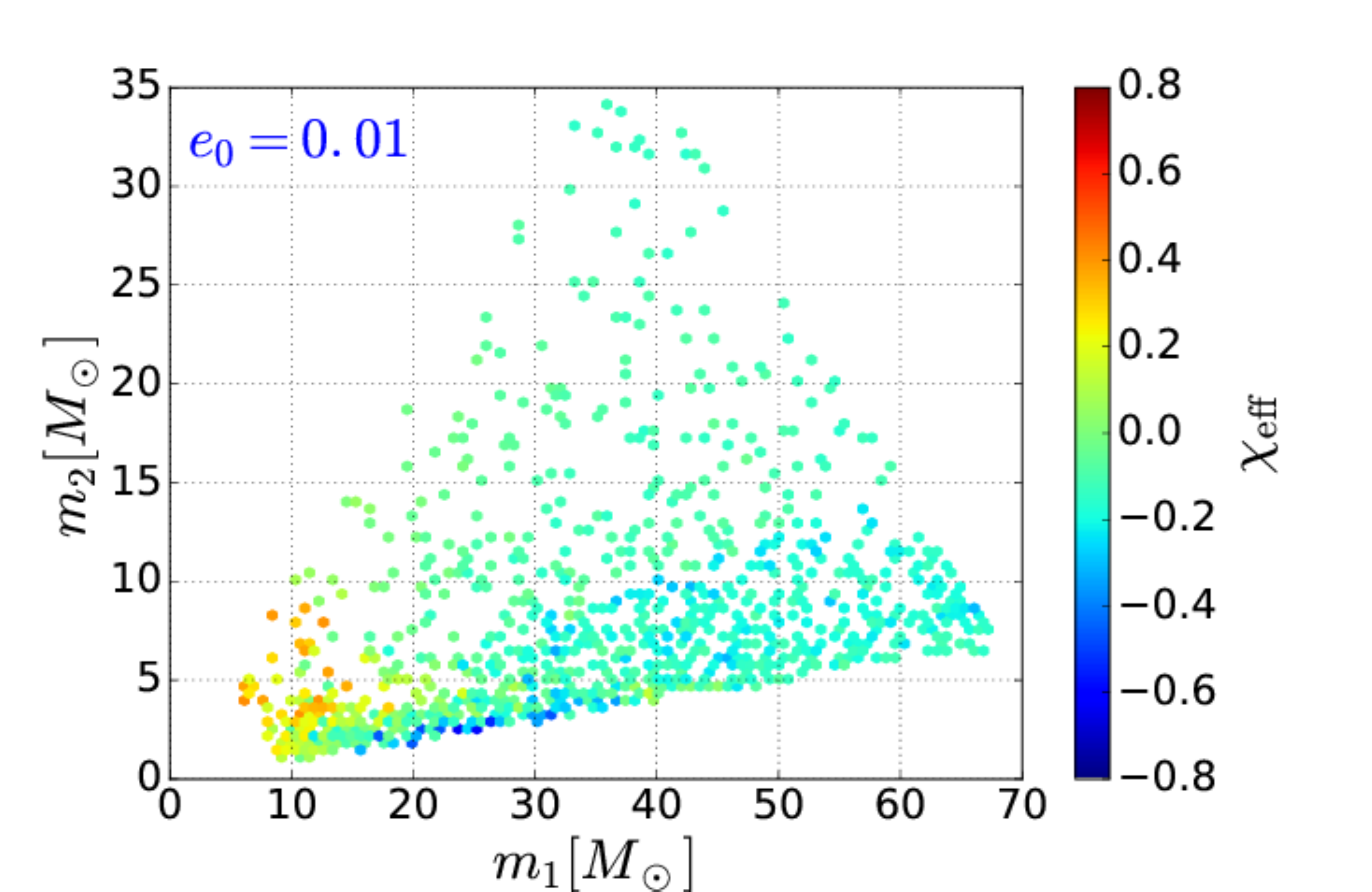}
\includegraphics[height=0.33\textwidth,  clip]{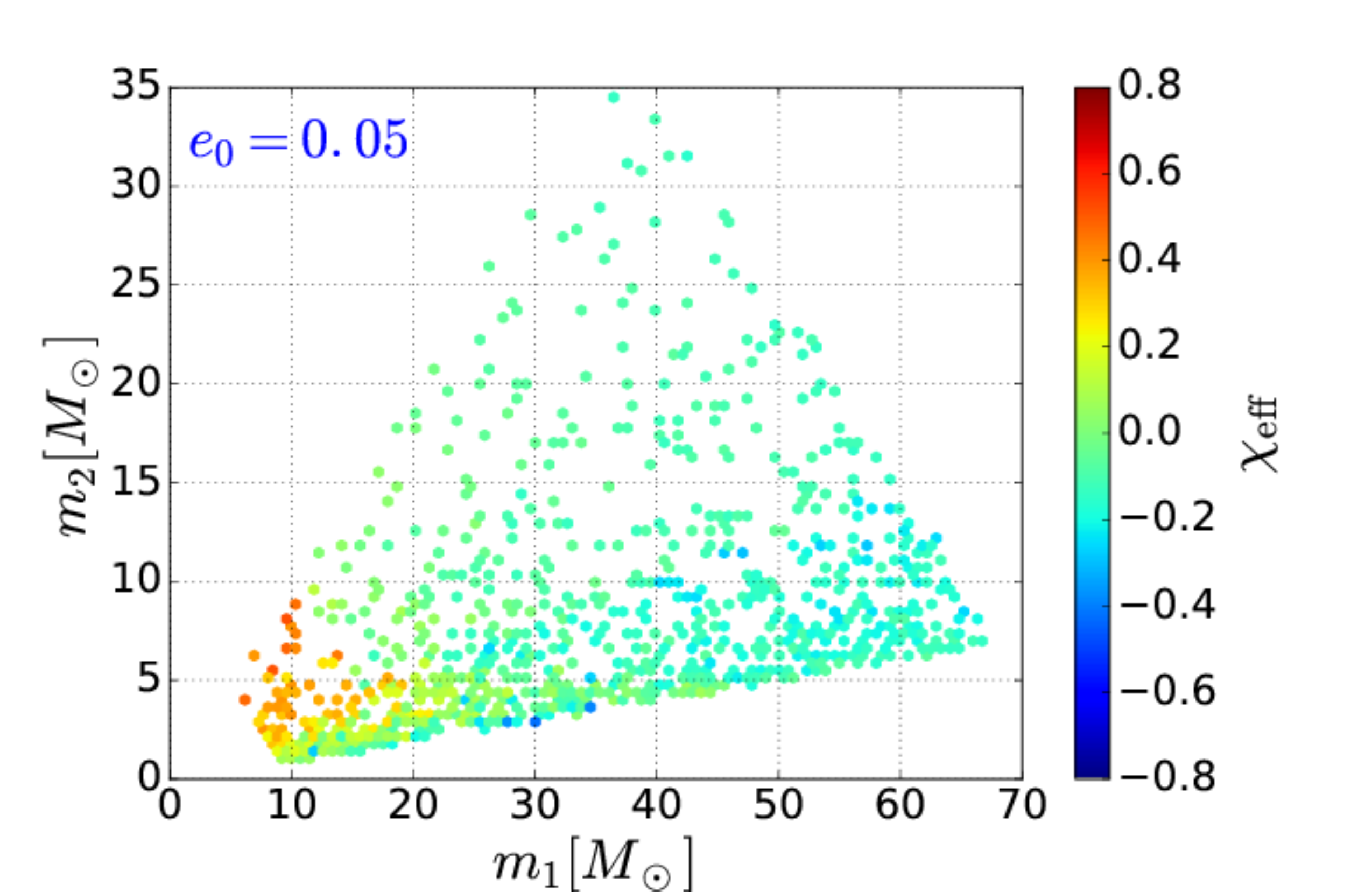}
}
\centerline{
\includegraphics[height=0.33\textwidth,  clip]{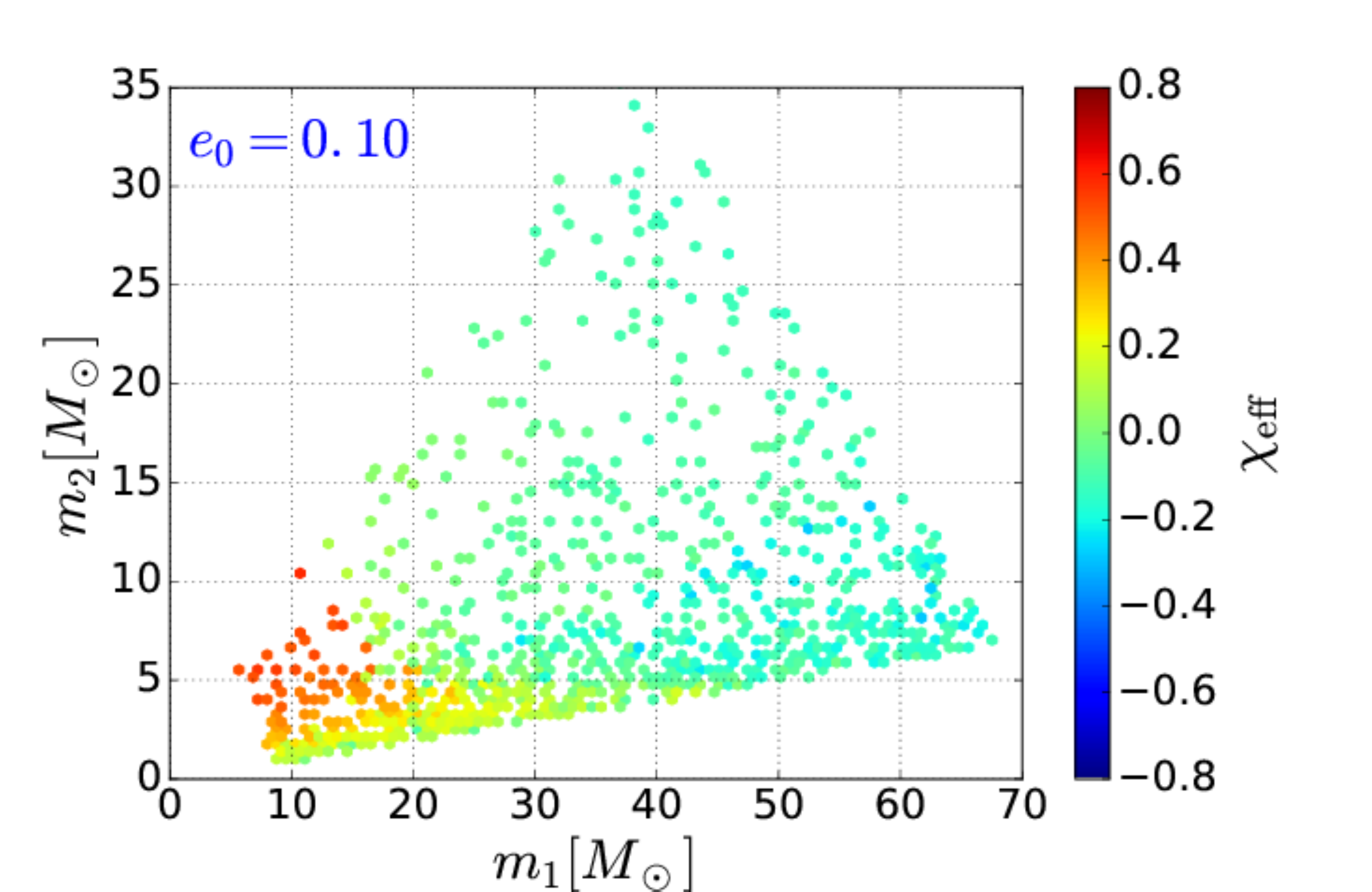}
\includegraphics[height=0.33\textwidth,  clip]{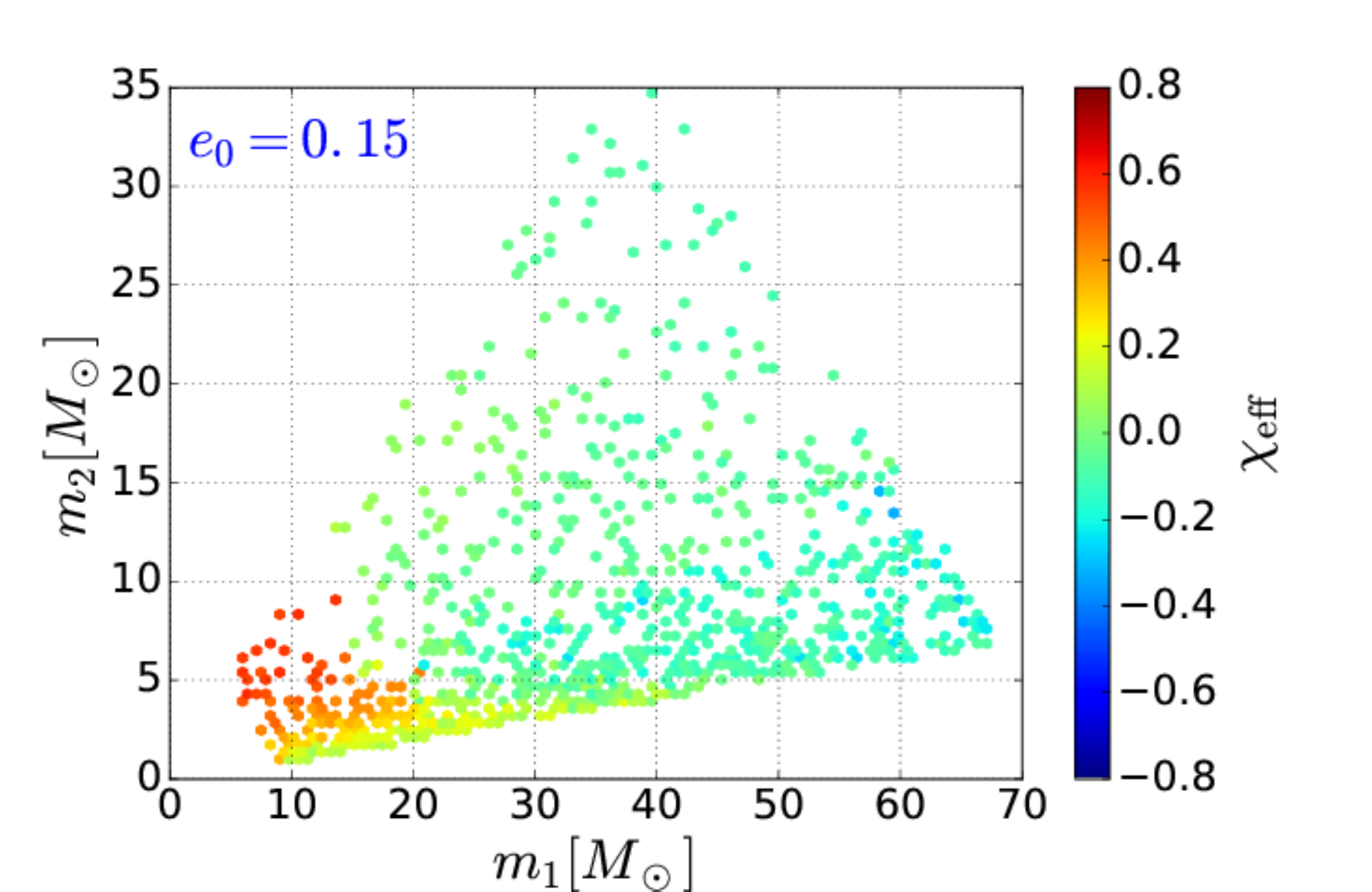}
}
\centerline{
\includegraphics[height=0.33\textwidth,  clip]{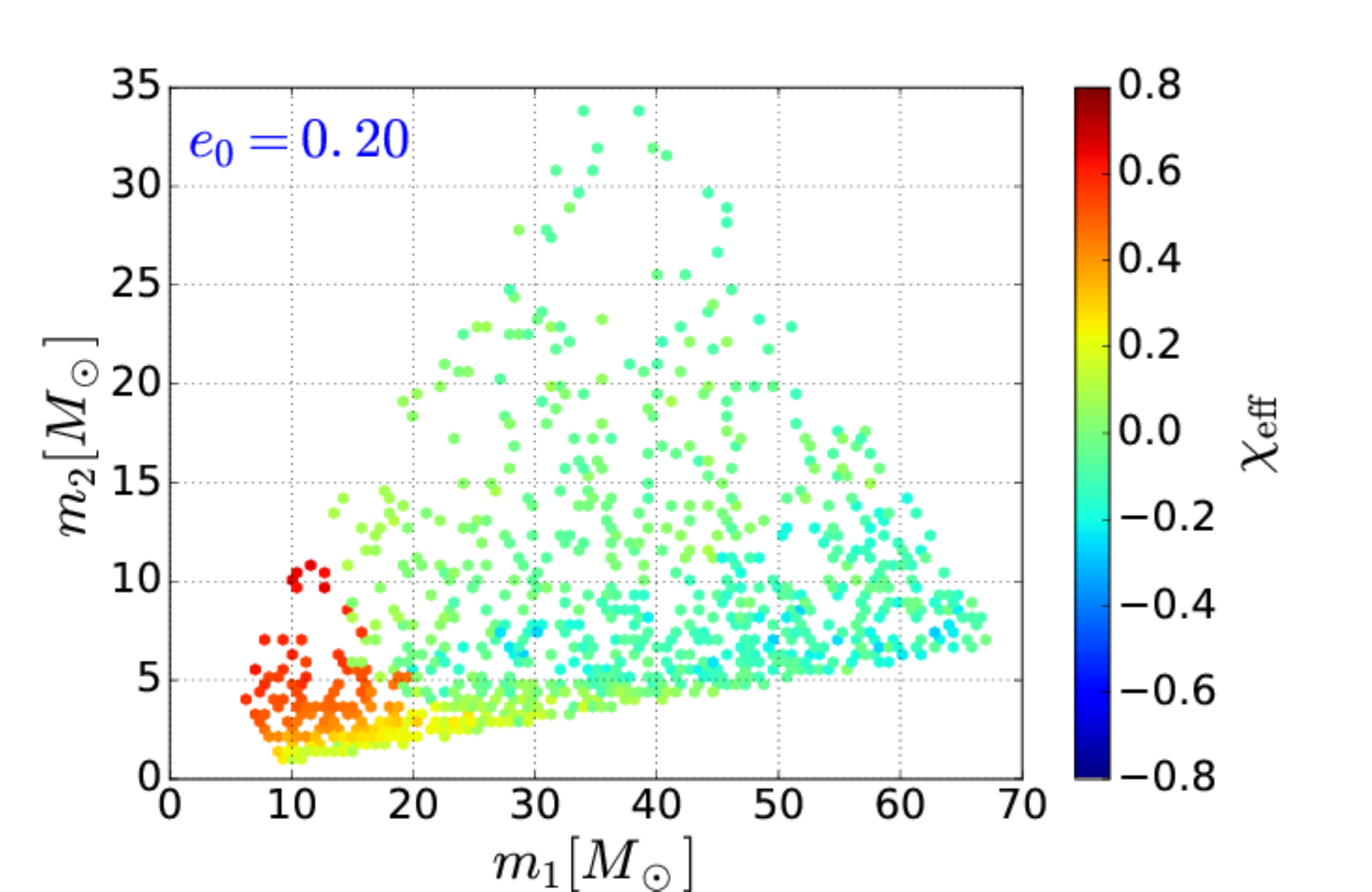}
\includegraphics[height=0.33\textwidth,  clip]{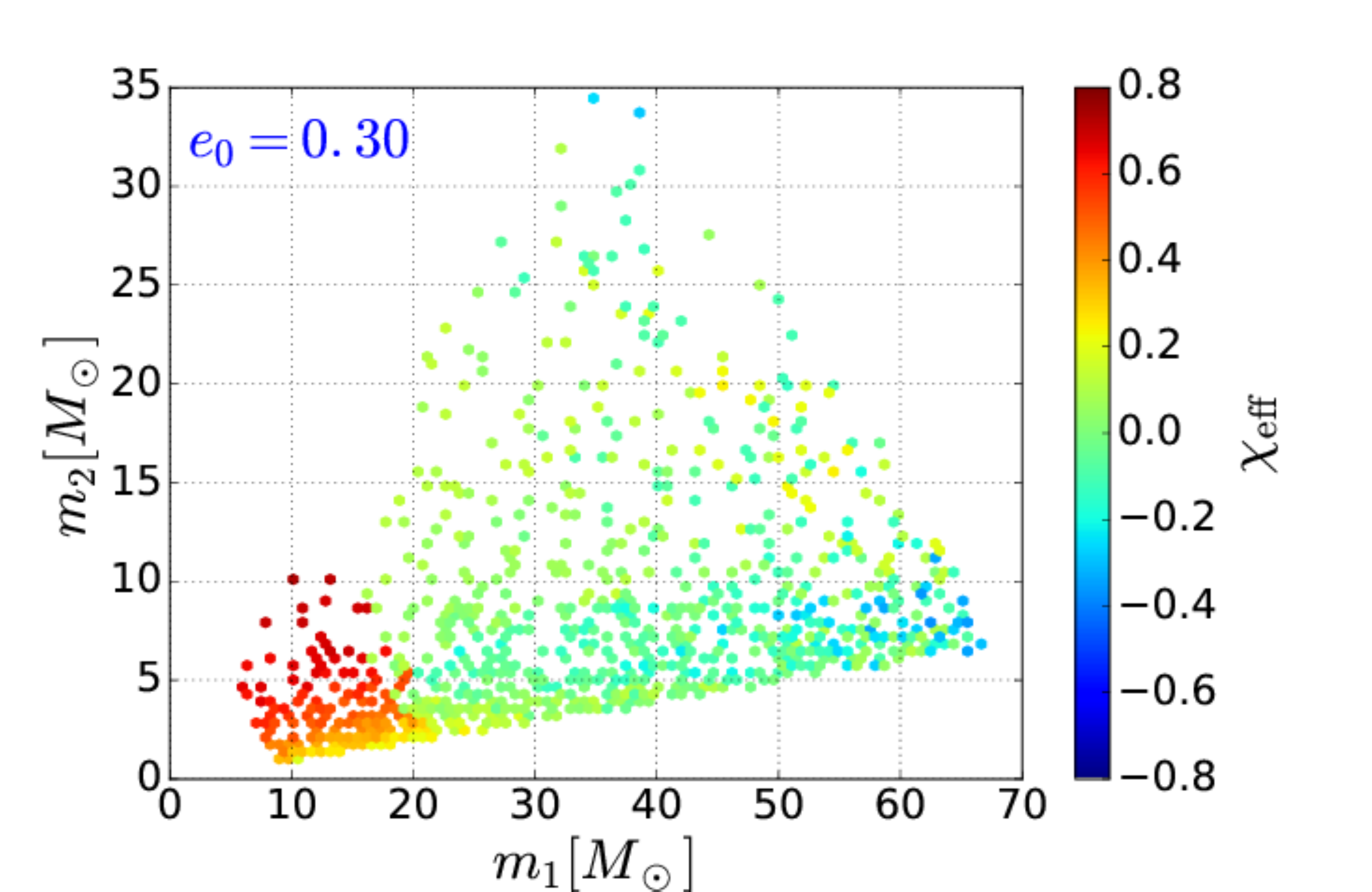}
}
\caption{Effective spin \(\chi_{\rm eff}\) with which eccentric signals are recovered (see Eq.~\eqref{chi_eff} in the main text). The magnitude of \(\chi_{\rm eff}\) indicates that spin-aligned SEOBNRv2 template banks significantly improve the recovery of non-spinning, eccentric waveforms for low total mass systems.}
\label{spin_and_ecc}
\end{figure*}

Up to this point we have discussed recovery of non-spinning template banks and aligned-spin template banks in parallel. We now investigate in further detail a different aspect of the impact of aligned-spin SEOBNRv2 template banks. To do so we compute the effective spin, \(\chi_{\rm eff}\), of the spin-aligned template waveforms that best recovered eccentric signals in our simulations, i.e.,

\beq
\label{chi_eff}
\chi_{\rm eff} = \frac{m^t_1}{M^t}\chi_1^z + \frac{m^t_2}{M^t}\chi_2^z - \frac{38\eta^t}{113}\left(\chi_1^z+\chi_2^z\right)\,,
\eeq

\noindent where \((m^t_1, m^t_2)\) are the template masses, \((\chi_1^z,\, \chi_2^z)\) are the dimensionless spins of the templates, and \(M^t=m^t_1+m^t_2\), \(\eta^t = m^t_1 m^t_2/M^2_t\). Figure~\ref{spin_and_ecc} presents the \(\chi_{\rm eff}\) for eccentric compact binary populations with \(e_0\leq0.3\). These results present the following global picture: template bank optimization with spin-aligned templates improves overlaps for eccentric populations with \(e_0\leq0.05\), and slightly increases recovery with respect to the non-spinning SEOBNRv2 bank. This is because the spin-aligned degrees of freedom: (i) compensate for the modeling errors of \(ax\)--waveforms in the \(e_0\rightarrow0\) limit;  and (ii) are able to reproduce the minor shortening effect of weakly eccentric signals. However, for \(e_0\geq0.1\) the modeling errors of quasi-circular \(ax\)--waveforms are small compared to the effect of eccentricity, and recovery is dominated by eccentricity. We notice that in this eccentricity regime, \(\chi_{\rm eff}\) only achieves significant values for low total mass systems, i.e., spin does not play a significant role in eccentric signal recovery. As we discussed above, this implies that non-spinning, eccentric populations with \(e_0\geq0.1\) define a manifold that is predominantly orthogonal to the quasi-circular, non-spinning and spin-aligned manifolds.

The results presented in this Section clearly indicate that matched-filtering algorithms tuned for quasi-circular waveforms will not be effectual at recovering stellar mass BBH and NSBH systems with astrophysically motivated values of eccentricity, i.e., \(e_0\sim 0.1\)~\cite{Anto:2015arXiv}. We have also shown that the two GW transients already detected by the aLIGO detectors could have had non-negligible values of residual eccentricity at \(f_{\rm GW}=14{\rm Hz}\), and still be detected with high \({\cal{FF}}\) values using spin-aligned SEOBNRv2 template banks. These results are the first in their kind in the literature.

\section{Conclusion}
\label{disc}

We have developed a waveform model for eccentric compact binaries that represents the inspiral, merger and ringdown, and that reproduces zero eccentricity binary waveforms much more accurately than previous eccentric waveform models. We have also demonstrated that our new model can accurately describe comparable mass-ratio, moderately eccentricity BBH NR simulations. With this model we studied the importance of including eccentricity in detecting eccentric NSBH and BBH systems with aLIGO. We showed that using the design sensitivity of aLIGO and a lower frequency cut--off of 15Hz, the IMR \(ax\)--model can reproduce the SEOBNRv2 model in the zero eccentricity limit with overlap values \({\cal{O}}\gtrsim 0.95\) over a wide range of the stellar mass BBH and NSBH parameter space that is accessible to aLIGO.

Using our IMR \(ax\) model we explore the detectability of eccentric compact binaries. Our results indicate that template banks of quasi-circular, spin-aligned SEOBNRv2 waveforms can recover GW150914 with \({\cal{FF}}\geq0.95\) if \(e_0\leq0.15\), and  GW151226 with \({\cal{FF}}\geq0.94\) if \(e_0\leq0.1\). We have also found that template banks of quasi-circular, spin-aligned waveforms can improve the recovery of low total mass moderately eccentric signals. Our results also indicate that low mass BBH and NSBH systems with astrophysically motivated values of eccentricity \((e_0\sim0.1)\) will be poorly recovered with available quasi-circular matched-filtering algorithms (\({\cal{FF}}\leq0.85\)). In order to detect these events, it is necessary to develop new data analysis algorithms that specifically target eccentric GW sources. 

A key assumption in the construction of our \(ax\)--model is that compact binaries circularize prior to merger. We explore the validity of this assumption and find that we can cover a large portion of the parameter space of compact binaries that aLIGO will be able to detect. In order to minimize the effect of inherent waveform inaccuracies in the \(ax\)--model, particularly in the context of parameter estimation studies,  we are exploring two ways to enhance its accuracy in the \(e\rightarrow 0\) limit. The first improvement deals with the hybridization between inspiral-PN model and gIRS merger--ringdown model: in its current version the \(ax\)--model consists of a simple hybridization between the PN--inspiral evolution and the gIRS model we have described in Section~\ref{liv}. The key for this procedure to work requires that both frameworks meet at an optimal frequency where they render the correct dynamical evolution. The results we have obtained in this work suggests that using up-to-date results from the self-force formalism and PN theory provides a robust framework to capture the inspiral dynamics of compact binaries with asymmetric mass-ratios. The enhanced inspiral evolution we have constructed is good to \textit{explore} the late time dynamics of BBHs, but it can only go so far. At the other end of the spectrum, the gIRS model is reliable in the vicinity of the light-ring. We can see in Figure~\ref{mer_fig} that this approach starts to deteriorate when we push the model several cycles prior to the merger event. Therefore, a critical correction to further improve the IMR \(ax\)--model is the development of a new merger-ringdown prescription that captures the true dynamical evolution \textit{several} cycles before merger, and which can provide a wider window of frequencies to hybridize the inspiral evolution with the merger phase. 

Our second planned improvement concerns the inspiral dynamics itself. Presently, 4-6PN terms in the binding energy of compact binaries \(E(x,\,\eta)^{\rm 6PN}\), cf. Eq.~\eqref{pre_hyb}, only include first order in symmetric mass-ratio corrections. We will further improve the inspiral dynamics by including terms at \emph{second order} in symmetric-mass ratio. Furthermore, building up on~\cite{Huerta:2014a,Kapadia:2016}, we will amend the energy flux prescription, \(\dot{E}(x,\,\eta)^{\rm 6PN}\), by constraining missing \(\eta^2\) corrections in the energy flux expression used in Eq.~\ref{pre_hyb}. 

We expect that combining the aforementioned improvements will provide an enhanced performance of the \(ax\)--model in the \(e_0\rightarrow0\) limit so that the overlap with SEOBNRv2 templates satisfies \({\cal{O}}\gtrsim0.99\) over the stellar mass BBH and NSBH parameter space accessible to aLIGO. The results we present in this article indicate that a consistent combination of higher-order PN calculations, self-force corrections and NR can enable the construction of accurate, computationally inexpensive waveform models that encode the dynamics of compact binary systems across the parameter space accessible to aLIGO--type detectors. These results further support the importance of deriving second order self-force effects~\cite{rosen:2006PhRvD,pound:2012PhRvL,gralla:2012PhRvD,pm:2014PhRvD,pojere:2014PhRvD,adampound:2015prd,binidamo:2016arXiv}. Previous studies have strongly relied on self-force calculations for waveform modeling, source detection and parameter estimation studies, and have exhibited their applicability for extreme and comparable mass-ratio systems~\cite{Amaro:2013GW,Huerta:2011vna,Huerta:2012,Huerta:2010,higherspin,Huerta:2009,Huerta:2011a,Huerta:2011b,wargar,Huerta:2014a,smallbody,Osburn:2016}. Moving forward, it is necessary to develop new waveform models that enable the description of compact binaries whose components have non-zero spin and which evolve on eccentric orbits. Using eccentric NR simulations both for calibration and validation purposes will enable the construction of robust waveform models that are adequate for detailed parameter estimation studies. This work should be pursued in the near future.

\section*{Acknowledgments}
We thank Mark Fredricksen, Campus Cluster Administrator at NCSA, for his help configuring UIUC's campus cluster to obtain some of the computations presented in this article. B.~A. and W.~R. gratefully acknowledges a Students Pushing Innovation (SPIN) internship from NCSA. We thank Gabrielle Allen, Haris Markakis and Ed Seidel for fruitful interactions and comments on the article. We thank Andrea Taracchini and Zhoujian Cao for reviewing this manuscript and providing suggestions to improve it. We also thank Sean McWilliams for comments on the IRS model. We gratefully acknowledge support for this research at CITA from NSERC of Canada, the Ontario Early Researcher Awards Program, the Canada Research Chairs Program, and the Canadian Institute for Advanced Research; at Caltech from the Sherman Fairchild Foundation and NSF grants PHY-1404569 and AST-1333520; at Cornell from the Sherman Fairchild Foundation and NSF grants PHY-1306125 and
AST-1333129; and at Princeton from NSF grant PHY-1305682 and the Simons Foundation.  Calculations were performed at the GPC supercomputer at the SciNet HPC Consortium~\cite{scinet}; SciNet is
funded by: the Canada Foundation for Innovation (CFI) under the auspices of Compute Canada; the Government of Ontario; Ontario Research Fund (ORF) -- Research Excellence; and the University of
Toronto. Further calculations were performed on the Briar\'ee cluster at Sherbrooke University, managed by Calcul Qu\'ebec and Compute Canada and with operation funded by the Canada Foundation for
Innovation (CFI), Minist\'ere de l'\'Economie, de l'Innovation et des Exportations du Quebec (MEIE), RMGA and the Fonds de recherche du Qu\'ebec - Nature et Technologies (FRQ-NT); on the Zwicky cluster at Caltech, which is supported by the Sherman Fairchild Foundation and by NSF award PHY-0960291; on the NSF XSEDE network under grant TG-PHY990007N; on the NSF/NCSA Blue Waters at the University of Illinois with allocation jr6 under NSF PRAC Award ACI-1440083. This research is part of the Blue Waters sustained-petascale computing project, which is supported by the National Science Foundation (awards OCI-0725070 and ACI-1238993) and the state of Illinois. Blue Waters is a joint effort of the University of Illinois at Urbana-Champaign and its National Center for Supercomputing Applications. This article has LIGO Document number P1600186. 
\clearpage
\appendix 
\begin{widetext}
\section{}
\label{ap1}

Higher-order PN calculations for eccentric binaries have been computed in terms of the mean motion \(n\) and \(e\) in Ref.~\cite{Arun:2009PRD}. In this Appendix we re-write those results in terms of the gauge-invariant quantity \(x=\left(M\omega\right)^{2/3}\) and \(e\). To do so we use the following relation between the mean motion \(n\), the gauge-invariant quantity \(x\) and \(e\)~\cite{Hinder:2010,Mishra:2015}:

\(M_\odot\)

\bea
\label{xhi_of_n} 
M n &=&\frac{x^{3/2}}{(1-e^2)^3}\Bigg[1-3 e^2+3 e^4-e^6+x \left(-3+6 e^2-3
e^4\right)+x^2 \left[-\frac{9}{2}+7 \eta +\left(-\frac{33}{4}-\frac{\eta
}{2}\right) e^2+\left(\frac{51}{4}-\frac{13 \eta }{2}\right)\eta
e^4\right]\nonumber\\&+&x^3 \Bigg[\frac{3}{2}+\eta  \left(\frac{457}{4}-\frac{123 \pi
^2}{32}\right)-7 \eta ^2+\left(-\frac{267}{4}+\eta  \left(\frac{279}{2}-\frac{123
\pi ^2}{128}\right)-40 \eta ^2\right) e^2+\left(-\frac{39}{2}+\frac{55 \eta
}{4}-\frac{65 \eta ^2}{8}\right) e^4\nonumber\\&+&\sqrt{1-e^2} \left(-15+6 \eta
+(-30+12 \eta ) e^2\right)\Bigg] \Bigg] + {\cal{O}}\left(x^{11/2}\right)\,.    
\eea

\noindent The time evolution of \(x\) is given by:

\beq
M\dot{x} = \dot{x}_{0\rm{PN}}x^5 + \dot{x}_{1\rm{PN}}x^6 + \dot{x}_{2\rm{PN}}x^7 + \dot{x}_{3\rm{PN}}x^8 +\, \dot{x}_{\rm{HT}}\,,
\label{rad_terms}
\eeq   

\noindent where \(\dot{x}_{\rm{HT}}\) stands for hereditary terms. \((\dot{x}_{0\rm{PN}},\, \dot{x}_{1\rm{PN}})\) can be found in~\cite{Hinder:2010}:

\beq
\label{x_0}
\dot{x}_{0\rm{PN}}=\frac{2 \left(37 e^4+292 e^2+96\right) \eta }{15 \left(1-e^2\right)^{7/2}}\,,
\eeq

\bea
\label{x_1}
\dot{x}_{1\rm{PN}}&=&\frac{\eta  \left(11717 e^6+171038 e^4+87720 e^2-28 \left(296 e^6+5061 e^4+5700
   e^2+528\right) \eta -11888\right)}{420 \left(1-e^2\right)^{9/2}}\,,
\eea

In this work, we have derived \(\dot{x}_{2\rm{PN}} ,\, \dot{x}_{3\rm{PN}}\) and \(\dot{x}_{\rm{HT}}\):

\bea   
\label{x_2}
\dot{x}_{2\rm{PN}} &=&-\frac{\eta}{45360 \left(1-e^2\right)^{11/2}}\Bigg[ -3 e^8 \Big(4 \eta  \Big(163688 \eta -271665\Big)+1174371\Big)\nonumber\\&+&16 e^2 \left(-21 \eta 
   \left(-76824 \sqrt{1-e^2}+182387 \eta +46026\right)-4033260
   \sqrt{1-e^2}+5802910\right)\nonumber\\&+&32 \left(-9 \eta  \left(-2016 \sqrt{1-e^2}+6608 \eta
   +15677\right)-45360 \sqrt{1-e^2}+11257\right)\nonumber\\&+&6 e^6 \left(7 \eta  \left(25200
   \sqrt{1-e^2}-1543544 \eta +2931153\right)-3 \left(147000
   \sqrt{1-e^2}+4634689\right)\right)\nonumber\\&+&12 e^4 \left(\eta  \left(2210544
   \sqrt{1-e^2}-13875505 \eta +17267022\right)-34 \left(162540
   \sqrt{1-e^2}+1921\right)\right)\Bigg]\,,
 \eea
 \bea  
\label{x_3}
\dot{x}_{3\rm{PN}} &=&\frac{\eta}{598752000 \left(1-e^2\right)^{13/2}}\Bigg[  25 e^{10} \Bigg\{2699947161-176 \eta  \bigg(4 \eta  \Big(2320640 \eta -2962791\Big)+16870887\bigg)\Bigg\}\nonumber\\&+&
32 e^2 \Bigg\{55 \eta  \bigg[270 \left(7015568 \sqrt{1-e^2}-9657701\right) \eta
   -8125851600 \sqrt{1-e^2}+38745 \pi ^2 \left(1121 \sqrt{1-e^2}+1185\right)\nonumber\\&-&901169500 \eta ^2+5387647438\bigg]    + 31050413856\sqrt{1-e^2}+358275866598\Bigg\}        \nonumber\\& + & 128 \Bigg\{-275 \eta  \bigg[81 \left(16073-17696 \sqrt{1-e^2}\right) \eta\nonumber\\&-&1066392 \sqrt{1-e^2}+46494 \pi ^2
   \left(\sqrt{1-e^2}-45\right)+470820 \eta ^2+57265081\bigg]
    -3950984268 \sqrt{1-e^2}+12902173599\Bigg\}   
    \nonumber\\&+&e^8 \Bigg\{162 \left(1240866000
   \sqrt{1-e^2}+19698134267\right)-1100 \eta  \bigg[16 \eta  \left(-3582684 \sqrt{1-e^2}+137570300 \eta -286933509\right)\nonumber\\&+&27 \left(6843728 \sqrt{1-e^2}+255717
   \pi ^2+173696120\right)\bigg]\Bigg\}
  \nonumber\\& + & 12 e^6 \Bigg\{55 \eta  \bigg[90 \left(52007648 \sqrt{1-e^2}+311841025\right) \eta \nonumber\\&+&3 \left(4305 \pi ^2 \left(14
   \sqrt{1-e^2}-19113\right)-5464335200 \sqrt{1-e^2}+767166806\right)          -17925404000 \eta ^2\bigg]\nonumber\\&+&742016570592 \sqrt{1-e^2}+6005081022\Bigg\} \nonumber\\&+&
   8 e^4 \Bigg\{55
   \eta  \bigg[270 \left(71069152 \sqrt{1-e^2}+6532945\right) \eta \nonumber\\&-&74508169680 \sqrt{1-e^2}+116235 \pi ^2 \left(1510 \sqrt{1-e^2}-4807\right)-23638717900
   \eta ^2+88628306866\bigg]         \nonumber\\&+&6 \left(332891836596 \sqrt{1-e^2}+8654689873\right)\Bigg\}
   \nonumber\\&+&40677120 \Big(891 e^8+28016 e^6+82736 e^4+43520 e^2+3072\Big) \log
   \left(\frac{x}{x_0}\frac{\left(1+\sqrt{1-e^2}\right)}{2\left(1-e^2\right)}\right) \Bigg]\,,
 \eea
 
 \bea
 \label{her}
 \dot{x}_{\rm{HT}}&=& \eta\, x^{13/2}\Bigg[\frac{256\pi}{5}\phi(e) + \left( \frac{256\pi}{1-e^2}\phi(e) + \frac{2}{3}\left(-\frac{17599\pi}{35}\psi_n(e) -\frac{2268\eta\pi}{5}\zeta_n(e)-\frac{788\pi e^2}{\left(1-e^2\right)^2}\varphi_e  \right)    \right)x  \nonumber\\&+& \frac{64}{18375}\Bigg(- 116761 \kappa +\Bigg(19600 \pi^2 - 59920 \gamma    - 
 59920 \log\left(\frac{4\,x^{3/2}}{x_0}\right)\Bigg)F(e)\Bigg)x^{3/2}\Bigg]\,.
 \eea

\noindent We have derived analytical relations for the various functions that appear in Equation~\eqref{her}:

\bea
\label{ke}
\phi(e) &=&\sum _{p=1}^{\infty } \frac{p^3}{4} \Bigg[\left(\left(-e^2-\frac{3}{e^2}+\frac{1}{e^4}+3\right) p^2+\frac{1}{3}-\frac{1}{e^2}+\frac{1}{e^4}\right) J_p(p e){}^2+\left(-3 e-\frac{4}{e^3}+\frac{7}{e}\right) p J_p'(p e) J_p(p e)\\\nonumber&+&\left(\left(e^2+\frac{1}{e^2}-2\right) p^2+\frac{1}{e^2}-1\right) J_p'(p e){}^2\Bigg]\,,\\
\label{kj}
\tilde{\phi}(e)&=&\sum _{p=1}^{\infty } \frac{p^2 \sqrt{1-e^2} }{2} \Bigg[\left(-\frac{2}{e^4}-1+\frac{3}{e^2}\right) p J_p(p e){}^2+\left(2 \left(e+\frac{1}{e^3}-\frac{2}{e}\right) p^2-\frac{1}{e}+\frac{2}{e^3}\right) J_p'(p e) J_p(p e)\\\nonumber&+&2 \left(1-\frac{1}{e^2}\right) p J_p'(p e){}^2\Bigg]\,,
\eea

\noindent where the notation \(\phi(e),\, \tilde{\phi}(e)\) has been chosen to coincide with that used in Ref.~\cite{Arun:2009PRD}, such that in Equation~\eqref{her}:

\beq
\label{another_phi}
\varphi_e=\frac{192}{985}\frac{\sqrt{1-e^2}}{e^2}\bigg[\sqrt{1-e^2}\phi(e) - \tilde{\phi}(e)\bigg]\,.
\eeq

\noindent In order to decrease the computational burden incurred by the numerical evaluation of Eqs.~\eqref{ke} and~\eqref{kj}, we have derived analytical expressions that reproduce the numerical results up to the twelfth significant figure in the range \(e\in[0,\,0.7]\). Setting \({\cal{E}}\equiv \left(1-e^2\right)^{-1/2}\), we can write our results as follows:

\bea
\label{ke_an}
\phi(e)& =& {\cal{E}}^{10}\Bigg\{1 + \frac{18970894028}{2649026657} e^2 + \frac{157473274}{30734301} e^4 +  \frac{48176523}{177473701} e^6 + \frac{9293260}{3542508891} e^8 - 
 \frac{5034498}{7491716851} e^{10} \nonumber\\&+& \frac{428340}{9958749469} e^{12}\Bigg\}\,,\\
 \label{kj_an}
\tilde{\phi}(e)& = &{\cal{E}}^{7}\Bigg\{1 +\frac{ 413137256}{136292703} e^2 + \frac{37570495}{98143337} e^4  - \frac{2640201}{993226448} e^6 - \frac{4679700}{6316712563} e^8 - 
 \frac{328675}{8674876481} e^{10}\Bigg\}\,,
\eea

\noindent In Eq.~\eqref{her} \(\gamma\) stands for Euler's constant. The functions \(F(e),\, \zeta_n,\, \phi_n\) given in Ref.~\cite{Arun:2009PRD} depend on the new functions we present in Equations~\eqref{psi_e}-~\eqref{kappa_t}. We have constructed these new analytical formulae ensuring that they reproduce the numerical data provided in Ref.~\cite{Arun:2009PRD} with an accuracy better than  $0.1\%$ for eccentricity values \(e\in[0\,,0.7]\):

\bea
\label{psi_e}
\psi(e) &=&  {\cal{E}}^{12}\left(1 - \frac{185}{21} e^2 - \frac{3733}{99} e^4 - \frac{1423}{104} e^6\right)\,,
\eea
\bea
\label{zed_e}
\zeta(e)&=& {\cal{E}}^{12}\left(1 + \frac{2095}{143} e^2 + \frac{1590}{59} e^4 + \frac{977}{113} e^6\right)\,,\eea
\bea
\label{k_e}
\kappa(e) &=&  {\cal{E}}^{14}\left(1 + \frac{1497}{79} e^2 + \frac{7021}{143} e^4 + \frac{997}{98} e^6 + \frac{463}{51} e^8 - \frac{3829}{120} e^{10} \right)\,,
\eea
\bea
\label{psi_t}
\tilde\psi(e) &=& {\cal{E}}^{9} \left(1 - \frac{2022}{305} e^2 - \frac{249}{26} e^4 - \frac{193}{239} e^6 + \frac{23}{43} e^8 - \frac{102}{463} e^{10}\right)\,,
\eea
\bea
\label{zed_t}
\tilde\zeta(e) &=& {\cal{E}}^{9} \left(1 + \frac{1563}{194} e^2 + \frac{1142}{193} e^4 + \frac{123}{281} e^6 - \frac{27}{328} e^8\right)\,,\nonumber\\
\eea
\bea
\label{kappa_t}
\tilde\kappa(e) &=& {\cal{E}}^{10}\left(1 + \frac{1789}{167} e^2 + \frac{5391}{340} e^4 + \frac{2150}{219} e^6 - \frac{1007}{320} e^8 + \frac{2588}{189} e^{10}\right)\,.
\eea

\noindent Regarding the evolution of the orbital eccentricity, we have used the 3PN accurate equations derived in Ref.~\cite{Arun:2009PRD}:

\beq
\label{ecc_ev}
M\dot{e}=  \dot{e}_{0\rm{PN}}x^4 + \dot{e}_{1\rm{PN}}x^5 + \dot{e}_{2\rm{PN}}x^6 + \dot{e}_{3\rm{PN}}x^7 +\, \dot{e}_{\rm{HT}}\,,
\eeq

\noindent where the \(e_{i\rm{PN}}\) with \(i=1,\,2,\,3\) are given by Eqs.~(6.19a,\, 6.19b), (C10, C11) of Ref.~\cite{Arun:2009PRD}, and the higher-order hereditary terms \(\dot{e}_{\rm{HT}}\) are given by~\cite{Arun:2009PRD}:

\bea
\label{ecc_ev_ht}
\dot{e}_{\rm HT}&=&
{32\over5}\,e\,\eta\,x^4\Biggl\{-\frac{985}{48}\pi\,x^{3/2}\,\varphi_{e}(e)+\pi\,x^{5/2}\bigg[
\frac{55691}{1344}\psi_{e}(e)+\frac{19067}{126}\eta\,\zeta_{e}(e)\bigg]\nonumber\\&+&x^3\Bigg[\left(\frac{89\,789\,209}{352\,800} -\frac{87\,419}{630}\ln2+\frac{78\,003}{560}\ln3\right)
\kappa_{e}(e)-\frac{769}{96}\left(
\frac{16}{3} \,\pi^2 -\frac{1712}{105}\,\gamma
-\frac{1712}{105}\ln\left[\frac{4\,x^{3/2}}{x_0}\right]\right)F_{e}(e)\Bigg]\Biggr\}\,.\nonumber\\
\eea

\noindent The functions \(\psi_{e}(e),\, \zeta_{e}(e),\,\kappa_{e}(e)\) and \(F_{e}(e)\) are provided in~\cite{Arun:2009PRD}, and also depend on the Eqs.~\eqref{psi_e}-~\eqref{kappa_t} we have derived in this paper.  We have verified that, as discussed in~\cite{Arun:2009PRD},  the arbitrary length scale \(x_0\) cancels out when adding 3PN terms for the orbital eccentricity evolution.

\section{}
\label{importance_hereditary}

\begin{figure*}[htp!]
\centerline{
\includegraphics[height=0.4\textwidth,  clip]{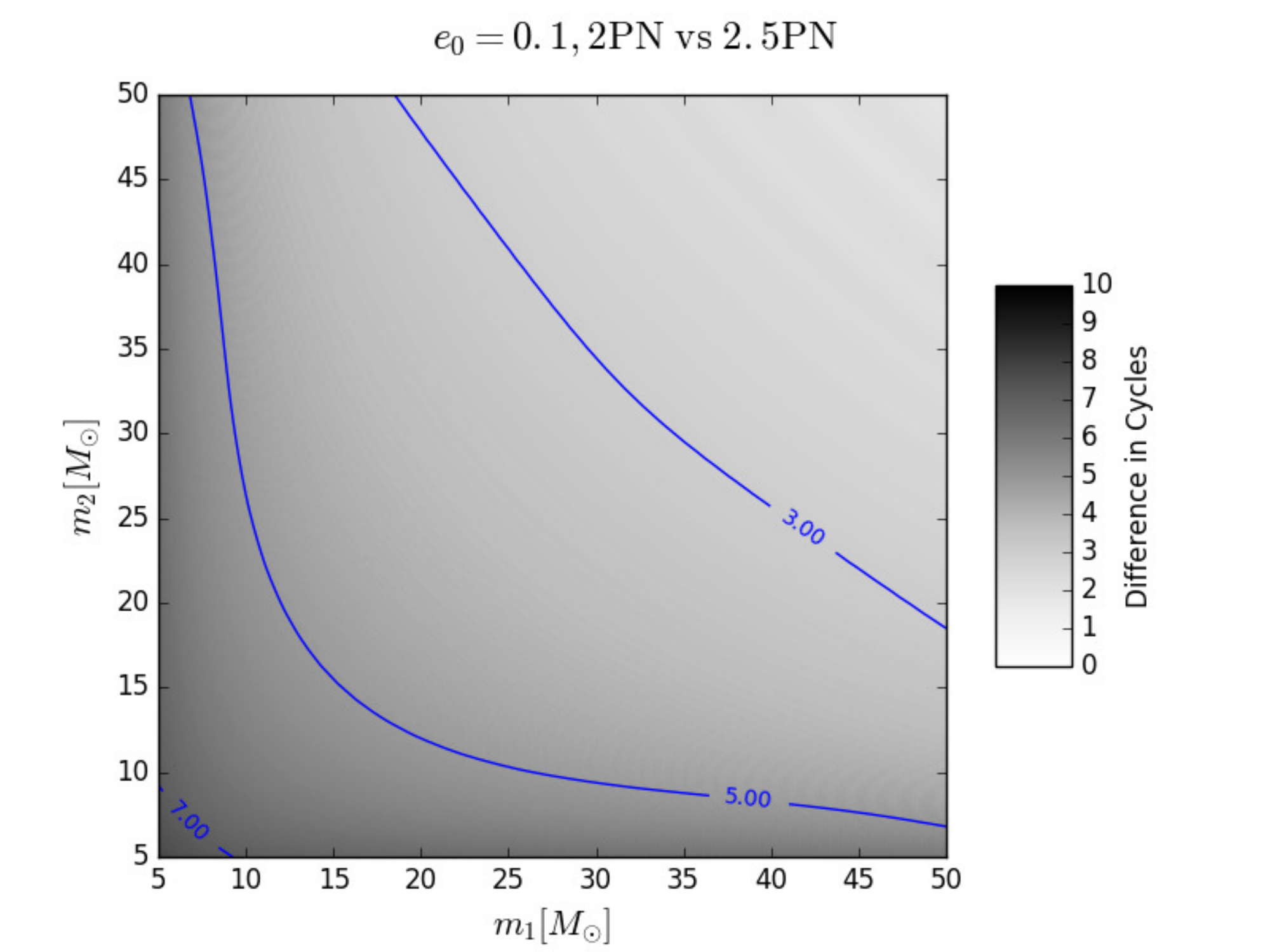}
\includegraphics[height=0.4\textwidth,  clip]{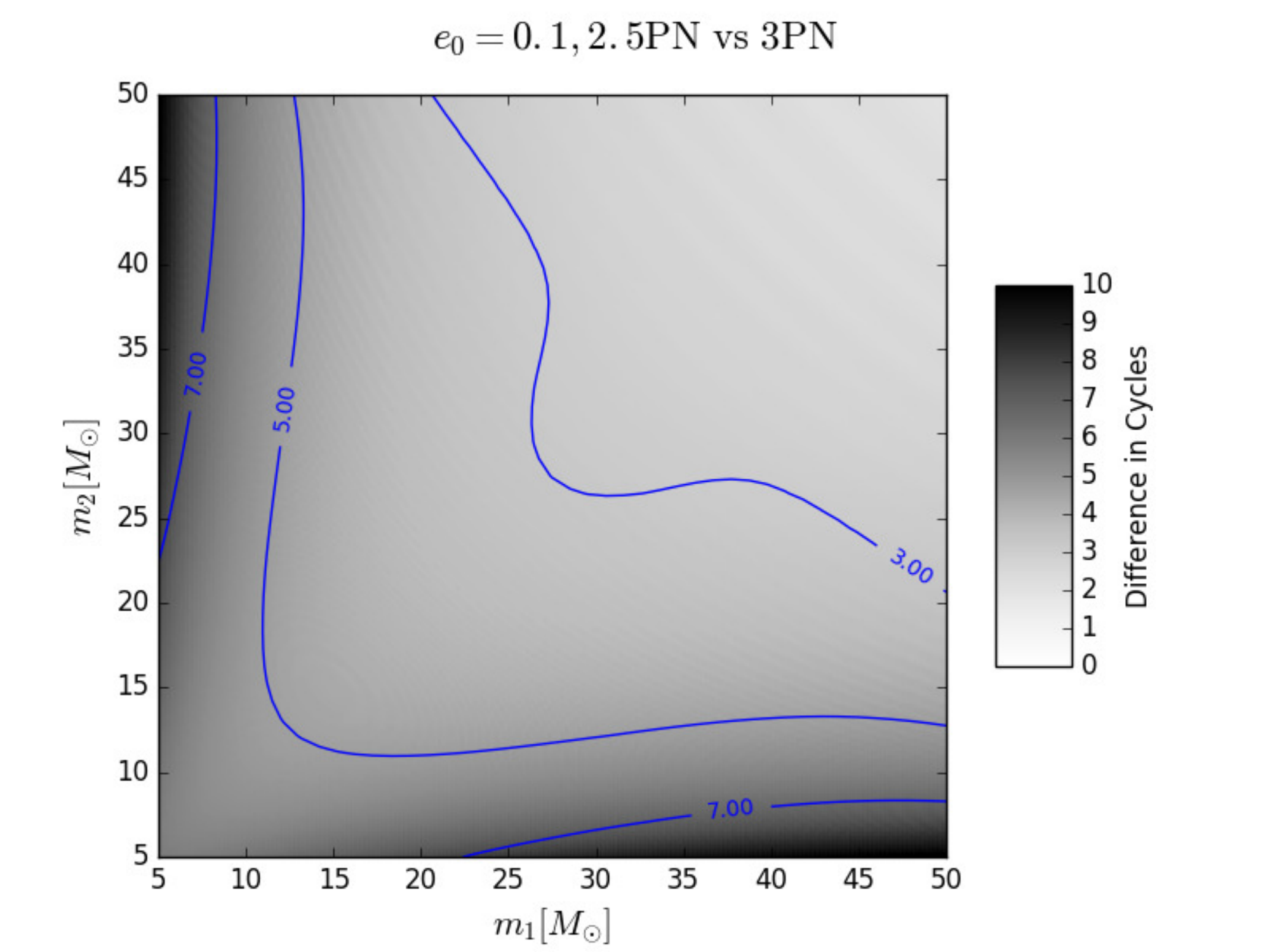}
}
\centerline{
\includegraphics[height=0.4\textwidth,  clip]{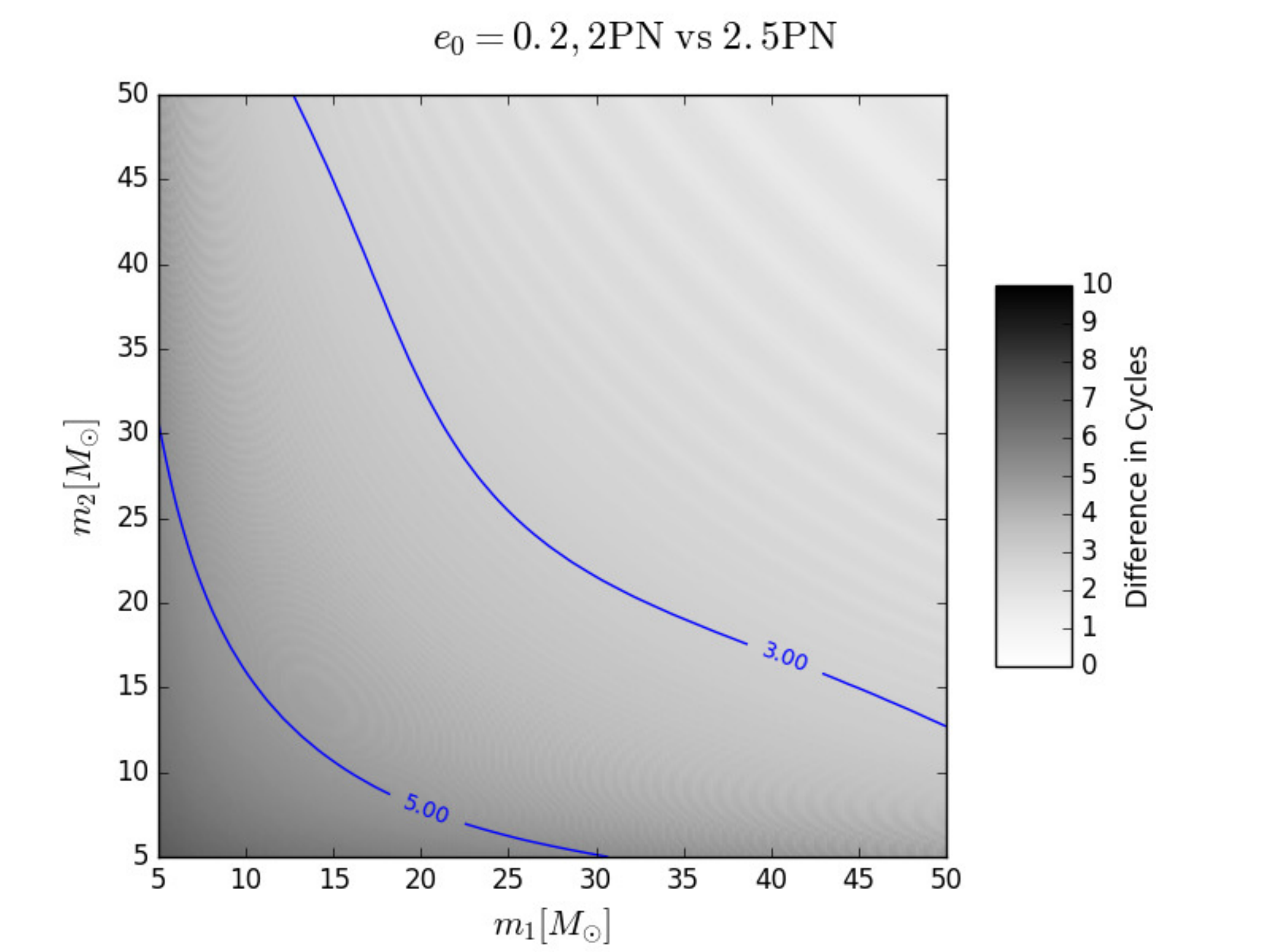}
\includegraphics[height=0.4\textwidth,  clip]{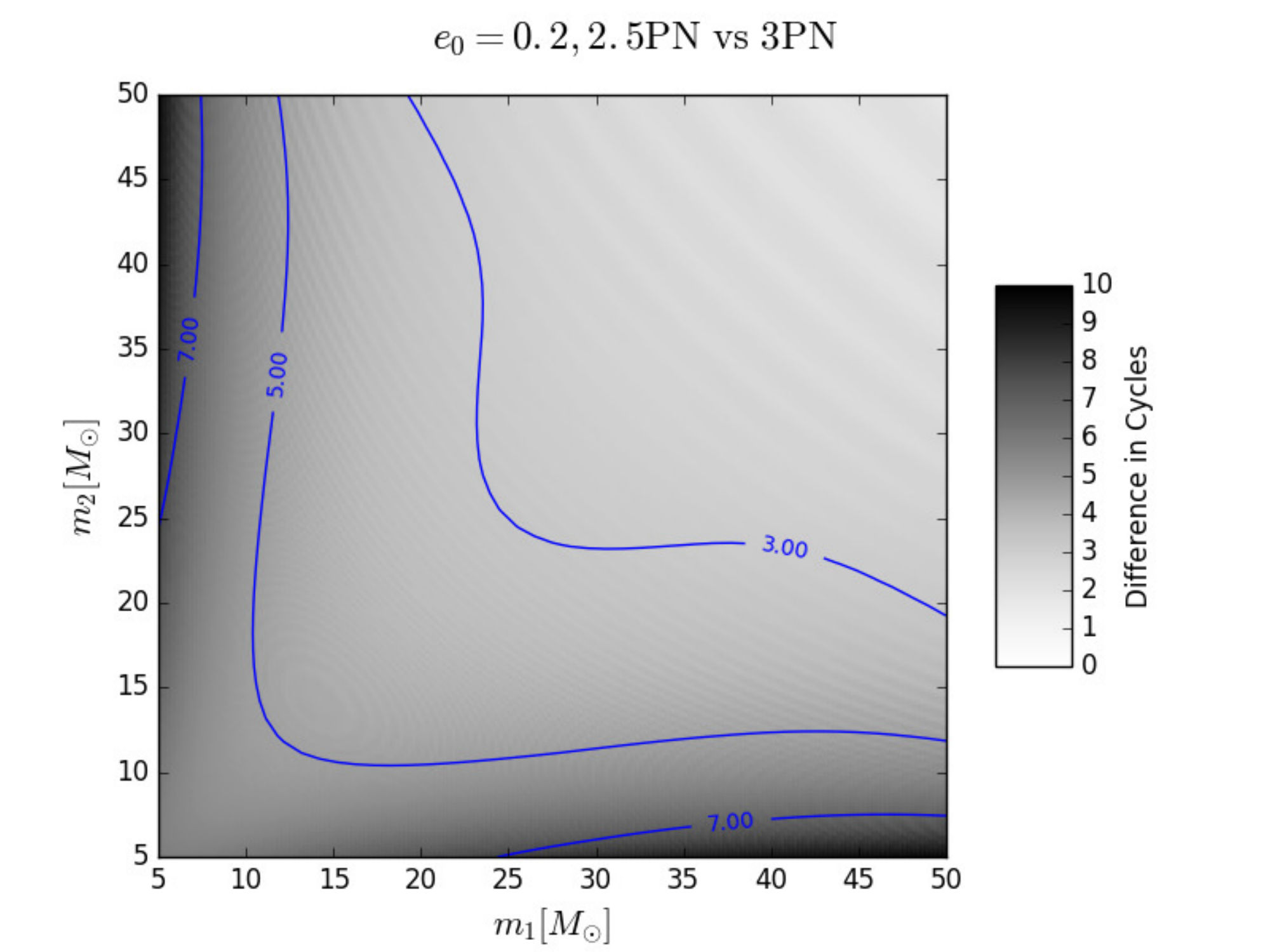}
}
\caption{Left column: difference in number of cycles using the pairwise comparison  \(\Delta {\cal{N}} = |{\cal{N}} (2.5{\rm PN}) - {\cal{N}} (2{\rm PN})|\). Right column: pairwise comparison between \(\Delta {\cal{N}} = |{\cal{N}} (3{\rm PN}) - {\cal{N}} (2.5{\rm PN})|\).}
\label{partial_her}
\end{figure*}

In this Appendix we quantify the importance of including higher-order hereditary contributions to describe the radiative dynamics of eccentric compact binary coalescence. As shown in Eqs.~\eqref{her} and ~\eqref{ecc_ev_ht} in Appendix~\ref{ap1}, the hereditary contributions we consider in this study correspond to non-linear corrections that enter the PN equations of motion at orders 1.5PN, 2.5PN and 3PN. It is important to emphasize that the hereditary contributions are gauge invariant at 1.5PN and 2.5PN orders. However, as we explicitly show in Eqs.~\eqref{her} and~\eqref{ecc_ev_ht}, the 3PN hereditary corrections have gauge dependent logarithms of the form \(\log(x/x_0)\), where \(x_0\) is a constant introduced to regularize ultra-violet divergences~\cite{Arun:2009PRD}. These pieces are of critical importance to provide a gauge-independent description of the radiative evolution of eccentric binaries up to 3PN order. This is because the instantaneous part of the fluxes also includes logarithms of the same type that are exactly cancelled by their 3PN hereditary counterparts. In summary: in order to provide a gauge-invariant description of the radiative dynamics of eccentric binaries at the highest PN order currently available, it is necessary to use the 3PN hereditary calculations we present in this article. 

In Figures~\ref{partial_her} and~\ref{partial_her_v2} we present results that shed light on the importance of including higher-order hereditary contributions. These results are obtained using 3PN accurate calculations for the equations of motion. On the other hand, we model the radiative piece using corrections up to 2PN, 2.5PN and 3PN order. We compute the number of cycles using Eq.~\eqref{cycles} for each case and then make pairwise comparisons, namely: \(\Delta {\cal{N}} = |{\cal{N}} (2.5{\rm PN}) - {\cal{N}} (2{\rm PN})|\) and \(\Delta {\cal{N}} = |{\cal{N}} (3{\rm PN}) - {\cal{N}} (2.5{\rm PN})|\). The case \(\Delta {\cal{N}} = |{\cal{N}} (3{\rm PN}) - {\cal{N}} (2.5{\rm PN})|\) is presented in Figure~\ref{cycles_15hz}. We use as the bare minimum a model that includes 2PN radiative corrections. We do this because this paper builds upon a model that already includes 2PN radiative corrections~\cite{Hinder:2010}, and the aim of this exercise is to assess the importance of including the new calculations presented in this work, namely at 2.5PN and 3PN order. 

Figures~\ref{partial_her},~\ref{partial_her_v2} and Figure~\ref{cycles_15hz} in the main text support the well known fact that eccentric PN expansions are characterized by poor convergence~\cite{improved,Levin:2011C}. In particular, we find that including up to 2.5PN hereditary corrections is definitely not a good strategy~\cite{improved}. On the other hand, incorporating both instantaneous and hereditary contributions to the highest order available is the preferred approach as discussed in the literature on the subject~\cite{improved, Arun:2009PRD}. This is expected because for the class of moderately eccentric sources considered in this work, once flux expressions are pushed to higher order, the size of eccentricity corrections will tend to diminish and will ultimately converge to the true inspiral evolution~\cite{improved}. Furthermore, recent work has shown that eccentric templates that only include 2PN radiative corrections will significantly hinder our ability to detect compact binaries with moderate values of eccentricity~\cite{Huerta:2014}. In different words, for astrophysically motivated sources that we can target with this model, it is important to ensure that the quasi-circular limit is reproduced at an acceptable level. This is the main motivation to compute 3PN accurate instantaneous and hereditary eccentricity corrections, and implement them in the IMR \(ax\) model.

\begin{figure*}[htp!]
\centerline{
\includegraphics[height=0.4\textwidth,  clip]{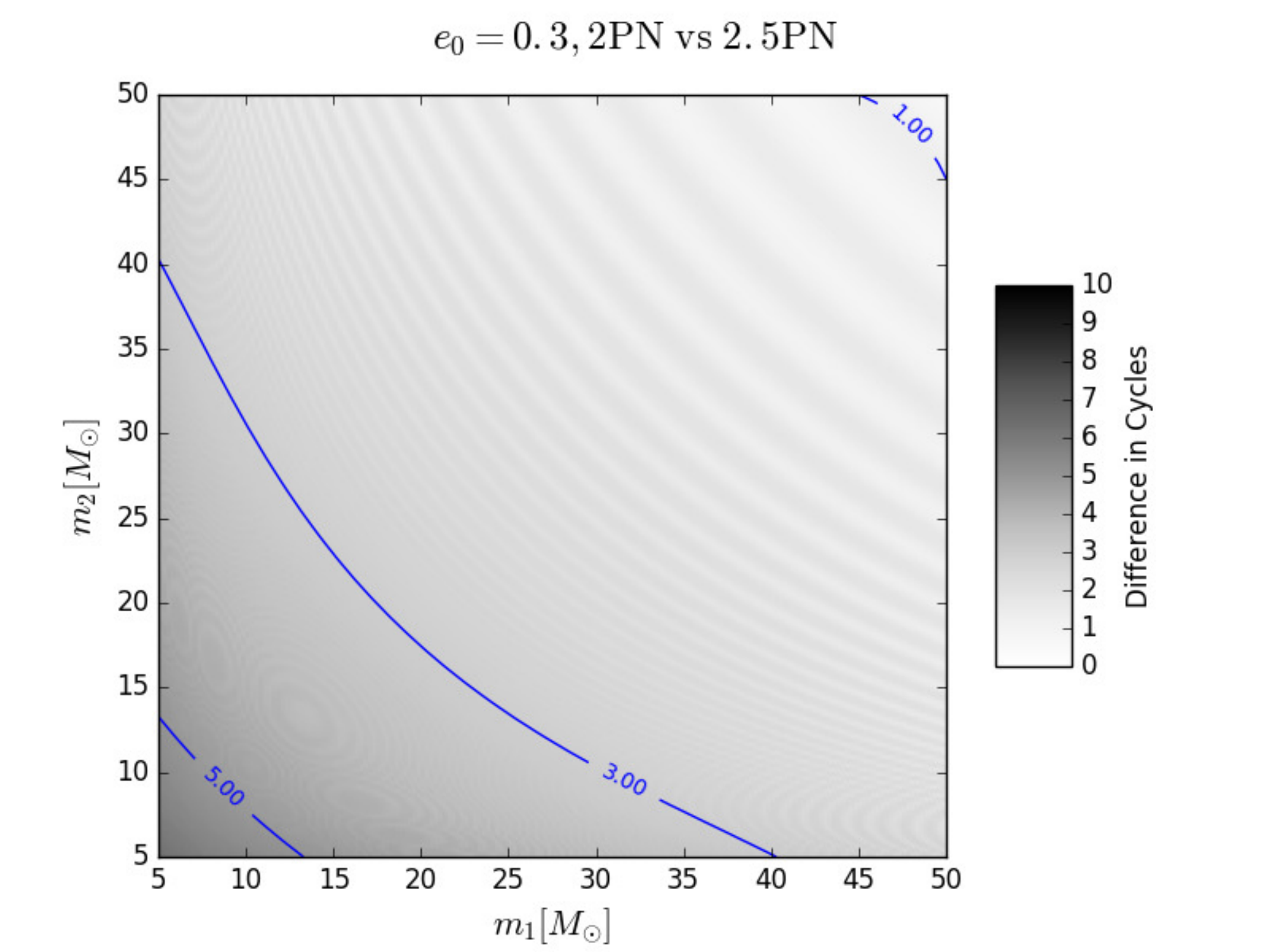}
\includegraphics[height=0.4\textwidth,  clip]{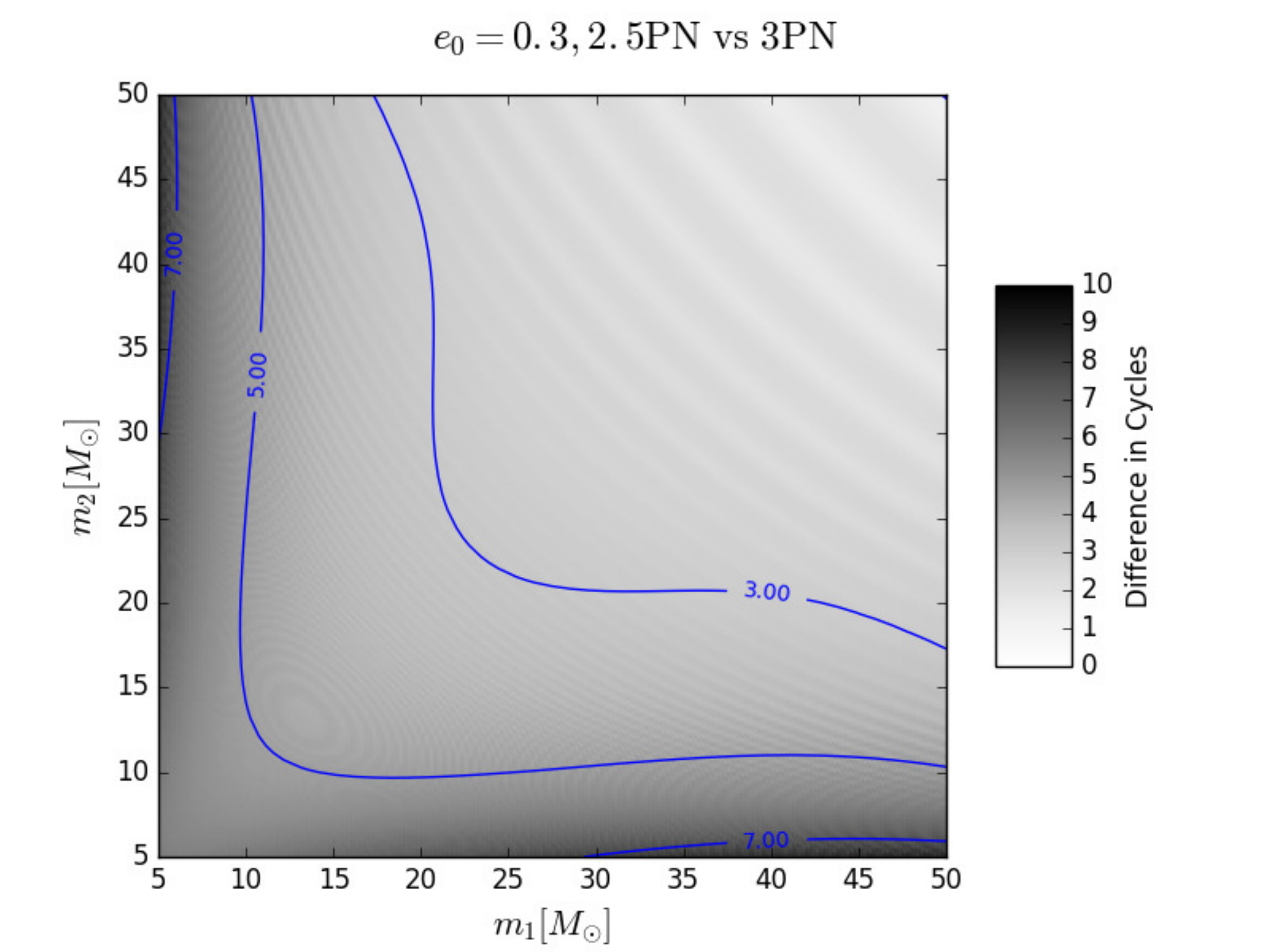}
}
\centerline{
\includegraphics[height=0.4\textwidth,  clip]{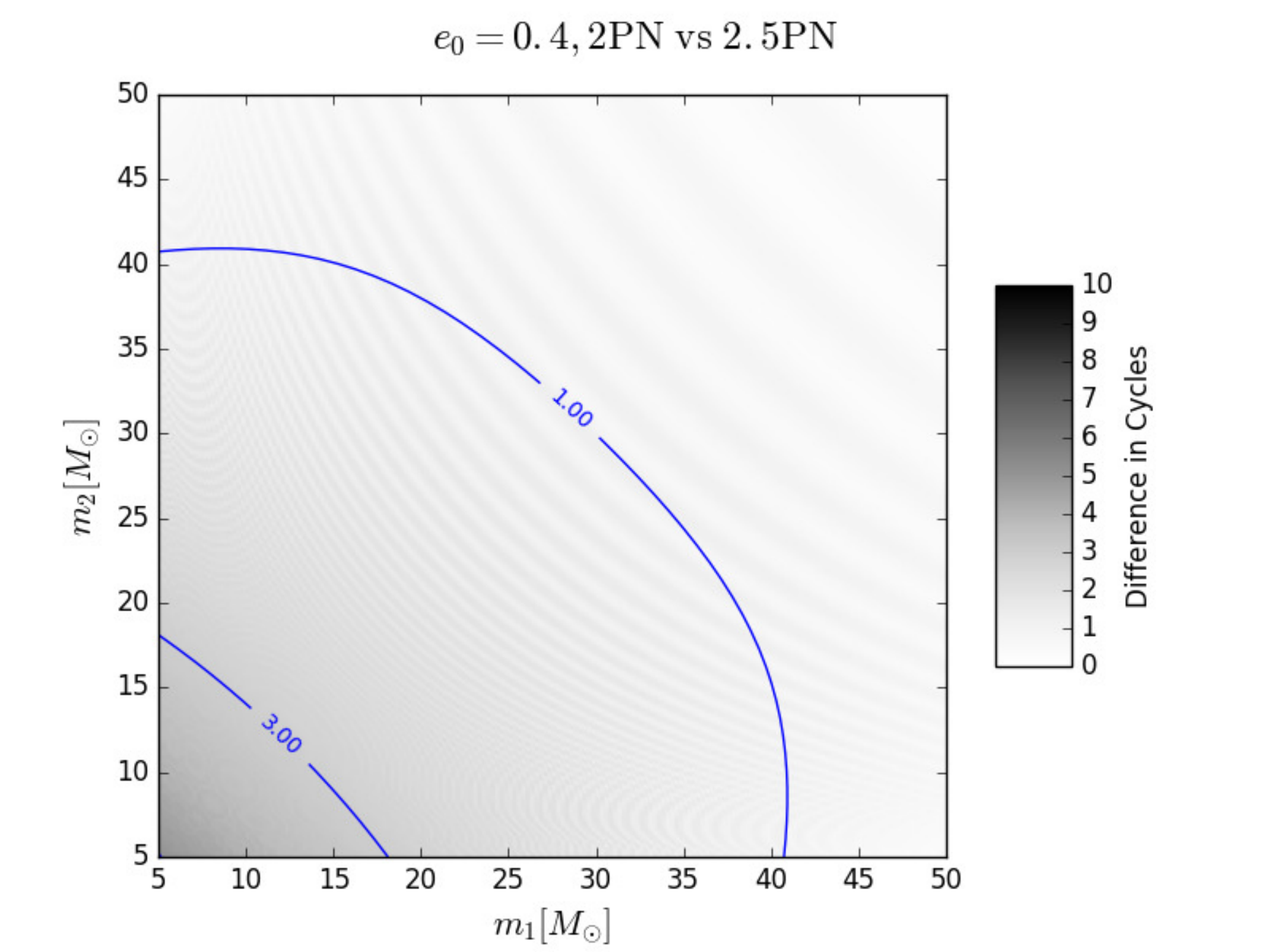}
\includegraphics[height=0.4\textwidth,  clip]{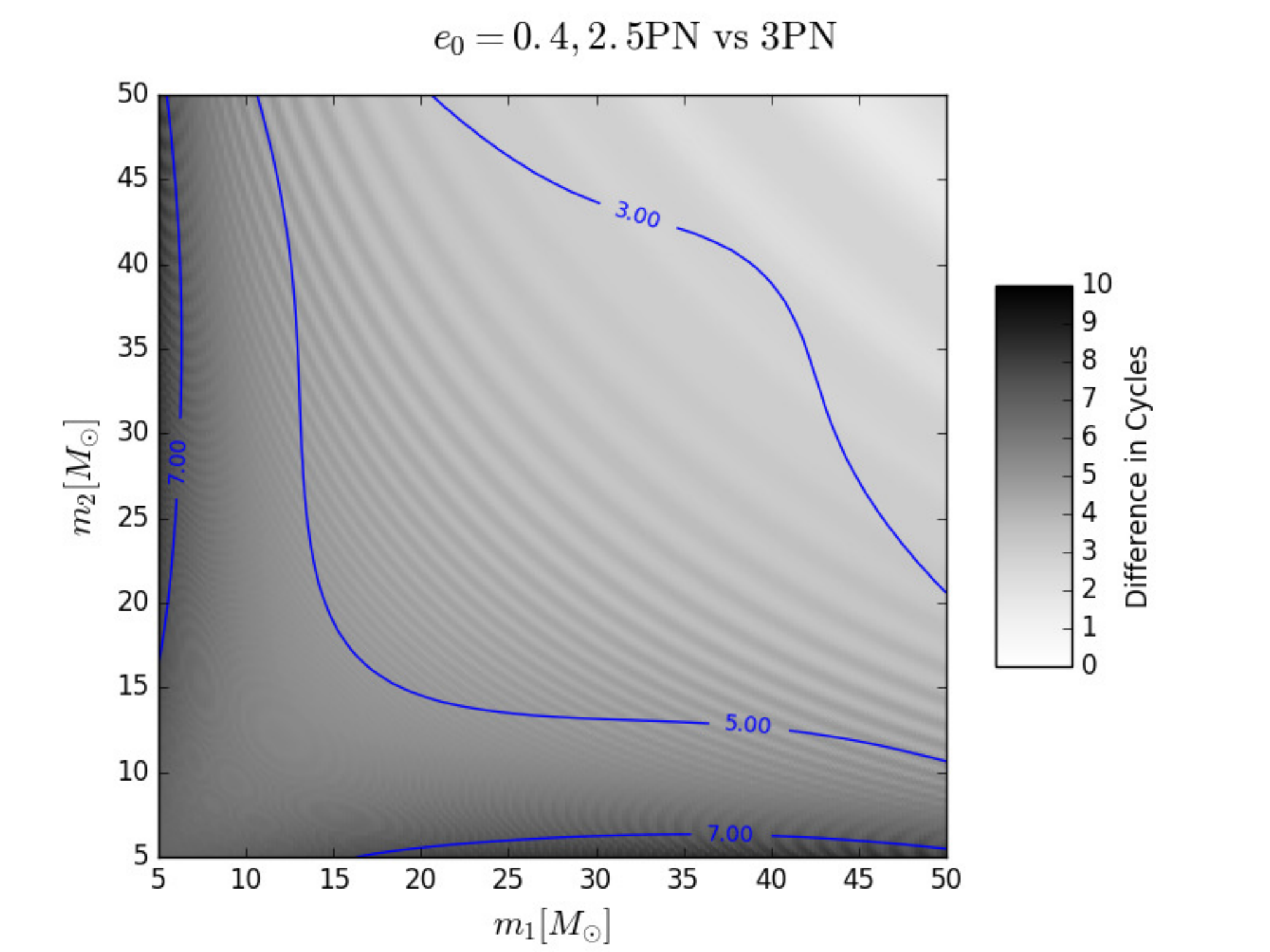}
}
\caption{As Figure~\ref{partial_her}, but now for \(e_0=\{0.3,\,0.4\}\).}
\label{partial_her_v2}
\end{figure*}

\noindent On the other hand, including only 3PN corrections in a template waveform is definitely not sufficient to reproduce the quasi-circular limit, in particular for asymmetric mass-ratio systems. To circumvent this problem we have amended the eccentric PN calculations with higher-order quasi-circular corrections using the self-force formalism and black hole perturbation theory. In Section~\ref{perform_eccentric}, we have shown that this approach renders a good description of moderately eccentric, comparable mass-ratio NR simulations. Looking forward, we plan to use a catalog of eccentric NR simulations to test and improve the accuracy of the IMR \(ax\) model across the parameter space detectable by aLIGO.

\section{}
\label{ho_pn_corrections}

In Section~\ref{ax_zero_e_limit} we presented a framework to increase the reliability of the \(ax\) model to describe binaries with asymmetric mass-ratios. This new prescription is given by Equation~\eqref{dxdtT4_byho}. We have derived the following coefficients for this expression:

\bea
\label{a4}
a_4&=& -5\, \eta\, \alpha_0 -\frac{97 \eta ^4}{3888}-\frac{18929389 \eta ^3}{435456}-\frac{3157 \pi ^2 \eta
   ^2}{144}+\frac{54732199 \eta ^2}{93312}-\frac{47468}{315} \eta  \log (x)-\frac{31495 \pi ^2 \eta
   }{8064}-\frac{856 \gamma  \eta }{315}\nonumber\\&+&\frac{59292668653 \eta }{838252800}-\frac{1712}{315} \eta  \log
   (2)+\frac{124741 \log (x)}{8820}-\frac{361 \pi ^2}{126}+\frac{124741 \gamma
   }{4410}+\frac{3959271176713}{25427001600}-\frac{47385 \log (3)}{1568}\nonumber\\&+&\frac{127751 \log (2)}{1470}\,,
\eea

\bea
\label{a92}   
 a_{9/2}&=&\frac{9731 \pi  \eta ^3}{1344}+\frac{42680611 \pi  \eta ^2}{145152}+\frac{205 \pi ^3 \eta }{6}-\frac{51438847
   \pi  \eta }{48384}-\frac{3424}{105} \pi  \log (x)-\frac{6848 \gamma  \pi }{105}+\frac{343801320119 \pi
   }{745113600}\nonumber\\&-&\frac{13696}{105} \pi  \log (2)  \,,
   \eea

 \bea
\label{a5}   
 a_5&=&\frac{155 \alpha_0 \eta ^2}{12}+\frac{1195 \alpha_0 \eta }{336}-6 \eta \alpha_1  -\frac{11567 \eta
   ^5}{62208}+\frac{51474823 \eta ^4}{1741824}+\frac{9799 \pi ^2 \eta ^3}{384}-\frac{9007776763 \eta
   ^3}{11757312}+\frac{216619}{189} \eta ^2 \log (x)\nonumber\\&-&\frac{126809 \pi ^2 \eta ^2}{3024}-\frac{2354 \gamma 
   \eta ^2}{945}+\frac{1362630004933 \eta ^2}{914457600}-\frac{4708}{945} \eta ^2 \log (2)+\frac{53963197
   \eta  \log (x)}{52920}+\frac{14555455 \pi ^2 \eta }{217728}\nonumber\\&+&\frac{3090781 \gamma  \eta
   }{26460}-\frac{847101477593593 \eta }{228843014400}-\frac{15795 \eta  \log (3)}{3136}+\frac{2105111 \eta 
   \log (2)}{8820}-\frac{5910592 \log (x)}{1964655}-\frac{21512 \pi ^2}{1701}\nonumber\\&-&\frac{11821184 \gamma
   }{1964655}+\frac{29619150939541789}{36248733480960}+\frac{616005 \log (3)}{3136}-\frac{107638990 \log
   (2)}{392931}  \,,
   \eea
   
   \bea
   \label{a112}
   a_{11/2} & =&-20 \pi  \eta \alpha_0  +\frac{49187 \pi  \eta ^4}{6048}-\frac{7030123 \pi  \eta ^3}{13608}-\frac{112955 \pi
   ^3 \eta ^2}{576}+\frac{1760705531 \pi  \eta ^2}{290304}-\frac{189872}{315} \pi  \eta  \log (x)\nonumber\\&-&\frac{26035
   \pi ^3 \eta }{16128}-\frac{3424 \gamma  \pi  \eta }{315}-\frac{2437749208561 \pi  \eta
   }{4470681600}-\frac{6848}{315} \pi  \eta  \log (2)+\frac{311233 \pi  \log (x)}{11760}+\frac{311233 \gamma 
   \pi }{5880}\nonumber\\&+&\frac{91347297344213 \pi }{81366405120}-\frac{142155}{784} \pi  \log (3)+\frac{5069891 \pi 
   \log (2)}{17640}\,,
   \eea

   \bea
   \label{s6}
   a_6 & = & -\frac{535 \alpha_0 \eta ^3}{36}+\frac{7295 \alpha_0 \eta ^2}{336}-\frac{248065 \alpha_0 \eta }{4536}+\frac{31
   \alpha_1 \eta ^2}{2}+\frac{239 \alpha_1 \eta }{56}-7 \alpha_2 \eta -7 \alpha_3 \eta  \log (x)-\alpha_3 \eta
   -\frac{155377 \eta ^6}{1679616}\nonumber\\&-&\frac{152154269 \eta ^5}{10450944}-\frac{1039145 \pi ^2 \eta
   ^4}{62208}+\frac{76527233921 \eta ^4}{94058496}-\frac{41026693 \eta ^3 \log (x)}{17010}+\frac{55082725 \pi ^2 \eta
   ^3}{217728}-\frac{2033 \gamma  \eta ^3}{1701}\nonumber\\&-&\frac{56909847373567 \eta ^3}{7242504192}-\frac{4066 \eta ^3 \log
   (2)}{1701}-\frac{271237829 \eta ^2 \log (x)}{127008}+\frac{92455 \pi ^4 \eta ^2}{1152}-\frac{4061971769 \pi ^2 \eta
   ^2}{870912}-\frac{21169753 \gamma  \eta ^2}{317520}\nonumber\\&+&\frac{3840832667727673 \eta ^2}{55477094400}-\frac{57915 \eta ^2
   \log (3)}{12544}-\frac{2724535 \eta ^2 \log (2)}{21168}-\frac{4387}{63} \pi ^2 \eta  \log (x)-\frac{12030840839 \eta 
   \log (x)}{37721376}+\frac{410 \pi ^4 \eta }{9}\nonumber\\&-&\frac{8774}{63} \gamma  \pi ^2 \eta +\frac{206470485307 \pi ^2 \eta
   }{1005903360}+\frac{362623282541 \gamma  \eta }{94303440}-\frac{12413297162366594971 \eta
   }{271865501107200}+\frac{3016845 \eta  \log (3)}{12544}\nonumber\\&-&\frac{17548}{63} \pi ^2 \eta  \log (2)+\frac{701463800861 \eta 
   \log (2)}{94303440}+\frac{366368 \log ^2(x)}{11025}+\frac{2930944 \log (2) \log (x)}{11025}-\frac{13696}{315} \pi ^2
   \log (x)\nonumber\\&+&\frac{1465472 \gamma  \log (x)}{11025}-\frac{155359670313691 \log (x)}{157329572400}-\frac{27392\, {\rm Zeta}
   (3)}{105}-\frac{256 \pi ^4}{45}-\frac{27392 \gamma  \pi ^2}{315}+\frac{1414520047 \pi ^2}{2619540}+\frac{1465472 \gamma
   ^2}{11025}\nonumber\\&-&\frac{155359670313691 \gamma
   }{78664786200}+\frac{1867705968412371074441833}{154211174411374080000}+\frac{5861888 \log
   ^2(2)}{11025}-\frac{37744140625 \log (5)}{260941824}-\frac{63722699919 \log (3)}{112752640}\nonumber\\&-&\frac{54784}{315} \pi ^2
   \log (2)+\frac{5861888 \gamma  \log (2)}{11025}-\frac{206962178724547 \log (2)}{78664786200}\,,\nonumber\\
   \eea
     
 \noindent where \({\rm Zeta}(3)\) stands for the Riemann zeta function with the given argument, and the coefficients \(\alpha_i\) with \(i=0,\,1,\,2,\,3\) are given by~\cite{barus}:
 
 \bea
 \alpha_0 &=& 153.8803\,,\\
 \alpha_1 &=& -55.13\,,\\
 \alpha_2 &=& 588\,,\\
 \alpha_3 &=& -1144\,.
 \eea
 
\end{widetext}

\clearpage

\bibliography{references}

\end{document}